\pdfoutput=1
\documentclass{aa}
\usepackage{pifont}
\usepackage{amsmath}
\usepackage{graphicx}
\usepackage{txfonts}
\usepackage{url}
\usepackage{subfigure}
\usepackage{longtable}
\usepackage{lscape}
\usepackage{natbib}
\usepackage[colorlinks,linkcolor=red,anchorcolor=green,citecolor=blue]{hyperref}

\usepackage[normalem]{ulem}

\newcommand{\HII}{H {\small{II}} }
\newcommand{\kms}{{\rm km~s}^{-1}}
\newcommand{\Msun} {M_\sun}

\newcommand{\mjyb}{{\rm mJy~beam}^{-1}}

\newcommand{\jyb}{{\rm Jy~beam}^{-1}}
\newcommand{\mjybkms}{{\rm mJy~beam}^{-1}{\rm km~s}^{-1}}

\begin{document}

   \title{Probing the initial conditions of high-mass star formation}
   \subtitle{III. Fragmentation and triggered star formation}
   \author{Chuan-Peng Zhang
          \inst{1,2,3}
          \and
          Timea Csengeri\inst{3}
          \and
          Friedrich Wyrowski\inst{3}
          \and
          Guang-Xing Li\inst{4,5}
          \and
          Thushara Pillai\inst{3}
          \and
          Karl M. Menten\inst{3}
          \and
          Jennifer Hatchell\inst{6}
          \and
          Mark A. Thompson\inst{7}
          \and
          Michele R. Pestalozzi\inst{8}
          }

    \institute{National Astronomical Observatories, Chinese Academy of Sciences, 100101 Beijing, P.R. China\\
    \email{cpzhang@nao.cas.cn}
    \and
    Max-Planck-Institut f\"ur Astronomie, K\"onigstuhl 17, D-69117 Heidelberg, Germany
    \and
    Max-Planck-Institut f\"ur Radioastronomie, Auf dem H\"ugel 69, D-53121 Bonn, Germany
    \and
    University Observatory Munich, Scheinerstrasse 1, D-81679 M\"unchen, Germany
    \and
    South-Western Institute for Astronomy Research, Yunnan University, Kunming, 650500 Yunnan, P.R. China
    \and
    School of Physics, University of Exeter, Stocker Rd, Exeter EX4 4QL, United Kingdom
    \and
    Centre for Astrophysics Research, University of Hertfordshire, College Lane, Hatfield, Herts AL10 9AB, United Kingdom
    \and
    Istituto di Fisica dello Spazio Interplanetario - INAF, via Fosso del Cavaliere 100, 00133, Roma, Italy
             }

   \date{Received XXX, XXX; accepted XXX, XXX}


   \abstract
   {Fragmentation and feedback are two important processes during the early phases of star formation.}
   {Massive clumps tend to fragment into clusters of cores and condensations, some of which form high-mass stars. In this work, we study the structure of massive clumps at different scales, analyze the fragmentation process, and investigate the possibility that star formation is triggered by nearby \HII regions.}
   {We present a high angular resolution study of a sample of massive proto-cluster clumps G18.17, G18.21, G23.97N, G23.98, G23.44, G23.97S, G25.38, and G25.71. Combining infrared data at 4.5, 8.0, 24, and 70\,$\mu$m, we use few-arcsecond resolution radio- and millimeter interferometric data taken at 1.3\,cm, 3.5\,mm, 1.3\,mm, and 870\,$\mu$m to study their fragmentation and evolution. Our sample is unique in the sense that  all the clumps have neighboring \HII regions. Taking advantage of that, we test triggered star formation using a novel method where we study the alignment of the centres of mass traced by dust emission at multiple scales.}
   {The eight massive clumps, identified based on single dish observations, have masses ranging from 228 to 2279\,$\Msun$ within an effective radius of $R_{\rm eff}$ $\sim$ 0.5\,pc. We detect compact structures towards six out of the eight clumps. The brightest compact structures within infrared bright clumps are typically associated with embedded compact radio continuum sources. The smaller scale structures of $R_{\rm eff}$ $\sim$ 0.02\,pc observed within each clump are mostly gravitationally bound and massive enough to form at least a B3-B0 type star. Many condensations have masses larger than 8\,$\Msun$ at small scale of $R_{\rm eff}$ $\sim$ 0.02\,pc. We find that the two lowest mass and lowest surface density infrared quiet clumps with $<300\,\Msun$ do not host any compact sources, calling into question their ability to form high-mass stars. Although the clumps are mostly infrared quiet, the dynamical movements are active at clump scale ($\sim$ 1\,pc).}
   {We studied the spatial distribution of the gas conditions detected at different scales. For some sources we find hints of external triggering, whereas for others we find no significant pattern that indicates triggering is dynamically unimportant. This probably indicates that the different clumps go through different evolutionary paths. In this respect, studies with larger samples are highly desired.}


   \maketitle
%

\section{Introduction}    
\label{sect_intro}

High-mass stars are well-known to form as part of stellar clusters \citep[e.g.,][]{Motte2018}, in which they are greatly outnumbered by their low-mass cousins. Thus, the formation mechanism for individual high-mass stars is closely related to the formation and evolution of star clusters, which begins with the gravitational collapse of dense gas clumps of a parsec scale.

Observing the density structure of collapsing dense clumps is very important for our understanding of the star formation process. The radial density profile holds information concerning how collapse occurs \citep{Larson1969,Penston1969,Shu1977,Vazquez2009,Ballesteros2011,Girichidis2014,Naranjo2015,Donkov2017,Li2018}, and the masses and densities of the fragments hold clues to how high-mass stars form \citep{McKee2002,McKee2003,Myers1998,Padoan2002,Bonnell2001,Klessen2001,Bonnell2002,Bonnell2004}. The observed fragments might also exhibit interesting spatial patterns: the fragments within a cluster often align along filamentary structures, e.g. G29.96e \citep{Pillai2011}, G28.34+p1 \citep{Zhang2009} and G11.11-p1 \citep{Wang2011,Wang2014}. In short, constraining the number of fragments, their masses, and densities is fundamental to understanding how high-mass stars form.

\begin{table*}[ht]
\caption{Parameters of PdBI observations.}
\label{tab_pdbi} \centering  \footnotesize 
\setlength{\tabcolsep}{1.6mm}{
\begin{tabular}{lcccccccc}
\hline \hline
Source &  ATLASGAL name	&	\multicolumn{2}{c}{Phase center (J2000)} 	&	\multicolumn{2}{c}{3.5mm beam \& rms}
&	\multicolumn{2}{c}{1.3mm beam \& rms}	 & Observational date  \\
&	&	h~~m~~s~~	&	$^\circ~~'~~''~~$	& \multicolumn{2}{r}{$''\,\times\,''$; $^\circ$ ~~~~ mJy/beam}	&	\multicolumn{2}{r}{$''\,\times\,''$; $^\circ$ ~~~~ mJy/beam}  &  \\
\hline
\multicolumn{2}{c}{CD configuration observations} \\
G18.17	&	G018.1750-0.2985	&	18 25 07.534	&	-13 14 32.75	&	4.75$\,\times\,$2.48;
21.27	&	0.23	&	2.46$\,\times\,$0.97;	19.98	&	1.67  & Mar.25 -- Apr.5, 2005\\
G18.21	&	G018.2150-0.3419	&	18 25 21.558	&	-13 13 39.56	&	4.94$\,\times\,$2.72;
18.23	&	0.20	&	2.47$\,\times\,$0.99;	19.15	&	1.81  & Mar.25 -- Apr.5, 2005\\
G23.97N	&	G023.9655+0.1382	&	18 34 28.833	&	-07 54 31.76	&	4.10$\,\times\,$2.82;
21.58	&	0.32	&	1.85$\,\times\,$1.07;	20.51	&	1.62  & Mar.28 -- Apr.6, 2005\\
G23.98	&	G023.9790+0.1498	&	18 34 27.823	&	-07 53 28.76	&	4.23$\,\times\,$2.95;
19.05	&	0.22	&	1.85$\,\times\,$1.09;	20.89	&	1.65  & Mar.28 -- Apr.6, 2005\\
G23.44	&	G023.4366-0.1828	&	18 34 39.253	&	-08 31 36.23	&	4.34$\,\times\,$2.91;
20.48	&	0.25	&	1.89$\,\times\,$1.05;	19.83	&	2.17	  & Mar.28 -- Apr.6, 2005\\
G23.97S	&	G023.9647-0.1094	&	18 35 22.160	&	-08 01 26.53	&	4.24$\,\times\,$2.68;
21.68	&	0.23	&	1.93$\,\times\,$1.01;	20.63	&	1.43	  & Mar.29 -- Apr.12, 2005\\
G25.38	&	G025.3826-0.1476	&	18 38 08.108	&	-06 46 54.93	&	4.18$\,\times\,$2.82;
21.85	&	0.39	&	1.76$\,\times\,$1.03;	21.27	&	1.34	  & Mar.29 -- Apr.12, 2005\\
G25.71	&	G025.7099+0.0448	&	18 38 03.184	&	-06 24 14.30	&	4.20$\,\times\,$2.83;
22.82	&	0.26	&	1.89$\,\times\,$1.05;	22.19	&	1.34  & Mar.29 -- Apr.12, 2005\\
\hline
\multicolumn{2}{c}{BCD configuration observations} \\
G23.44	&	G023.4366-0.1828	&	18 34 39.253	&	-08 31 36.23	&	2.20$\,\times\,$1.54;
19.62	&	0.13	&	0.82$\,\times\,$0.57;	17.32	&	0.75  & Mar.14, Mar.16, 2006\\
G23.97S	&	G023.9647-0.1094	&	18 35 22.160	&	-08 01 26.53	&	2.16$\,\times\,$1.53;
22.87	&	0.13	&	0.83$\,\times\,$0.59;	26.90	&	0.64  & Mar.14, Mar.16, 2006\\
G25.38	&	G025.3826-0.1476	&	18 38 08.108	&	-06 46 54.93	&	2.17$\,\times\,$1.55;
22.66	&	0.20	&	0.78$\,\times\,$0.60;	23.19	&	0.66	  & Feb.27, Mar.17, 2006\\
G25.71	&	G025.7099+0.0448	&	18 38 03.184	&	-06 24 14.30	&	2.21$\,\times\,$1.54;
24.73	&	0.16	&	0.83$\,\times\,$0.58;	24.19	&	0.66	  & Feb.27, Mar.17, 2006\\
\hline
\end{tabular}}
\end{table*}

Triggered star formation is still an open question, since the role and the mechanism of external influence on the collapse process is unclear. Several potential mechanisms exist, including ``Collect and collapse'' \citep{whit1994,dale2007}, and ``radiatively driven implosion'' \citep{Sandford1982}. In ``collect and collapse'', during the expansion of an \HII region, the diffuse gas surrounding it is collected into dense structures which then collapse \citep{N131,Zhang2016}. In ``radiatively driven implosion'', pre-existing overdensities in the molecular gas are enhanced when an ionization front sweeps away the less dense gas and begins to compress the overdensities from all sides, inducing collapse \citep{Henney2009}.  \HII regions interact with their neighboring molecular clouds by pushing away the lower density material faster than the higher density cloud cores embedded inside \citep{s51}. This may lead to bright rims and pillars \citep{Klein1980,Sugitani1989,Elmegreen2011}. However, it is still unclear how important these processes are for the gravitational collapse of the clumps and the subsequent star formation. In this paper, we address this question using a novel method where we study the alignment of the centre of mass traced by dust emission at multiple scales.

\citet{Kauffmann2010} defined an empirical threshold of $m(r) > 870\,{\Msun}\,(r/{\rm pc})^{1.33}$ for massive star formation, where $m(r)$ is the clump mass and the effective radius $r = {\rm FWHM} / (2\sqrt{\rm ln2})$. In our previous work \citep{Zhang2017b}, we have presented the \textit{mass-size} relations for all clumps, cores, and condensations at 3.5\,mm, 1.3\,mm, 870\,$\mu$m, 850\,$\mu$m, and 450\,$\mu$m. We found that our samples are potential intermediate- or high-mass star-forming candidates. We fit these parameters using a power law that yields $M \propto r^{1.68}$. \citet{Li2017} predict that if the ambient turbulence is characterised by a constant energy dissipation rate $\epsilon$, gravitationally bound structures will follow $M \approx \epsilon^{2/3} \eta^{-2/3}G^{-1}r^{5/3}$, where $\eta$ is a parameter for turbulent dissipation, and is close to unity, $G$ is the gravitational constant, and $r$ is the size. Thus the \textit{mass-size} scaling is determined by the energy dissipation rate $\epsilon$. When $\epsilon$ is a constant through the medium, $M\propto r^{5/3}\propto r^{1.67}$. Therefore, our observational results support a scenario where molecular gas in the high-mass star formation regions is supported by turbulence with an almost constant energy dissipation rate, and gas fragments like clumps and cores are structures which are dense enough to be dynamically detached from the ambient medium. A more detailed discussion is presented in \citet{Li2017} and \citet{Zhang2017b}.

\begin{table*}[ht]
\caption{Parameters of VLA observations.}
\label{tab_vla} \centering \footnotesize
\begin{tabular}{lcccc}
\hline \hline
Source &	\multicolumn{2}{c}{Phase center (J2000)} 
&	\multicolumn{2}{c}{NH$_3$ (1,1) beam \& rms}	   \\
&	h~~m~~s~~	&	$^\circ~~'~~''~~$	&	\multicolumn{2}{r}{$''\,\times\,''$; $^\circ$ ~~~~ mJy/beam}   \\
\hline
\multicolumn{2}{c}{D configuration} \\
G18.17	&	18 25 07.534	&	-13 14 32.75	&	4.61$\,\times\,$3.24;   13.50	&	$\sim$3.3  \\
G18.21	&	18 25 21.558	&	-13 13 39.56	&	4.64$\,\times\,$3.24;   15.06	&	$\sim$3.6  \\
G23.97N	&	18 34 28.833	&	-07 54 31.76	&	4.03$\,\times\,$3.34;   -5.64	&	$\sim$3.2  \\
G23.98	&	18 34 27.823	&	-07 53 28.76	&	4.03$\,\times\,$3.33;   -7.06	&	$\sim$3.3  \\
G23.44	&	18 34 39.253	&	-08 31 36.23 	&	4.18$\,\times\,$3.42;   -7.07	&	$\sim$3.6  \\ 
G23.97S	&	18 35 22.160	&	-08 01 26.53	&	4.04$\,\times\,$3.34;   -3.95	&	$\sim$2.8  \\
G25.38	&	18 38 08.108	&	-06 46 54.93	&	4.10$\,\times\,$3.41;   -7.48	&	$\sim$2.6  \\
G25.71	&	18 38 03.184	&	-06 24 14.30	&	4.06$\,\times\,$3.43;   -6.29	&	$\sim$2.6  \\
\hline
\end{tabular}
\end{table*}

\begin{figure*}
\centering
\subfigure[]{\includegraphics[width=0.65\textwidth, angle=0]{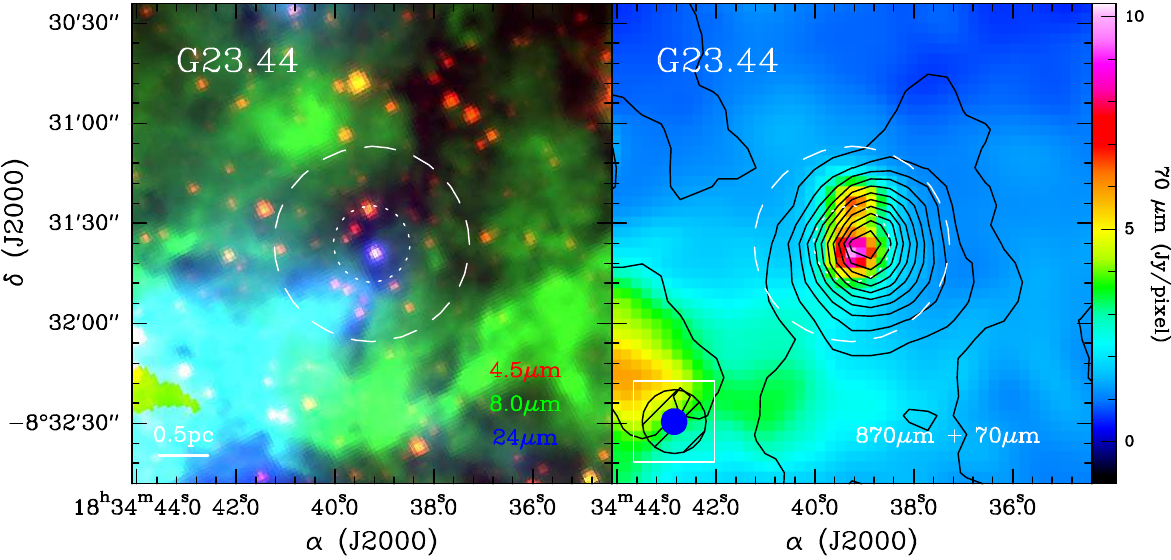}}
\subfigure[]{\includegraphics[width=0.33\textwidth, angle=0]{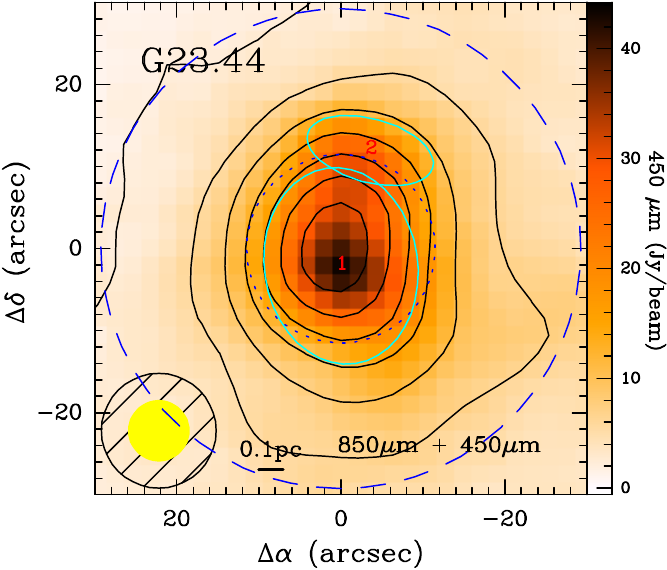}}
\subfigure[]{\includegraphics[width=0.33\textwidth, angle=0]{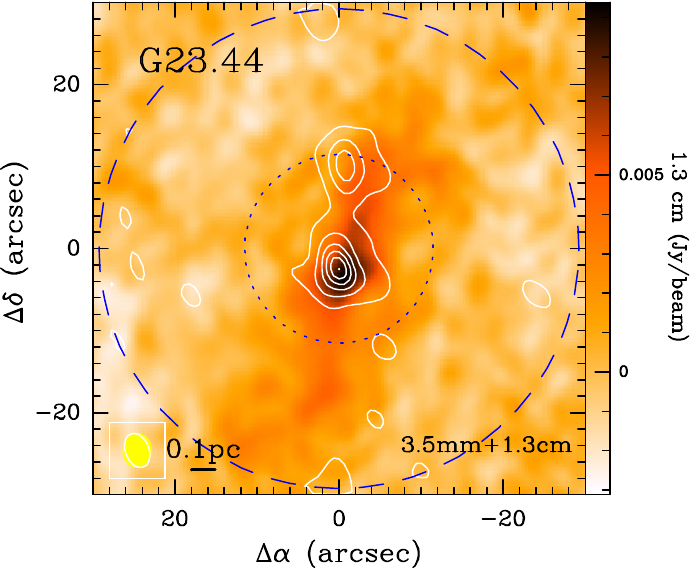}}
\subfigure[]{\includegraphics[width=0.33\textwidth, angle=0]{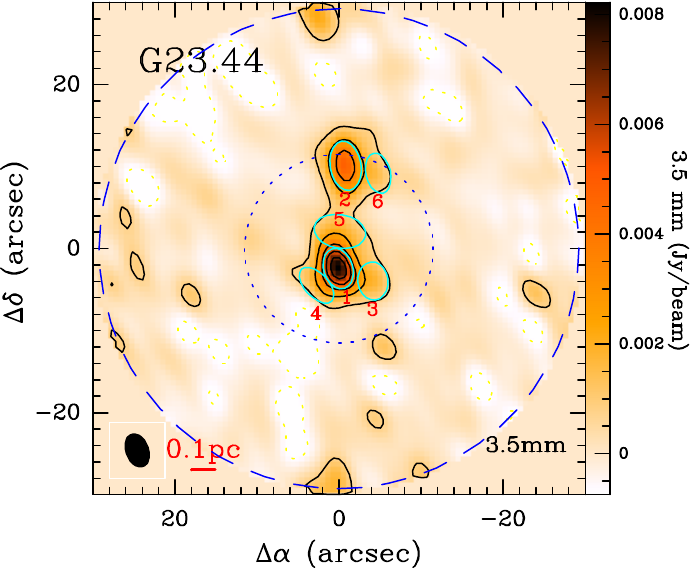}}
\subfigure[]{\includegraphics[width=0.33\textwidth, angle=0]{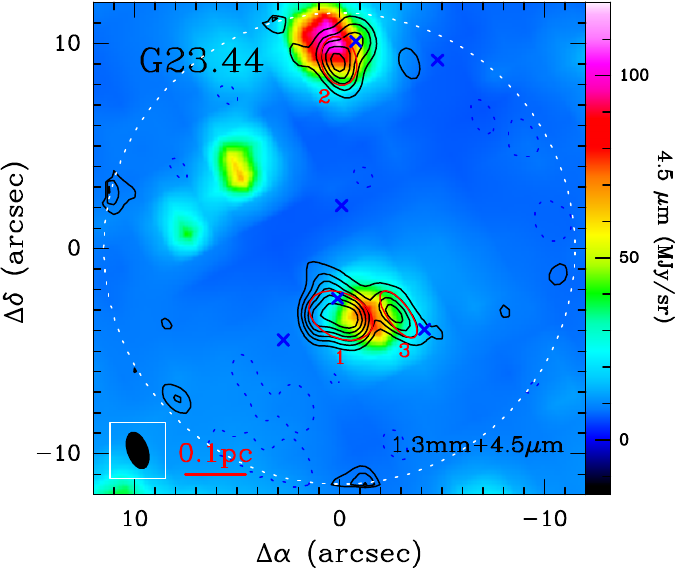}}
\subfigure[]{\includegraphics[width=0.33\textwidth, angle=0]{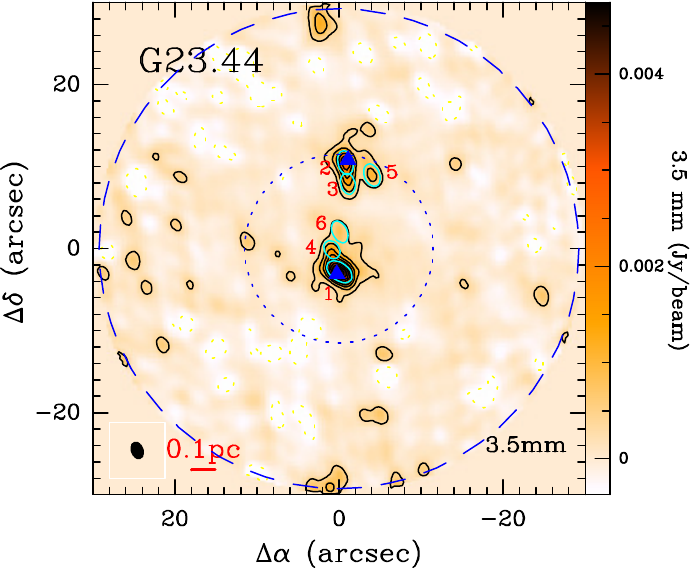}}
\subfigure[]{\includegraphics[width=0.33\textwidth, angle=0]{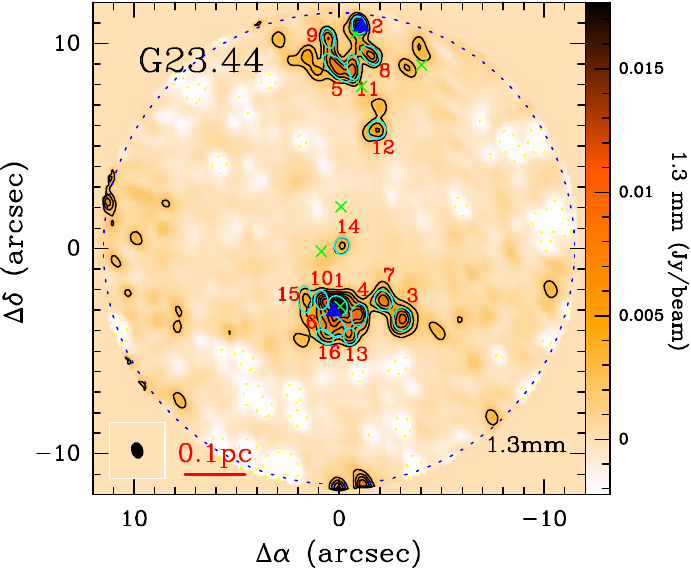}}
\caption{G23.44: {\it (a)} (left) Three-color image of 4.5\,$\mu$m (blue), 8.0\,$\mu$m (green), and 24\,$\mu$m (red); (right) 870\,$\mu$m contours overlaid on a 70\,$\mu$m color scale. The 870\,$\mu$m contour levels start at 6$\sigma$ in steps of 10$\sigma$ ($\sigma$ = 54\,$\mjyb$). {\it (b)} 850\,$\mu$m contours overlaid on a 450\,$\mu$m color scale. The 850\,$\mu$m contour levels start at 6$\sigma$ in steps of 12$\sigma$ ($\sigma$ = 83\,$\mjyb$). {\it (c)} 3.5\,mm contours overlaid on a 1.3\,cm color scale. For CD configuration observations, {\it (d)} the 3.5\,mm contour levels start at -3$\sigma$, 3$\sigma$ in steps of 6$\sigma$ ($\sigma$ = 0.25\,$\mjyb$), and {\it (e)} the 1.3\,mm contour levels start at -3$\sigma$, 3$\sigma$ in steps of 2$\sigma$ ($\sigma$ = 2.17\,$\mjyb$). For BCD configuration observations, {\it (f)} the 3.5\,mm contour levels start at -3$\sigma$, 3$\sigma$ in steps of 4$\sigma$ ($\sigma$ = 0.13\,$\mjyb$), and {\it (g)} the 1.3\,mm contour levels start at -3$\sigma$, 3$\sigma$ in steps of 2$\sigma$ ($\sigma$ = 0.75\,$\mjyb$) superimposed on a 24\,$\mu$m color scale. The ellipses with numbers indicate the positions of extracted sources. The crosses in the 1.3\,mm (B)CD configuration observations indicate the peak positions of the corresponding cores and condensations in the 3.5\,mm (B)CD configuration observations. The dashed and dotted circles in each subfigure indicate the primary beam scales of 3.5 and 1.3\,mm PdBI tracks, respectively. The blue triangle ``$\blacktriangle$'' indicates the position of 6668 MHz methanol masers. Versions of this figure for sources G18.17 (see Figure\,\ref{Fig_G18.17}), G18.21 (see Figure\,\ref{Fig_G18.21}), G23.97N (see Figure\,\ref{Fig_G23.97N}), G23.98 (see Figure\,\ref{Fig_G23.98}), G23.97S (see Figure\,\ref{Fig_G23.97S}), G25.38 (see Figure\,\ref{Fig_G25.38}), and G25.71 (see Figure\,\ref{Fig_G25.71}) are available online only.}
\label{Fig_G23.44}
\end{figure*}

In this work, we adopt a commonly used terminology for clumps, dense cores, and condensations having a physical FWHM size of $\sim$1, $\sim$0.1, and $\sim$0.01\,pc, respectively \citep{Williams2000,Wang2014,Pokhrel2017}. Based on the derived effective radii of our sample, we thus refer to the objects observed with single-dish telescopes at 450, 850, and 870\,$\mu$m observations as clumps, the objects in the 1.3\,cm VLA observations and in the 3.5 and 1.3\,mm PdBI CD configuration observations as cores, and the objects in the 3.5 and 1.3\,mm PdBI BCD configuration observations as condensations (see datasets in Section\,\ref{sect_data}). We report the fragmentation properties of eight high-mass star-forming regions, some of which are infrared quiet suggesting that they are in an early evolutionary stage. In Section\,\ref{sect_data}, we describe the sample selection. In Section\,\ref{sect_results}, we list observational results at different scales and different wavelengths. In Section\,\ref{sect_analysis}, infrared extinction, clump extraction, and kinetic temperature are presented and analyzed. In Section\,\ref{sect_discu}, we mainly discuss dynamics, virial stability, fragmentation features, evolutionary stages, and potential triggered star formation of the clumps. In Section\,\ref{sect summary}, we give a summary.

\section{Observations and data reduction}
\label{sect_data}

\subsection{SCAMPS sources --- sample selection}\label{scamps}
Eight massive clumps (G18.17, G18.21, G23.97N, G23.98, G23.44, G23.97S, G25.38, and G25.71) were selected from the SCUBA Massive Pre/Protocluster core Survey \citep[SCAMPS;][]{Thompson2005,pillai2007}. The SCAMPS survey takes a wide-field ($10'\,\times\,10'$) census of clouds in 32 high-mass star-formation regions at 850\,$\mu$m and 450\,$\mu$m with James Clerk Maxwell Telescope (JCMT). The rationale behind the SCAMPS strategy is that, as high-mass stars are found in clusters and many ultra-compact (UC) \HII regions are located in clusters, one expects to find even earlier phases of massive star formation in the vicinity of such UC \HII regions. In addition, as sub-millimeter continuum emission is optically thin and sensitive to both warm and cold dust, it provides an unbiased census of all massive pre- and protocluster clumps in the vicinity of the already formed massive young stellar objects (YSOs). The parameters including coordinates, observational dates, and distances of the selected eight clumps are listed in Table\,\ref{tab_pdbi} and Table\,\ref{tab_infrared}. The observational parameters of the clumps at 850 and 450\,$\mu$m are listed in Table\,\ref{tab_atlasgal}.

\subsection{ATLASGAL sources}

These eight massive clumps were also identified in the Atacama Pathfinder Experiment (APEX) Telescope Large Area Survey of the Galaxy (ATLASGAL\footnote{The Atacama Pathfinder Experiment (APEX)  project is a collaboration between the Max-Planck-Gesellschaft, the European Southern Observatory (ESO) and the Universidad de Chile.}; \citealt{Schuller2009}). The survey was carried out at 870\,$\mu$m with a resolution of 19.2$''$, and has a goal of investigating massive clumps associated with high-mass star-forming regions. ATLASGAL has been described in detail by \citet{Csengeri2014,Urquhart2014}. The observational parameters of the clumps at 870\,$\mu$m are listed in Table\,\ref{tab_atlasgal}.

\subsection{PdBI observations}

The eight clumps were observed with the IRAM\footnote{IRAM is supported by INSU/CNRS (France), MPG (Germany) and IGN (Spain).} Plateau de Bure Interferometer (PdBI) in the 3.5 and 1.3\,mm simultaneously during 2004 and 2006. The 3.5\,mm receivers were tuned to 86.086\,GHz in single sideband (SSB) mode. The 1.3\,mm receivers were tuned to 219.560\,GHz in double side-band (DSB) mode. Both receivers used two 320 MHz wide backends for continuum observation. In addition, we also obtained data for five molecular lines, i.e., NH$_2$D (85.926\,GHz), HC$^{15}$N (86.055\,GHz), SO (86.094\,GHz), H$^{13}$CN (86.340\,GHz), and C$^{18}$O (219.560\,GHz) with narrow correlator windows affording adequate high spectral resolution. The spectral data will be presented in a forthcoming paper.

For the clumps G18.17, G18.21, G23.97N, G23.98, G23.44, G23.97S, G25.38, and G25.71, the observations were first carried out with both C and D configurations between March and April 2005. The typical synthesized beam size for CD configuration observation is about $4.2''\times2.8''$ at 3.5\,mm and $1.8''\times1.0''$ at 1.3\,mm. Then, the four brightest clumps (G23.44, G23.97S, G25.38, and G25.71) were observed again in the B configuration between February and March 2006. The typical synthesized beam size for images produced from the combined C+D configuration data is about $4.2''\times2.8''$ at 3.5\,mm and $1.8''\times1.0''$ at 1.3\,mm. The typical synthesized beam size for images produced from the combined B+C+D configuration data is about $2.2''\times1.5''$ at 3.5\,mm and $0.8''\times0.6''$ at 1.3\,mm. For calibrations, the quasar B1741-038 was used as phase calibrator, and the quasar 3C273 and the evolved star MWC 349 were used as absolute flux calibrators.

The CLIC and MAPPING modules in the IRAM software package GILDAS\footnote{\url{http://www.iram.fr/IRAMFR/GILDAS/}} were used for the data calibration and image processing. The $uv$-tables from the line-free channels were imaged and cleaned to obtain continuum images with uniform weighting to optimize spatial resolution for G18.17, G18.21, G23.44, G23.97S, G25.38, and G25.71, and with natural weighting to increase the signal to noise ratio for G23.97N and G23.98. A region twice the size of primary beam was imaged. Clean components were searched in the whole image without using a cleaning mask. The primary beam is 58.5$''$ at 86.086\,GHz, and 23.0$''$ at 219.560\,GHz, respectively. The images were corrected for primary beam attenuation. The phase center, beam size, and rms noise of the images not corrected for primary beam response are listed in Table\,\ref{tab_pdbi}.

\subsection{VLA observations}
\label{sect_vla_obs}

The same eight clumps were also observed at 1.3\,cm with NRAO\footnote{The National Radio Astronomy Observatory is a facility of the National Science Foundation operated under cooperative agreement by Associated Universities, Inc.} Very Large Array (VLA) D-configuration on November 2005 (project ID AW0669). The spectra $J, K =$ (1, 1) and (2, 2) transitions of NH$_3$ were simultaneously covered, using the 2-IF spectral line mode of the correlator, with 6.25\,MHz bandwidth and 127\,channels of width 49\,kHz (0.617\,$\kms$) each. the NH$_3$ (1,\,1) and (2,\,2) transitions have five and three hyperfine structure (HFS) lines, respectively, and the frequencies of the strongest HFS lines are at 23.694 and 23.723\,GHz, respectively. Due to the narrow bandwidth of 6.25\,MHz at around 23.7\,GHz, it can just cover at most four HFS lines in five of NH$_3$ (1,\,1). The primary beam is about 2$'$, and the typical synthesized beam size is about $3.0''\times2.5''$ at 1.3\,cm. The spectra and continuum were calibrated with primary beam correction.

The raw data was exported to MIRIAD\footnote{\url{http://www.cfa.harvard.edu/sma/miriad/}} for calibration. The absolute flux calibrator was J1331+305, the antenna gain and phase calibrator was B1743-038, and the bandpass calibrator was J1256-057. The phase centers were the same as for the PdBI observations, listed in Table\,\ref{tab_pdbi}. Natural weighting was used for data calibration and cleaning. The whole primary beam was imaged and cleaned. No polygon was introduced to avoid any biased cleaning.

With the MIRIAD software, we summed up the line-free channels to produce a ``pseudo'' continuum database which was subtracted from the $uv$-data in the $uv$-plane. The line-free continuum was self-calibrated, and the gain solutions were applied to the spectral line data. The image data cubes were exported to CLASS and GREG in GILDAS for further continuum and spectral line analysis. The phase center, beam size, and rms noise of the images not corrected for primary beam response are listed in Table\,\ref{tab_vla}.

\begin{table*}
\caption{The properties of infrared data.}
\label{tab_infrared} \centering \footnotesize
\begin{tabular}{lrrrrcccc}
\hline \hline
Source\tablefootmark{a} &  $S_{\rm 4.5\,\mu m}$  &  $S_{\rm 8.0\, \mu m}$  &  $S_{\rm 24\, \mu m}$  &  $S_{\rm 70\,
\mu m}$  & Infrared\tablefootmark{b} & \HII\tablefootmark{c}  &  Distance  & Ref. for dist.  \\
            & mJy & mJy & mJy & Jy  & & & kpc & \\
\hline
G18.17    & 1.3(0.2)    & 916.3(68.0)  & 214.9(56.6)    & 2.5(0.1)    & quiet   &  no   &   3.73 & 1, 2 \\
G18.21    & 13.6(0.8)   & 27.9(1.7)    & 773.3(14.2)    & 4.6(0.8)    & quiet   &  no   &   3.60 & 1, 2	\\
G23.97N   & ---         & ---          & ---            & 28.9(0.4)   & quiet   &  no   &  4.68  & 1, 2	\\
G23.98    & 4.9(0.6)    & 167.8(6.5)   & 8.7(2.6)       & ---         & quiet   &  no   &  4.68  & 1, 2 	\\
G23.44-l  & 14.9(2.0)   & 34.3(0.2)    & 1683.5(165.5)  & 150.3(0.5)  & bright  &  yes  &   5.88 & 3	\\
G23.44-u  & 25.7(2.2)   & 23.6(0.2)    & 300.1(68.3)    & 115.6(0.5)  & quiet   &  no   &   5.88 & 3	\\
G23.97S   & 10.2(0.4)   & 5.6(0.8)     & 709.6(3.6)     & 179.8(0.4)  & quiet\tablefootmark{d}   &  yes\tablefootmark{d}  &  4.70  & 1, 2	\\
G25.38-l  & 45.7(0.5)   & 50.1(3.6)    & 252.3(38.6)    & 28.7(0.4)   & quiet   &  no   &  5.60  & 4, 5, 6 	\\
G25.38-u  & ---         & ---          & ---            & ---         & quiet   &  no   &  5.60  & 4, 5, 6 	\\
G25.71-l  & 154.4(2.0)  & 272.0(8.0)   & 6197.1(122.6)  & 399.9(0.6)  & bright  &  yes  &  9.50  & 6, 7	\\
G25.71-u  & 138.7(2.2)  & 2453.2(24.9) & 16109.0(83.8)  & 182.3(0.5)  & bright  &  yes  &  9.50  & 6, 7	\\
\hline
\end{tabular}
\tablefoot{
\tablefoottext{a}{The ``-l'' and ``-u'' following source names indicate the lower and upper clusters (e.g. No.1 and No.2 at 3.5\,mm in Figure\,\ref{Fig_G23.44}(d)) in corresponding clumps.}
\tablefoottext{b}{Mainly based on a threshold of MIPS 24\,$\mu$m flux $S_{\rm 24\, \mu m}$ = 15.0\,Jy at a distance of 1.7\,kpc \citep{Motte2007}, this flux limit can be rescaled to the distance of the sources in this table.}
\tablefoottext{c}{Compact \HII region candidate, judged by whether there is a corresponding 1.3\,cm continuum at its sensitivity.}
\tablefoottext{d}{G23.97S is actually associated with the very weak infrared source at the north of the bright one (see Figures\,\ref{Fig_G23.97S}(e)), so it is an infrared quiet source. However, 1.3\,cm continuum was detected so it belongs to an \HII region.}\\
\textbf{References for distance.} \tablefoottext{\rm 1}{\citet{Wienen2012}}; \tablefoottext{\rm 2}{\citet{Reid2009}};
\tablefoottext{\rm 3}{\citet{Brunthaler2009}}; \tablefoottext{\rm 4}{\citet{ande2009}};
\tablefoottext{\rm 5}{\citet{Ai2013}}; \tablefoottext{\rm 6}{\citet{Urquhart2013}};
\tablefoottext{\rm 7}{\citet{Lockman1989}}.
}
\end{table*}

\section{Results}
\label{sect_results}

\subsection{Clumps on large scale}

In Figure\,\ref{Fig_G23.44}(a), we present the large-scale emission of the clumps as imaged by APEX ATLASGAL at 870\,$\mu$m (see also Figures\,\ref{Fig_G18.17}(a) to \ref{Fig_G25.71}(a) in Appendix). The 870\,$\mu$m emission is an excellent tracer of cool and dense gas. Each peak position of the 870\,$\mu$m emission is close to a strong UC \HII region. We find that most of the UC \HII regions have  corresponding compact millimeter wavelength sources, detected at 870, 850, and 450\,$\mu$m. The emission distributions at 870, 850, and 450\,$\mu$m have similar morphological correlations as shown in Figure\,\ref{Fig_G23.44}(b) (see also Figures\,\ref{Fig_G18.17}(b) to \ref{Fig_G25.71}(b) in Appendix).

To measure the physical parameters of the clumps at different wavelengths, we used the \texttt{Gaussclumps} algorithm \citep{Stutzki1990} to extract their peak positions, sizes, and peak fluxes (see details in Section\,\ref{sect_extraction}). We take $R_{\rm eff} = \rm FWHM/(2\sqrt{\rm ln2})$ as the source effective radius, where FWHM is the Gaussian source size. The effective radii $R_{\rm eff}$ of the clumps range from 0.17 to 0.89\,pc (see Table\,\ref{tab_atlasgal}). Based on these measurements, we further derived other parameters including integrated flux, mass and density, shown in Table\,\ref{tab_atlasgal}. The fitted FWHMs at 450\,$\mu$m are indicated with green ellipses in Figure\,\ref{Fig_G23.44}(b).

\subsection{Cores and condensations on small scale}

From the sizes and masses measured at 870, 850, and 450\,$\mu$m (Table\,\ref{tab_atlasgal}), we expect these clumps are candidate hosts of forming or recently formed high mass stars. To unveil the population of dense cores, we show the higher angular resolution observations at 3.5 and 1.3\,mm with the B, C, and D configurations of PdBI in Figures\,\ref{Fig_G23.44}(d)(e). The physical parameters of the cores at the 3.5 and 1.3\,mm are listed in Tables\,\ref{tab_3.5mm} and \ref{tab_1.3mm}. The derived effective radii $R_{\rm eff}$ of the cores and condensations range from 0.027 to 0.102\,pc (see Table\,\ref{tab_3.5mm}), and from 0.010 to 0.044\,pc (see Table\,\ref{tab_1.3mm}), respectively.

At the current sensitivity and $\sim$1$-$2$''$ resolution, the clumps G18.17, G18.21, G23.97N, and G23.98 are probably resolved, as they only host isolated and weak cores at 1.3\,mm. However, clumps G23.44, G23.97S, G25.38, and G25.71 show compact structures and high flux densities in the CD configuration data. It is likely that these observations have not completely resolved these densest cores. Based on this, the higher angular resolution BCD configuration observations were used to investigate these four sources further.

In Figures\,\ref{Fig_G23.44}(f)(g), we show the small-scale fragmented condensations as imaged from the 3.5 and 1.3\,mm BCD configuration data towards clump G23.44 (see also clumps G23.97S, G25.38, and G25.71 in Appendix Figures\,\ref{Fig_G23.97S}(f)(g), \ref{Fig_G25.38}(f)(g), and \ref{Fig_G25.71}(f)(g), respectively). The best mass sensitivity is 0.5\,$\Msun$ with a spatial scale of about 2063 AU corresponding to the source G23.97S-6 (see Table\,\ref{tab_1.3mm}) in these observations. The physical parameters determined from these data are also listed in Tables\,\ref{tab_3.5mm} and \ref{tab_1.3mm}. The clumps fragment further at this higher resolution into clusters of small cores and condensations. The densest and most massive condensations are potential candidate hosts of high-mass protostars. We will further investigate their chemistry, dynamics, and luminosity in a forthcoming work. Their physical parameters, such as mass, size, and density, will be analyzed in the following sections.

\section{Analysis}
\label{sect_analysis}

\subsection{Clump, core, and condensation extractions}
\label{sect_extraction}

Assuming that the flux density of each sub-source is approximated by a Gaussian distribution, the \texttt{Gaussclumps} procedure \citep{Stutzki1990,Kramer1998} in the GILDAS software package was used to characterize clumps, cores, and condensations at all different scales and wavelengths in our sample. We consider sources with peak intensity above 5$\sigma$ before primary beam correction. The derived parameters are listed in Tables\,\ref{tab_atlasgal}, \ref{tab_3.5mm}, and \ref{tab_1.3mm}. Their statistics shows that 27 and 18 compact sources were identified in the eight clumps in the 3.5 and 1.3\,mm CD configuration observations, respectively. This gives on average 3.4 and 2.3 sources per clump corresponding to a population of cores and condensations, respectively, similar to that observed in high-mass clumps by \citet{Csengeri2017}. Despite their prominent peak at 870\,$\mu$m, towards the clumps G23.97N, and G23.98 we do not find strong compact continuum sources at 3.5 and 1.3\,mm. While G23.97N may host a relatively weak single source, the G23.98 field is devoid of compact sources.

\subsection{Infrared emission}
\label{sect_infrared}

The earliest phase of (high-mass) star formation takes place in cold and dense material, prior to the appearance of luminous embedded objects. (Near) Infrared emission which traces warm dust is a useful tool to study the evolution at these early phases. We investigated objects associated with our sample of clumps by combining infrared data including the GLIMPSE 4.5\,$\mu$m, 8.0\,$\mu$m \citep{benj2003,chur2009}, MIPSGAL 24\,$\mu$m \citep{care2009,Gutermuth2015}, and PACS (or Hi-GAL) 70\,$\mu$m \citep{Poglitsch2010,Molinari2016,Marton2017} images and catalogs. For the sources not available in the archival catalogs, we used aperture photometry in the MOsaicker and Point source EXtractor (MOPEX) package \citep{Makovoz2005} to compute fluxes (listed in Table\,\ref{tab_infrared}) at 4.5, 8.0, 24, and 70\,$\mu$m. Using these measurements, we followed the criteria of \citet{Motte2007} and \citet{Russeil2010} (towards Cygnus X and NGC 6334 -- NGC 6357, respectively) for identifying infrared bright and quiet point sources based on the MIPSGAL 24\,$\mu$m images. We scaled the flux density threshold of $S_{\rm 24 \, \mu m}$ = 15.0\,Jy at a distance of 1.7\,kpc \citep{Motte2007} to the distances of our sources as $S_{\rm 24 \, \mu m}$ < $(\frac{\rm 1.7\,kpc}{D})^2 \times 15.0$\,Jy to identify infrared quiet clumps, where $D$ is the distances of our sources, and $S_{\rm 24 \, \mu m}$ is the flux density obtained from the MIPSGAL survey, as listed in Table\,\ref{tab_infrared}. We find that most of the sources fall below the infrared quiet criterion, but four clumps are infrared bright (see also Figures\,\ref{Fig_G18.17}(a) to \ref{Fig_G25.71}(a)). Point-source like emission at 70\,$\mu$m is also a reasonably good indicator of embedded YSOs. Using the Hi-GAL 70\,$\mu$m data \citep{Molinari2016}, we find that roughly half of the sources are associated with bright 70\,$\mu$m point sources, while the clumps G23.98 and G25.38-u do not host any 70\,$\mu$m point source. This suggests that star formation may be active in the infrared bright cores within some of the clumps. We discuss the infrared features of individual clumps in Appendix \ref{sect_individual}.

The saturated infrared emission seen in Figure\,\ref{Fig_G23.44}(a) originates from the neighboring UC \HII regions that are in the vicinity of our targets (see also Section\,\ref{scamps}). These certainly harbor more evolved objects than our source sample, and in Section\,\ref{sect_trigger} we test whether feedback from these (UC-)\HII regions influences the evolution of our targets.

\subsection{Kinetic temperature}
\label{sect_rot}

\begin{figure}[ht]
\centering
\includegraphics[width=0.47\textwidth, angle=0]{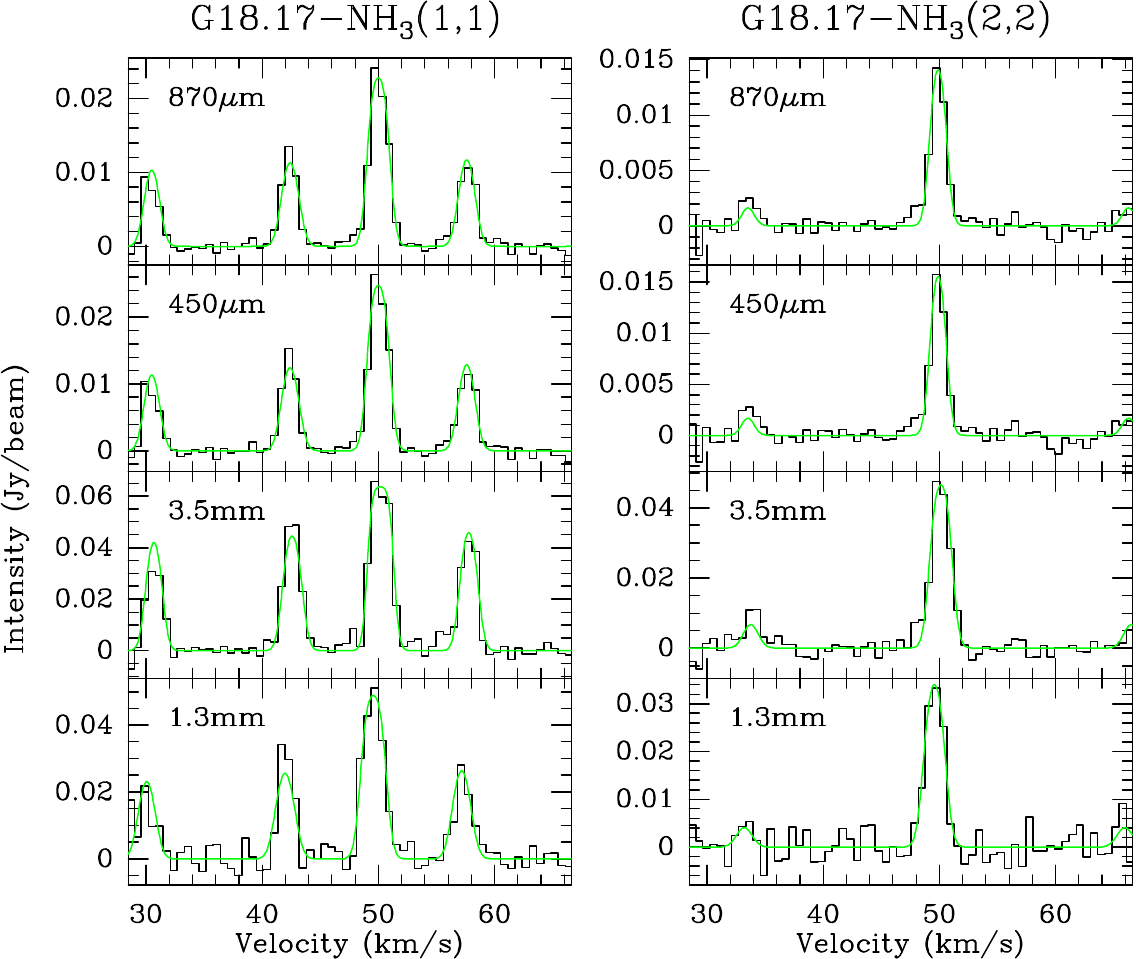}
\caption{NH$_3$ (1,\,1) and (2,\,2) example spectra overlaid with their HFS fits for the fragment G18.17-No.1 (see Tables\,\ref{tab_atlasgal}, \ref{tab_3.5mm}, and \ref{tab_1.3mm}). Top to bottom: the lines are averaged within different fragment scales, respectively, at 870\,$\mu$m, 450\,$\mu$m, 3.5\,mm (CD configuration observations), and 1.3\,mm (CD configuration observations). }
\label{Fig_nh3}
\end{figure}

\begin{figure}[ht]
\centering
\subfigure[]{\includegraphics[width=0.45\textwidth,angle=0]{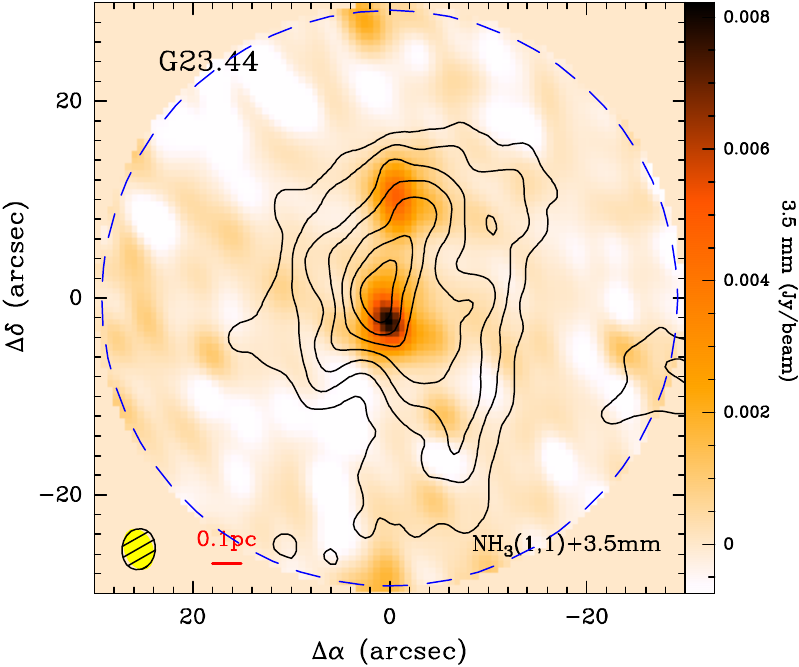}}
\subfigure[]{\includegraphics[width=0.45\textwidth,angle=0]{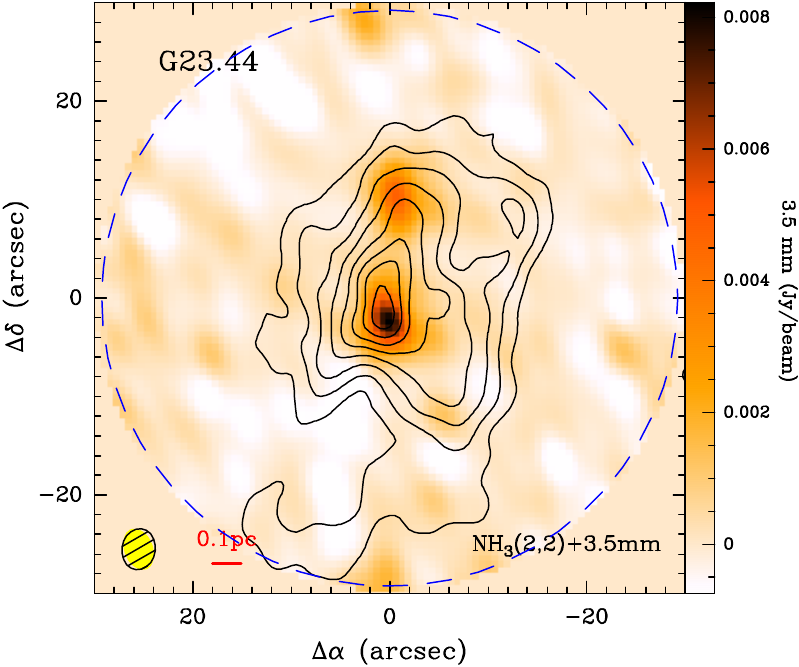}}
\caption{NH$_3$ (1,\,1) and (2,\,2) integrated-intensity contours overlaid on a 3.5\,mm continuum with velocity range covering only the main line, respectively. The contour levels start at -3$\sigma$ in steps of 3$\sigma$ for NH$_3$ (1,\,1) with $\sigma =$ 15.8 $\mjybkms$ and NH$_3$ (2,\,2) with $\sigma =$ 12.3 $\mjybkms$. The synthesized beam size of each subfigure is indicated at the bottom-left corner. Other sources are presented in Figures\,\ref{Fig_nh3-11_appendix} and \ref{Fig_nh3-22_appendix}.}
\label{Fig_nh3-11}
\end{figure}

\begin{figure}[ht]
\centering
\includegraphics[width=0.45\textwidth,angle=0]{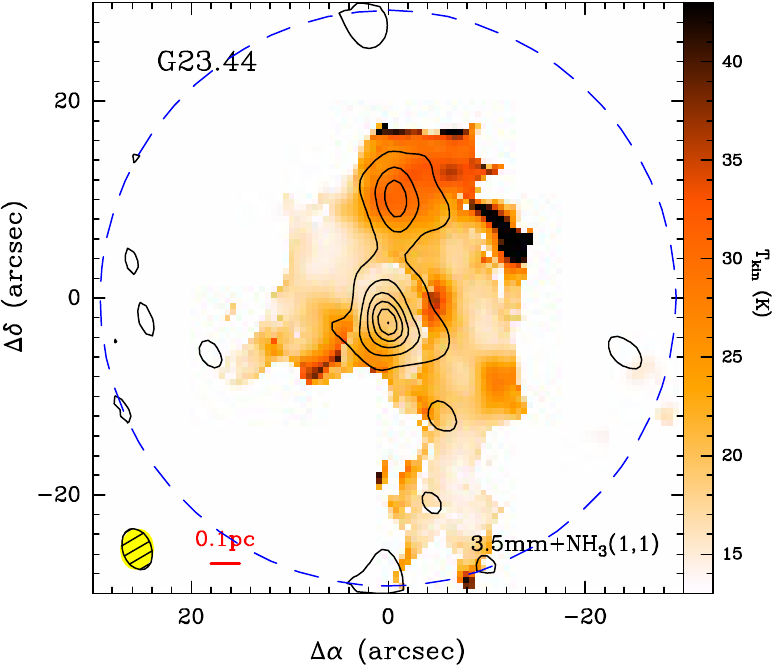}
\caption{The kinetic temperature $T_{\rm kin}$ maps overlaid by 3.5\,mm continuum contours. The 3.5\,mm contour levels start at 3$\sigma$ in steps of 6$\sigma$ ($\sigma$ = 0.25\,$\mjyb$). The synthesized beam size is indicated at the bottom-left corner. Other sources are presented in Figure\,\ref{Fig_temperature_appendix}.}
\label{Fig_temperature}
\end{figure}

The rotational temperature of the ammonia transitions was estimated using the ratio of the brightness temperatures between the NH$_3$ (1,\,1) and (2,\,2) transitions (e.g., Figures\,\ref{Fig_nh3}, and \ref{Fig_nh3-11}), along with the NH$_3$ (1,\,1) optical depth following \citet{Ho1983}. The optical depths of NH$_3$ (1,\,1) were derived from fitting the hyperfine structure (HFS) lines and obtaining the ratio of the main to the satellite line strengths. Considering that the extracted sources have different scales, we average these spectra of NH$_3$ (1,\,1) and (2,\,2) within the measured Gaussian FWHM of each source to get a high signal to noise ratio spectrum.

The kinetic temperature $T_{\rm kin}$ can be estimated from the rotational temperature $T_{\rm rot}$ derived using NH$_3$ (1,\,1) and (2,\,2) transitions. Based on large velocity gradient (LVG) calculations, \citet{Ott2011} provide a reasonable approximation for the conversion from $T_{\rm rot}$ to $T_{\rm kin}$ as
  \begin{eqnarray}
T_{\rm kin}=6.06 \times {\rm e}^{0.061T_{\rm rot}} \, .
  \end{eqnarray}
The derived kinetic temperatures are listed in Tables\,\ref{tab_atlasgal}, \ref{tab_3.5mm}, and \ref{tab_1.3mm}. The kinetic temperature maps overlaid by 3.5\,mm continuum contours are presented in Figure\,\ref{Fig_temperature}. The temperatures range from 12.5 to 63.1\,K with mean and median of 28.3 and 23.6\,K, respectively. Only 10.2\% of all sources have kinetic temperatures above 40\,K. Such high temperatures are observed mainly towards the cores associated with compact \HII regions in clumps G23.44 and G23.97S\footnote{Ammonia is not a good probe of the material close to the heating source; instead it shows the conditions in the outer layer of the envelope. Therefore, the temperatures in clumps containing compact \HII regions may be underestimated.}. Furthermore, it is obvious that the obtained rotational temperatures at large scale are lower than those at small scale.

\subsection{Mass and density calculations}
\label{sect_mass}

It is assumed that the dust emission is optically thin and the gas to dust ratio is 100 by mass. For all clumps, cores, and condensations, we adopt the kinetic temperatures derived from their rotational temperatures between NH$_3$ (1,\,1) and (2,\,2) as an approximate dust temperature (see Section\,\ref{sect_rot}). The source masses are calculated using dust opacities of 0.002 cm$^2$\,g$^{-1}$ at 3.5\,mm and 0.009 cm$^2$\,g$^{-1}$ at 1.3\,mm for thin ice mantles at a gas density of 10$^6$ cm$^{-3}$ \citep{Ossenkopf1994,Pillai2011}. In addition, dust opacities of 0.0185, 0.0182, and 0.0619 cm$^2$\,g$^{-1}$ are adopted for 870, 850, and 450\,$\mu$m, respectively \citep{Ossenkopf1994}. The total mass $M$ of each source can therefore be calculated \citep{Kauffmann2008} via
	\begin{eqnarray}
	\begin{aligned}
	\label{equa_mass}
	\left(\frac{M}{\Msun}\right) =  & 0.12 \left({\rm  e}^{14.39\left(\frac{\lambda}{\rm mm}\right)^{-1}\left(\frac{T_{\rm dust}}{\rm K}\right)^{-1}}-1\right) \times &  \\
        & \left(\frac{\kappa_{\nu}}{\rm cm^{2}g^{-1}}\right)^{-1} \left(\frac{S_{\nu}}{\rm Jy}\right) \left(\frac{D}{\rm kpc}\right)^{2}
	\left(\frac{\lambda}{\rm mm}\right)^{3}, &
	\end{aligned}
	\end{eqnarray}
where $\lambda$ is the observational wavelength, $T_{\rm dust}$ is the dust temperature, $\kappa_{\nu}$ is the dust opacity, $S_{\nu}$ is the integrated flux, and $D$ is the distance to the Sun. Moreover, the column density, volume density and surface density are derived as $N_{\rm H_2} \sim M/{\rm FWHM}^2$, $n_{\rm H_2} \sim M/{\rm FWHM}^3$, and $\Sigma = M/(\pi R_{\rm eff}^2)$, respectively, where $R_{\rm eff} = \rm FWHM/(2\sqrt{\rm ln2})$ is the source effective radius, and FWHM is the Gaussian source size. These corresponding parameters are listed in Tables\,\ref{tab_atlasgal}, \ref{tab_3.5mm}, and \ref{tab_1.3mm}. The mass uncertainties are mainly derived by error propagation from the rotational temperature. Considering the interferometric filtering, the derived masses are lower limits.

In addition, we distinguish the contribution from the dust and the free-free emission at 3.5\,mm based on the SED fitting for the clumps G23.44, G23.97S, G25.38, and G25.71 (see Section\,\ref{sect_sed} and Fig.\,\ref{Fig_SED} in Appendix). For the most massive objects within the clumps G18.17, G18.21, G23.97N, and G23.98, the expected free-free emission level at 3.5\,mm is at least half orders of magnitude lower than the detected dust emission at 3.5\,mm according to these detection limits at 6 and 1.3\,cm, therefore, we assume that their measured integrated fluxes are purely from dust emission.

In Table\,\ref{tab_atlasgal}, there is some difference among the masses derived from 870, 850, and 450\,$\mu$m. While the flux calibration errors are similar at each wavelength, the measured Gaussian sizes from \texttt{Gaussclumps} are different as different wavelengths trace different scales and structures. In this work, we adopt the masses and sizes derived mainly at 870\,$\mu$m for further investigation.

\section{Discussion}
\label{sect_discu}

\subsection{Gas dynamics}
\label{sect_dynamics}

\begin{figure}[ht]
\centering
\subfigure[]{\includegraphics[width=0.45\textwidth,angle=0]{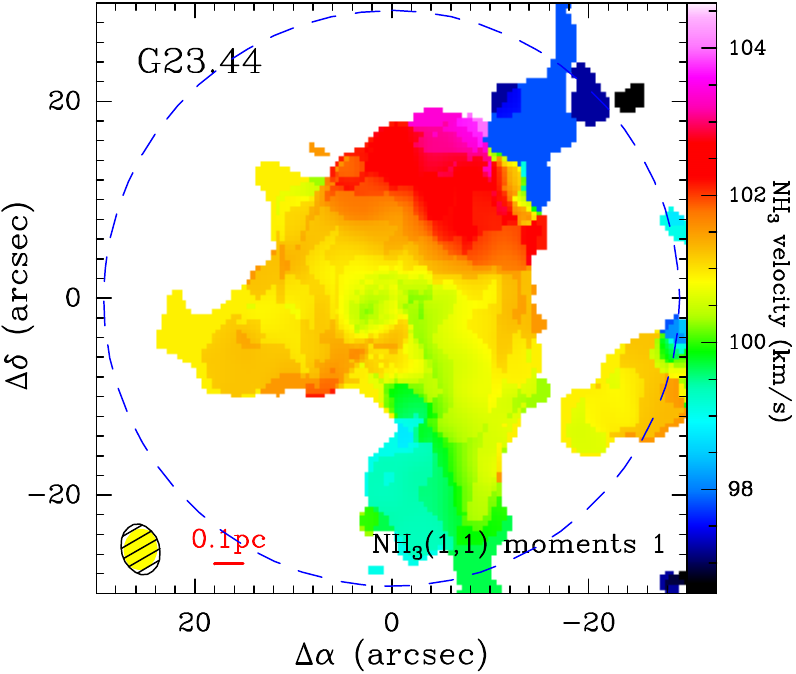}}
\subfigure[]{\includegraphics[width=0.45\textwidth,angle=0]{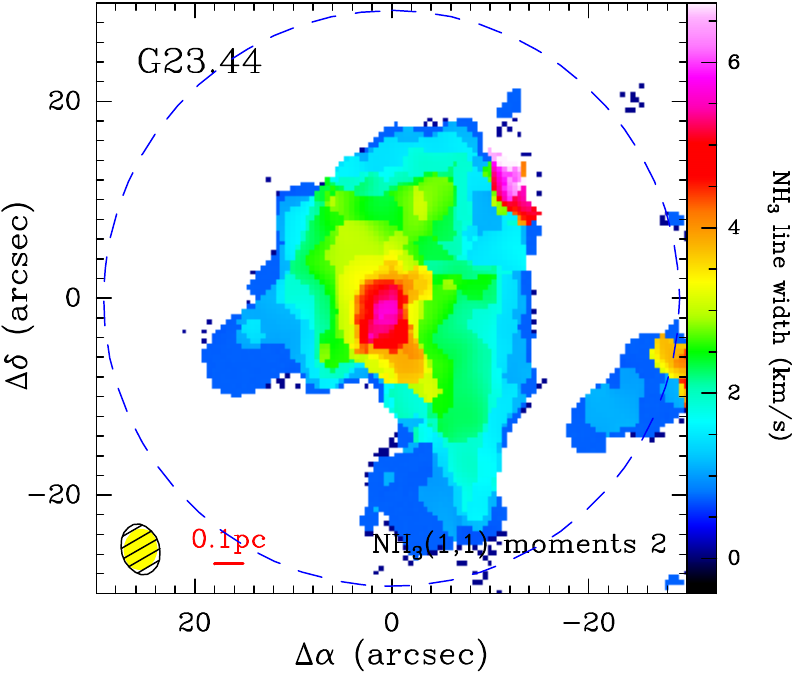}}
\caption{(a) Velocity distribution (moment\,1) and (b) velocity dispersion (moment\,2) of NH$_3$ (1,\,1) line. The synthesized beam size of each subfigure is indicated at the bottom-left corner. Other sources are presented in Figures\,\ref{Fig_mom1_appendix} and \ref{Fig_mom2_appendix}.}
\label{Fig_mom1}
\end{figure}

Gas dynamics shapes the morphology of molecular clouds at large scale, and plays a major role in stabilizing the entire core against fragmentation \citep{Csengeri2011a,Csengeri2011b,Naranjo2015}. Molecular clouds in early stage often exhibit complex gas dynamical phenomenons, such as outflow, rotation, convergent flow, and collision \citep{Schneider2010,Zhang2009,Beuther2013,Zhang2017c}.

NH$_3$ (1,\,1) is a good tracer of relatively diffuse and extended molecular clouds, and often used as dynamical tracer \citep[e.g.,][]{Galvan2009,Zhang2014}. In Figure\,\ref{Fig_mom1}, we present velocity distributions (moment\,1) of NH$_3$ (1,\,1). It is very obvious that there exists a steep velocity gradient for the eight molecular clumps, such as G18.17 and 23.97N in east-west direction, G23.44, G25.38, and G25.71 in north-south direction, and the other sources (G23.98 and G23.97S) having two velocity components crossing into together. It is likely that the molecular clumps are colliding each other, or simply overlapping in the plane of the sky but are physically separated in the third spatial dimension. We also present velocity dispersion (moment\,2) of NH$_3$ (1,\,1) line. We found that the 3.5\,mm emission is associated with the dynamically dominated regions. In general, the dynamical movements are active at clump scale ($\sim$ 1\,pc), although most of the central cores are in a very early stage.

\subsection{Virial stability}
\label{sect_virial}

\begin{figure}[ht]
\centering
\includegraphics[width=0.45\textwidth, angle=0]{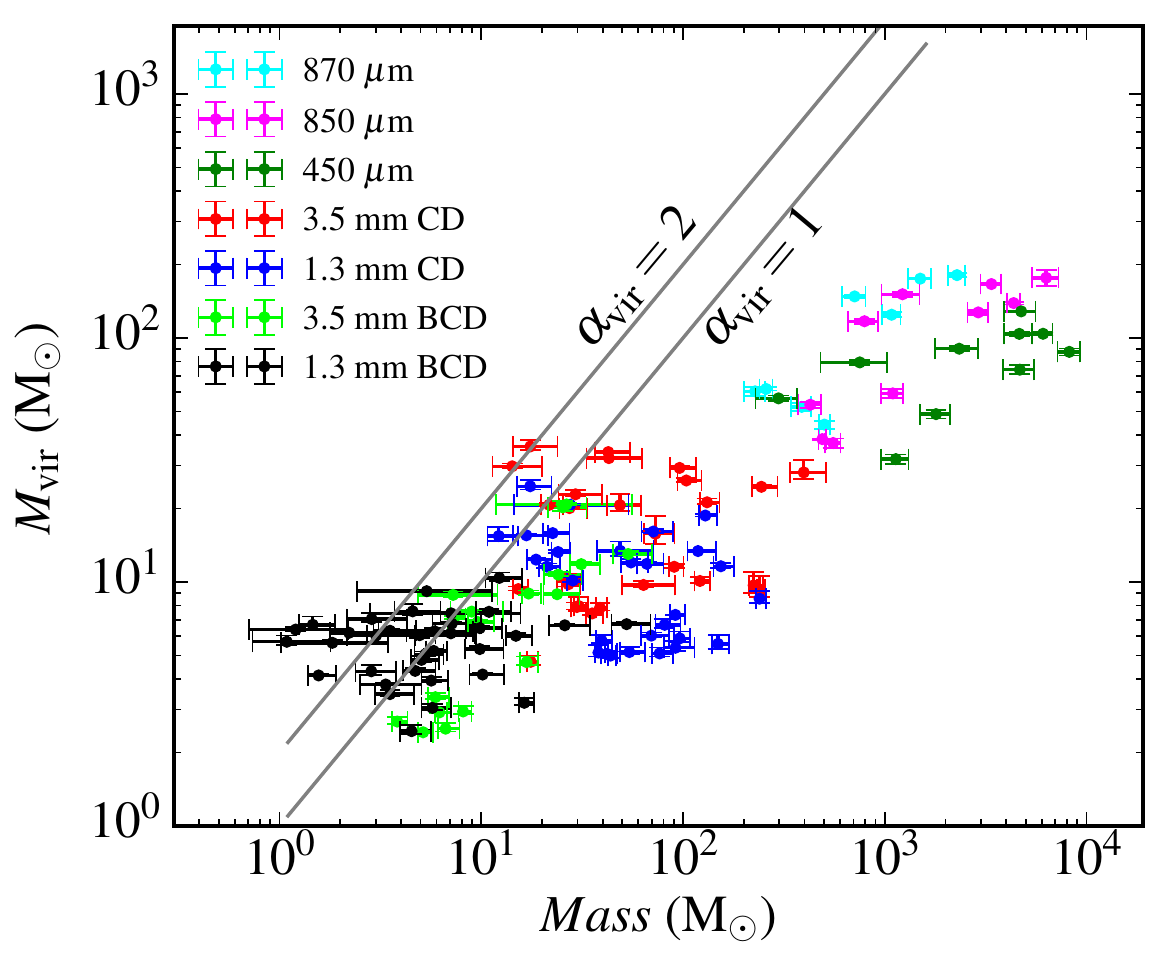}
\caption{\textit{Virial mass-dust mass} distributions of all sources at different wavelengths. The two straight grey lines show thresholds of virial parameters $\alpha_{\rm vir}=1$ and $\alpha_{\rm vir}=2$. }
\label{Fig_virial}
\end{figure}

The virial theorem can be used to test whether a source is unstable to collapse. To apply the virial theorem, a density profile needs to be specified. We assume a simple spherical source with a density distribution of $\rho\propto r^{-2}$ \citep[e.g.,][]{Shu1977}, which we take as an indication that self-gravity is dominating the collapse \citep{Li2018}. There are a few previous explanations of this profile: according to \citet{Shu1977}, an initial density profile of $\rho \propto r^{-2}$ is established before protostellar collapse due to pressure support, and a somewhat shallower density profile is established at the center after the formation of the protostar. The $r^{-2}$ profile in \citet{Shu1977} is hydrostatic in nature and exhibits zero infall. This is contradicted by the fact that clumps like ours typically show some degree of infall. It has also been argued that a free-fall collapse can produce such a steep density profile \citep{Larson1969,Penston1969,Vazquez2009,Ballesteros2011,Girichidis2014,Naranjo2015,Donkov2017} and the infall velocity is also close to the free-fall velocity \citep{Murray2015}. One might be temped to interpret the $r^{-2}$ profile using free-fall models \citep[e.g.][]{Schneider2016}, but a free-fall model is unlikely since the infall velocities of the observed clumps are lower than the free-fall value \citep[e.g.][]{Wyrowski2016}.

If we ignore magnetic fields and bulk motions of the gas, the virial mass of a source can be estimated with the formula \citep{MacLaren1988,Evans1999}:
	\begin{eqnarray}
	\label{equa_virial-mass}
    M_{\rm vir} \simeq 126\, R_{\rm eff}\, \Delta v^{2}_{\rm tur}\, (\Msun),
	\end{eqnarray}
where $R_{\rm eff}$ is the source effective radius in pc, and $\Delta v_{\rm  tur}$ is the turbulent line width in $\kms$ (see also ${\rm NH_3\,(1,1)}$ velocity dispersion in Figure\,\ref{Fig_mom1}),	
     	\begin{eqnarray}
	\label{equa_tur}
    \Delta v^2_{\rm  tur} = \Delta v^{2}_{\rm NH_3\,(1,1)}-\Delta v^{2}_{\rm ther} = \Delta v^{2}_{\rm NH_3\,(1,1)}-\frac{{\rm 8ln(2)} k T_{\rm kin}}{m_{\rm NH_3\,(1,1)}},
	\end{eqnarray}
where $k$ is the Boltzmann constant and $m_{\rm NH_3\,(1,1)}$ is the mass of the ammonia molecule. The corresponding data are listed in Tables\,\ref{tab_atlasgal}, \ref{tab_3.5mm}, and \ref{tab_1.3mm}. For the condensations in the 1.3\,mm BCD configuration observations, the adopted line widths are estimated using the NH$_3$\,(1,\,1) line observations, which have a synthesized beam size of around 4 factor larger. There is no simple way of assigning NH$_3$\,(1,\,1) emission to each 1.3\,mm fragment, as it is contaminated by emission from neighboring fragments. Some fragments have strong dynamical motions such as outflow and infall at a larger scale than our 1.3\,mm beam. Therefore, the derived virial masses and virial parameters could be overestimated. The distribution between virial mass and dust mass is shown in Figure \ref{Fig_virial}. We find that the clumps and cores at large scale are below the threshold with $\alpha_{\rm vir} < 1$, suggesting that they are gravitationally bound, while a few part of the condensations at small scale are above the threshold with $\alpha_{\rm vir} > 1$, indicating that they are gravitationally unbound and may be dispersing. This suggests that some of the condensations are unable to form stars. If we adopt $\alpha_{\rm vir}=2$ as a threshold for the critical virial parameter of a hydrostatic equilibrium sphere \citep{Kauffmann2013}, most clumps and fragments are gravitationally bound. This is consistent with the conclusion we arrived at by studying the radial density profile.

\begin{figure*}[ht]
\centering
\subfigure[]{\includegraphics[width=0.33\textwidth, angle=0]{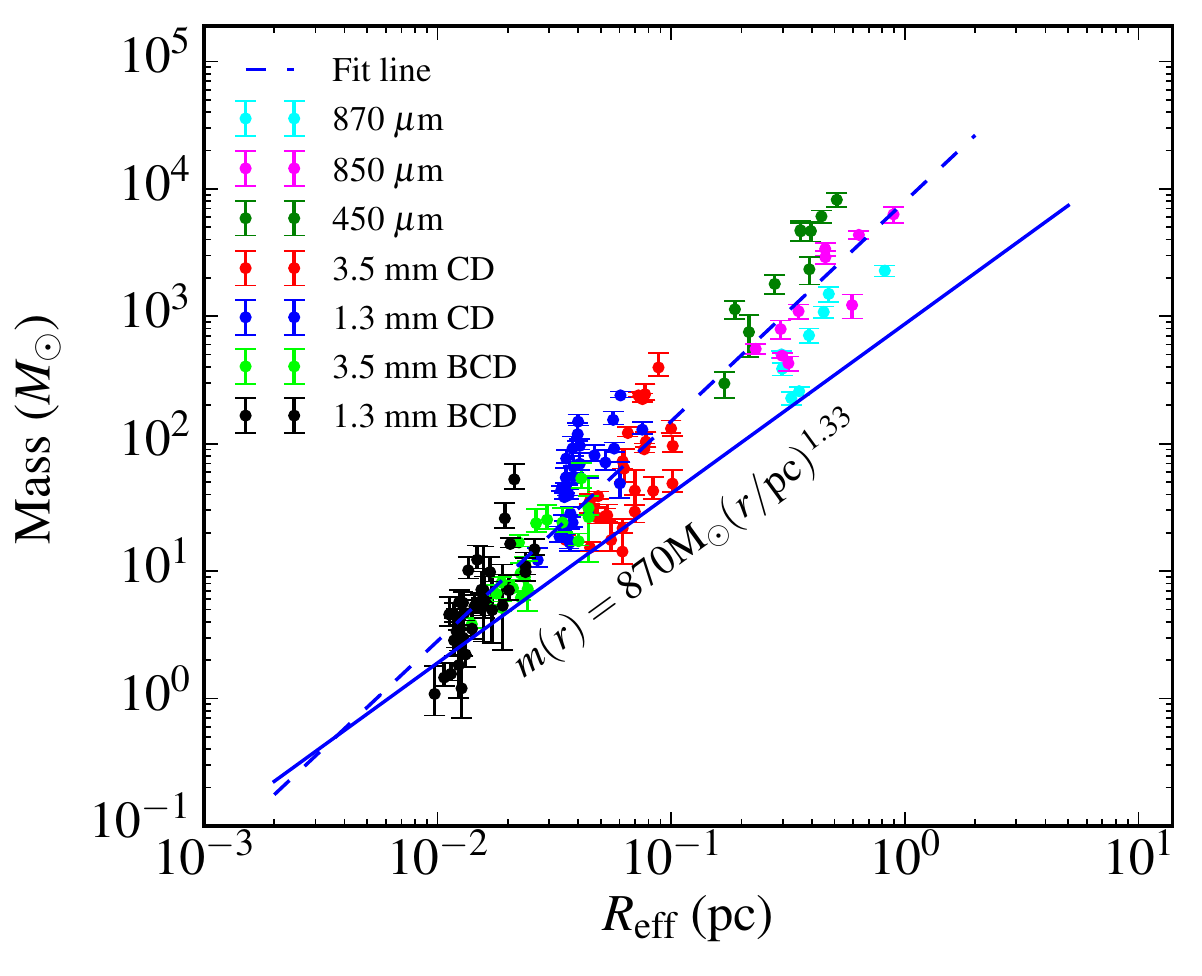}}
\subfigure[]{\includegraphics[width=0.3223\textwidth, angle=0]{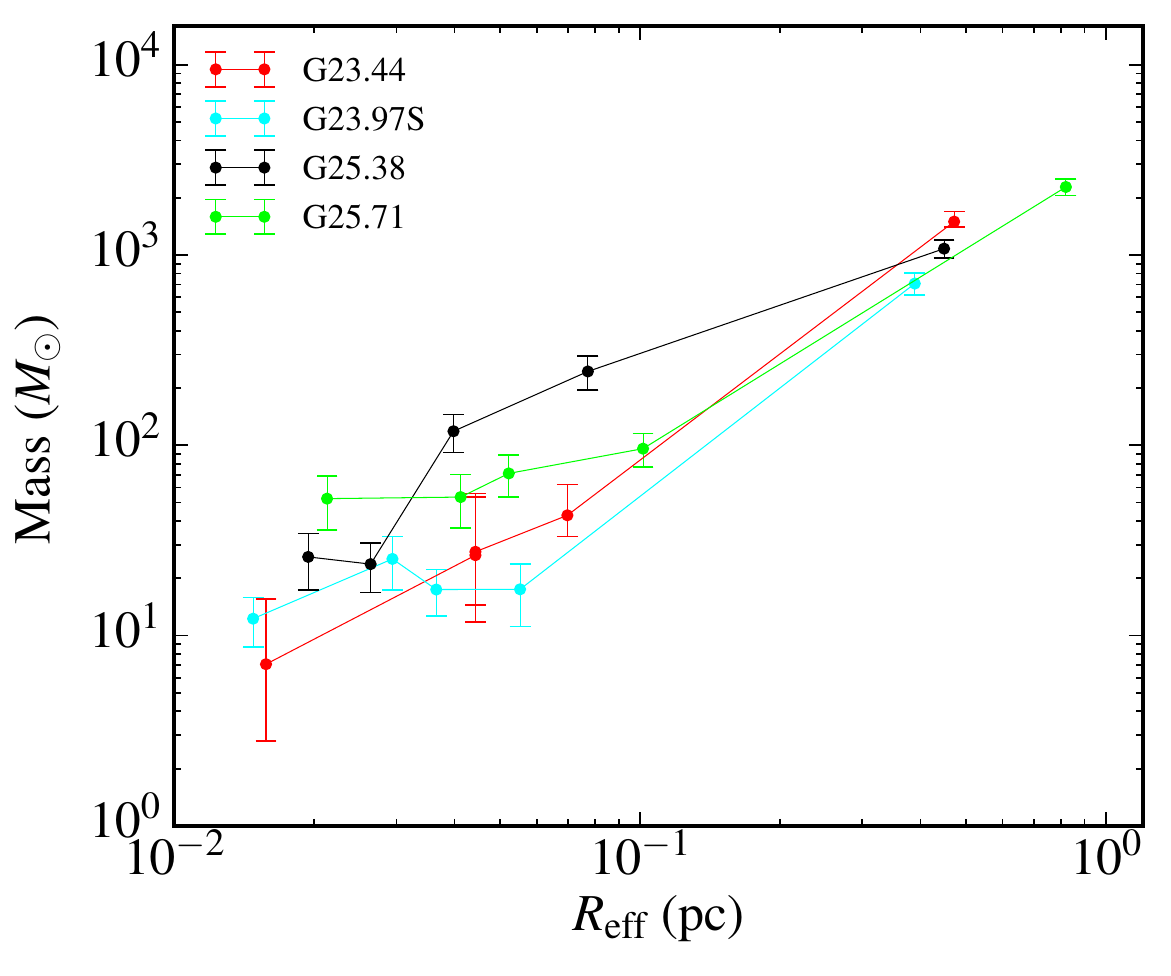}}
\subfigure[]{\includegraphics[width=0.337\textwidth, angle=0]{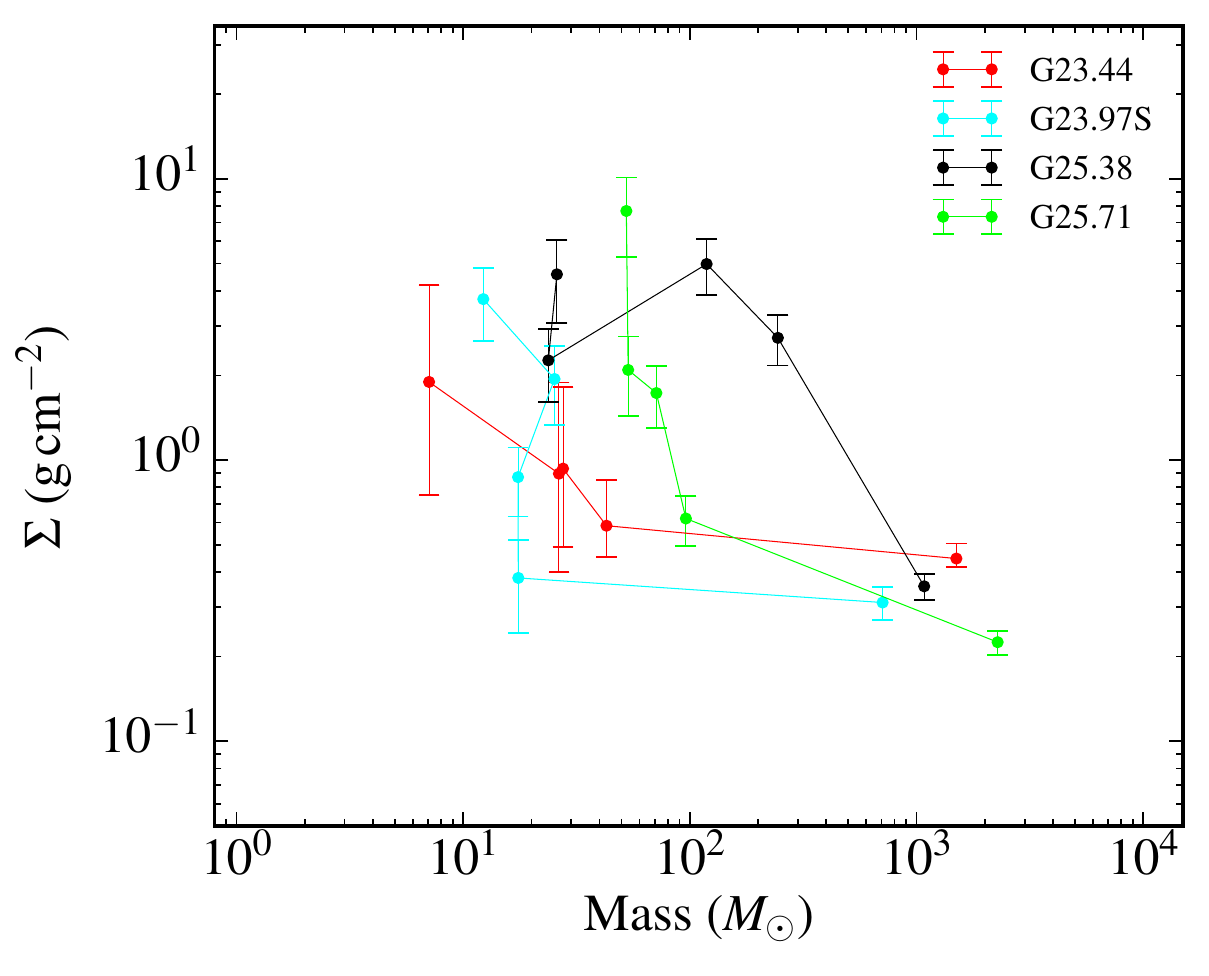}}
\caption{(a) \textit{Mass-size} distribution of all sources at different wavelengths, (b) \textit{mass-size} and (c) \textit{surface density-mass}  distributions only for the most massive objects within each clump. The straight line shows the empirical threshold for high-mass star formation \citep{Kauffmann2010}. The dashed line shows the result of a linear power-law fit to the whole sample: ${\rm log}({\rm Mass}(R_{\rm eff})/{\Msun}) = 3.86 + 1.68 {\times} {\rm log}(R_{\rm eff}/ {\rm pc})$ \citep[see details in][]{Zhang2017b}.}
\label{Fig_mass-size-all}
\end{figure*}

\subsection{Fragmentation}
\label{sect_fragmentation}

Jeans fragmentation tends to produce numerous small fragments rather than a single massive object. The thermal Jeans mass $M_{\rm J}$ depends on gas temperature $T$ and gas number density $n$ as $M_{\rm J} = 6.285 T^{1.5} n^{-0.5}$ \citep{Palau2013}. For a fragment at a gas temperature of 18\,K and gas number density of 10$^{6}$ cm$^{-3}$, the thermal Jeans mass is $\sim 0.5$\,$\Msun$. The measured masses in Tables\,\ref{tab_3.5mm} and \ref{tab_1.3mm} are much larger than the value of 0.5\,$\Msun$ suggesting that the fragments are gravitationally unstable. Other physical ingredients, such as radiation feedback, magnetic fields, turbulence and dynamic processes could play a role in forming cores above the thermal Jeans mass, thereby influencing the cloud fragmentation properties \citep[refer to a review by][]{Tan2014}.

At larger scales with the lower resolution at 870\,$\mu$m, each clump shows just one or two relatively compact components. The masses of the eight clumps range from 228 to 2279\,$\Msun$, which provides sufficient material for the formation of stellar clusters \citep{Blitz1991,Blitz1999}. We find that towards the two lowest mass and lowest surface density clumps ($\Sigma=$\,0.14\,g\,cm$^{-2}$), no compact sources are found, and we exclude the presence of compact fragments at our sensitivity limit of around 10\,$\Msun$. The lowest surface density where we start to see compact sources is $\Sigma>$\,0.23\,g\,cm$^{-2}$. In comparison, the study of \citet{Csengeri2017} identifies massive clumps with $\Sigma>$\,0.5\,g\,cm$^{-2}$ as potential high-mass protostellar candidates.

We find that the brightest fragments within the infrared bright clumps are always associated with compact 1.3\,cm radio emission. The observed radio emission is likely associated with the deeply embedded UC \HII regions. Altogether we identify four clumps (G23.44-1, G23.97S-1, G25.71-1, and G25.71-2 at 850\,$\mu$m) as hosting compact \HII regions. In general, these clumps have a rather low number of fragments up to spatial scales of $\sim$3$-$4\arcsec. All of these fragments are associated with deeply embedded mid-infrared point sources. This picture is, however, somewhat different for the infrared quiet clumps. Two out of the four have no compact sources, and the other two sources, G18.17 and G18.21, only show weak 24\,$\mu$m point sources towards the position of the fragments. From the low temperature estimates given by the NH$_3$ inversion transitions, these cores are likely to be cold, and thus potentially in an early phase of star formation. The core masses range up to 35\,$\Msun$ suggesting that they already harbour sufficient mass to form at least one B3$-$B0 type star assuming 20$-$40\% efficiency \citep{Tanaka2016}.

\subsection{Mass-size distributions}
\label{sect_mass_size_density}

In Figure\,\ref{Fig_mass-size-all}, we show the \textit{mass-size} distribution for all and for only the most massive objects potentially forming high-mass stars in panel (a) and (b), respectively \citep[see also other discussion in our previous work of][]{Zhang2017b}. The panel (b) in Figure\,\ref{Fig_mass-size-all} shows that the fragmentation slope gradually flattens from large scale to small scale and from 870\,$\mu$m to 1.3\,mm, while the \textit{surface density-mass} relation in the panel (c) gradually steepens. This indicates that as fragmentation size decrease, the fragment mass remains roughly constant while the surface density increases sharply. The central dense condensations at a scale of around 0.02\,pc can contain most of the mass of the source with little in the surrounding envelope. For the relatively young sources, e.g., G18.17 and G18.21, the mass does not concentrate on a single core, but instead in several cores with comparable masses (see also Figures\,\ref{Fig_G18.17} and \ref{Fig_G18.21}). For the relatively evolved sources, e.g., G24.44, G23.97S, G25.38, and G25.71, however, the mass is concentrated in a single condensation (see also Figures\,\ref{Fig_G23.44}, \ref{Fig_G23.97S}, \ref{Fig_G25.38}, and \ref{Fig_G25.71}).

\subsection{Evolutionary stages}
\label{sect_evolution}

\begin{figure*}
\centering
\includegraphics[width=0.59\textwidth, angle=0]{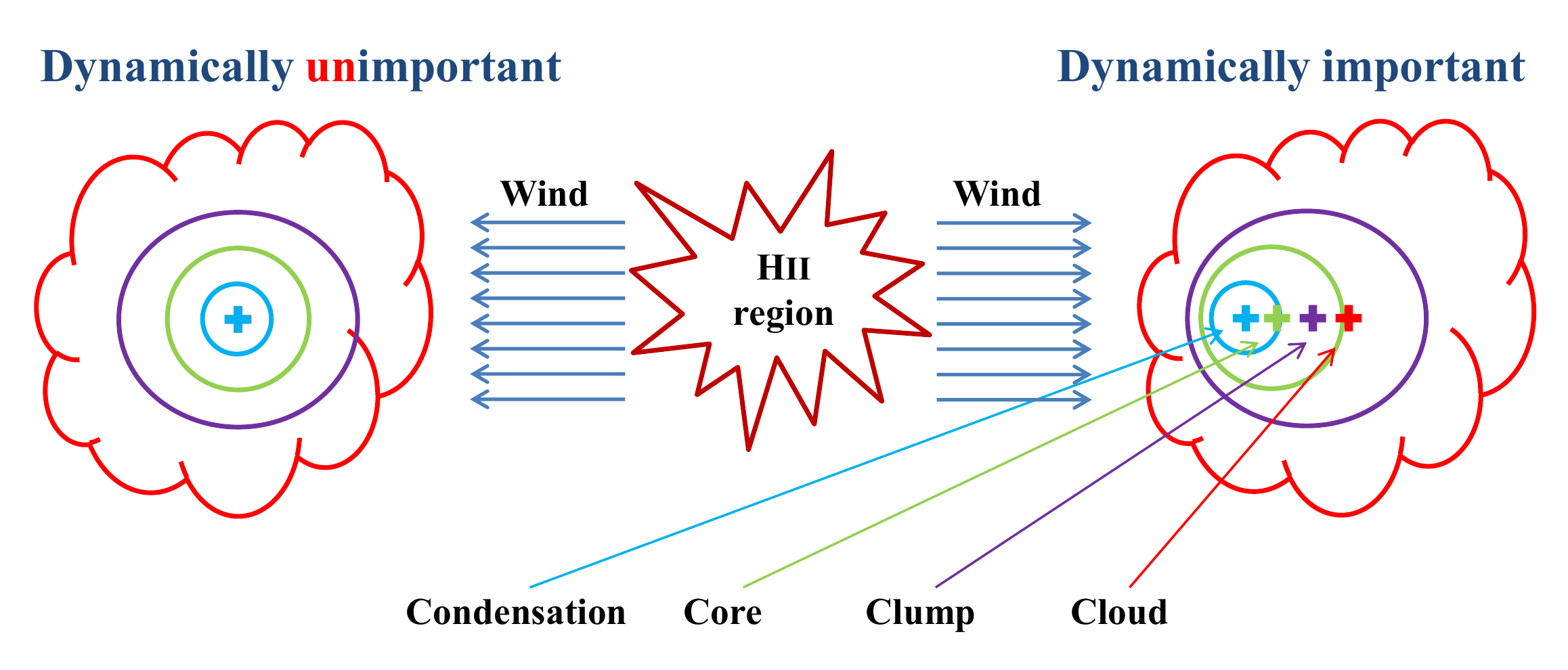}
\caption{A schematic illustration depicting the feedback of radiation from neighboring UC \HII regions onto molecular clouds. The feedback could induce a position offset of centroid for the embedded clumps, cores, and condensations. The pluses ``+'' in different colors indicate corresponding position center of each object.}
\label{Fig_trigger}
\end{figure*}

Millimeter emission on scales of $>$0.5\,pc is usually associated with near- and mid-infrared emission from embedded protostars. In this work, 870, 850, and 450\,$\mu$m emission seems to correlate well with the infrared emission distribution morphologically, except for clumps G23.97N and G23.98. The strong infrared emission tracing a protostar is contained within a cold clump traced by the millimeter emission. The infrared-quiet cores could contain early protostars too faint to be detected in the IR at these distances.

The 24\,$\mu$m emission can provide a hint of the evolutionary stage of star formation. Based on the 24\,$\mu$m flux density, most of the sample are infrared quiet. In particular, three sources (G23.97N, G23.98, and G25.38-u) seem to be in the earliest phase, corresponding either to starless or prestellar clumps, because we find no indication for on-going star formation despite their considerable mass and surface density. In addition, there exist offsets between the condensations and infrared peaks. On the other hand, the brightest fragments that are associated with a mid-infrared point source are resolved multiple components at our highest resolution, most of which are difficult to associate with a 24\,$\mu$m counterpart. This case is similar to the prestellar core with weak infrared emission near the protostar IRAS 16293 \citep{Lis2016}. Therefore, we suggest that some of the condensations will probably form a subsequent generation of high- and intermediate-mass stars.

Methanol masers (CH$_3$OH) are classified into two groups: Class I and Class II \citep{Menten1991}. The latter typically coincide in position with hot molecular cores, compact \HII regions, and near-IR sources \citep[e.g.][]{Minier2001,Ellingsen2006,Fontani2010}, which clearly gives an indication of their evolutionary status. Checking the 6-GHz methanol multibeam maser (MMB) catalogue \citep{Green2010,Breen2015}, we find that three of our sources harbour class II methanol masers: G23.44, G23.97S, and G25.71 (see Figures\,\ref{Fig_G23.44}, \ref{Fig_G23.97S}, and \ref{Fig_G25.71}, respectively), which are all associated with 1.3\,cm continuum emission. The Class II masers are located at the peak positions of the most massive condensations in clusters. It seems that the Class II methanol maser sources are the ones associated with free-free emission. The association with radio free-free emission is an indicator of a deeply embedded UC \HII region, suggesting that that fragment has already a high-mass star with ionising radiation. Therefore, the sources G23.44, G23.97S, and G25.71 may be at the compact \HII region stage, but are enveloped in a thick and cold dust envelope.

It is possible that a triggered hierarchical evolutionary sequence may exist, starting with the UC \HII regions and following an early star-forming dense cores and condensations. In the following section, we will look for evidence of triggered star formation.

\subsection{Triggered star formation by neighboring \HII regions?}
\label{sect_trigger}

\begin{figure*}
\centering
\includegraphics[width=0.245\textwidth, angle=0]{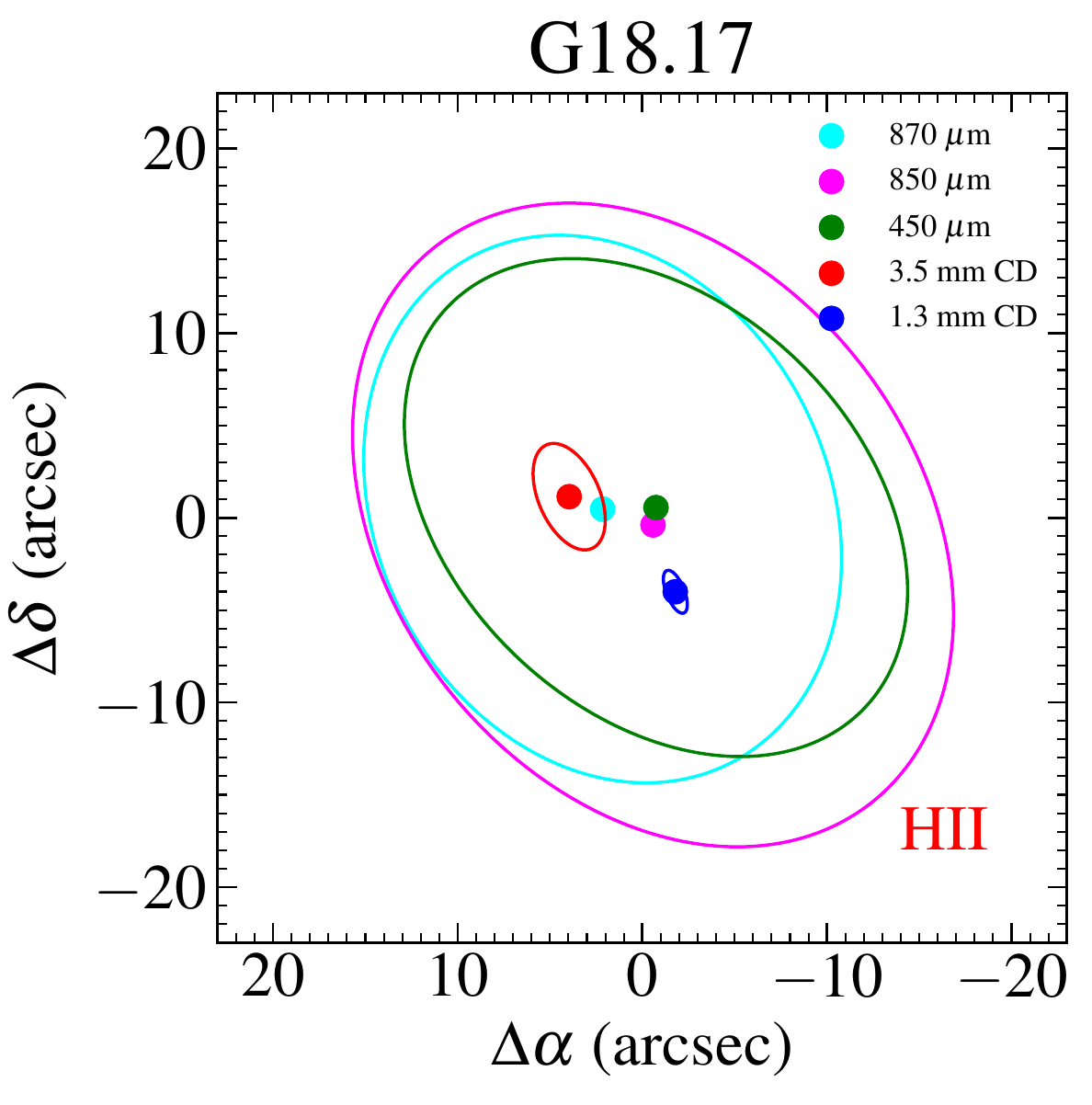}
\includegraphics[width=0.245\textwidth, angle=0]{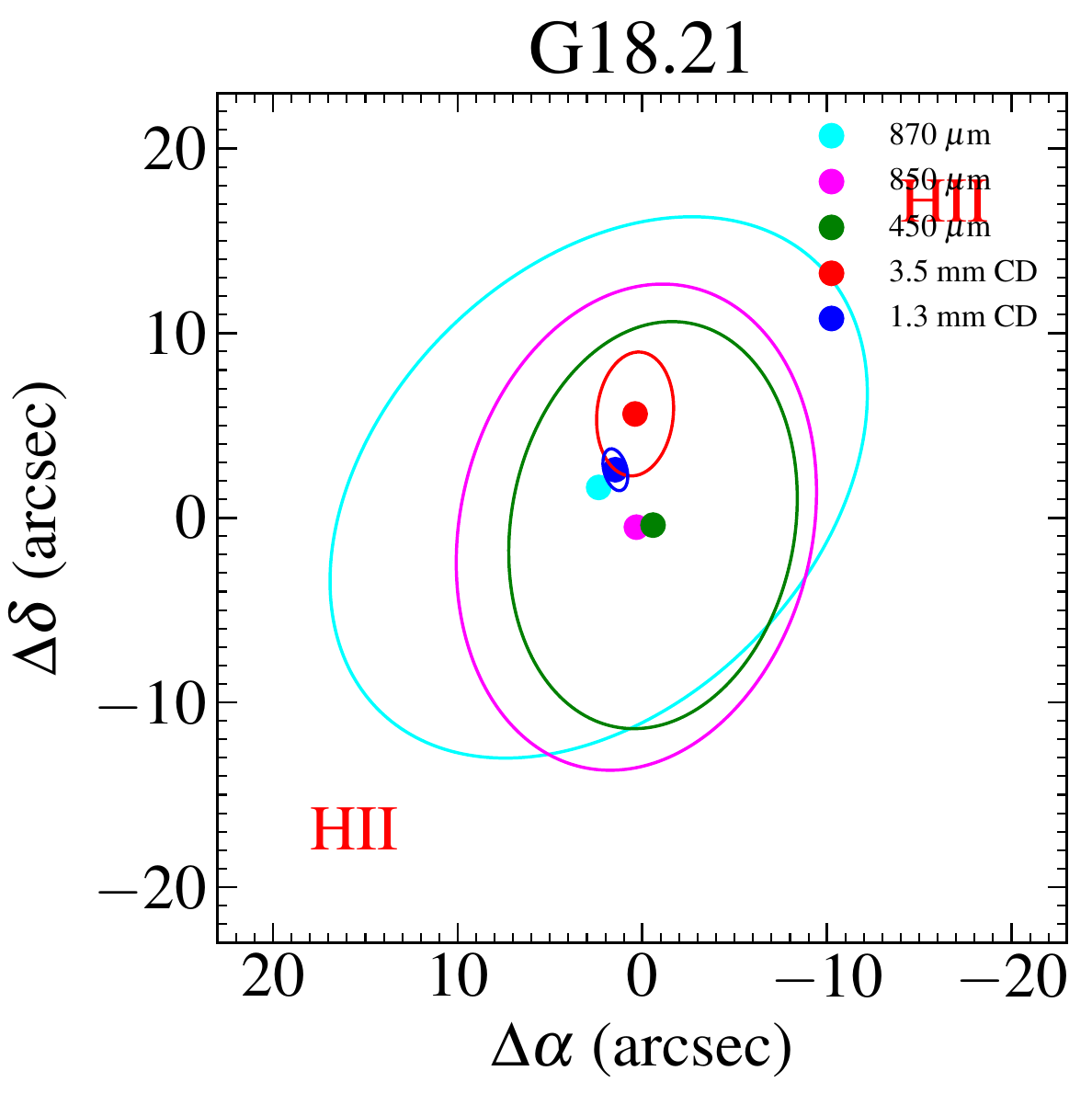}
\includegraphics[width=0.245\textwidth, angle=0]{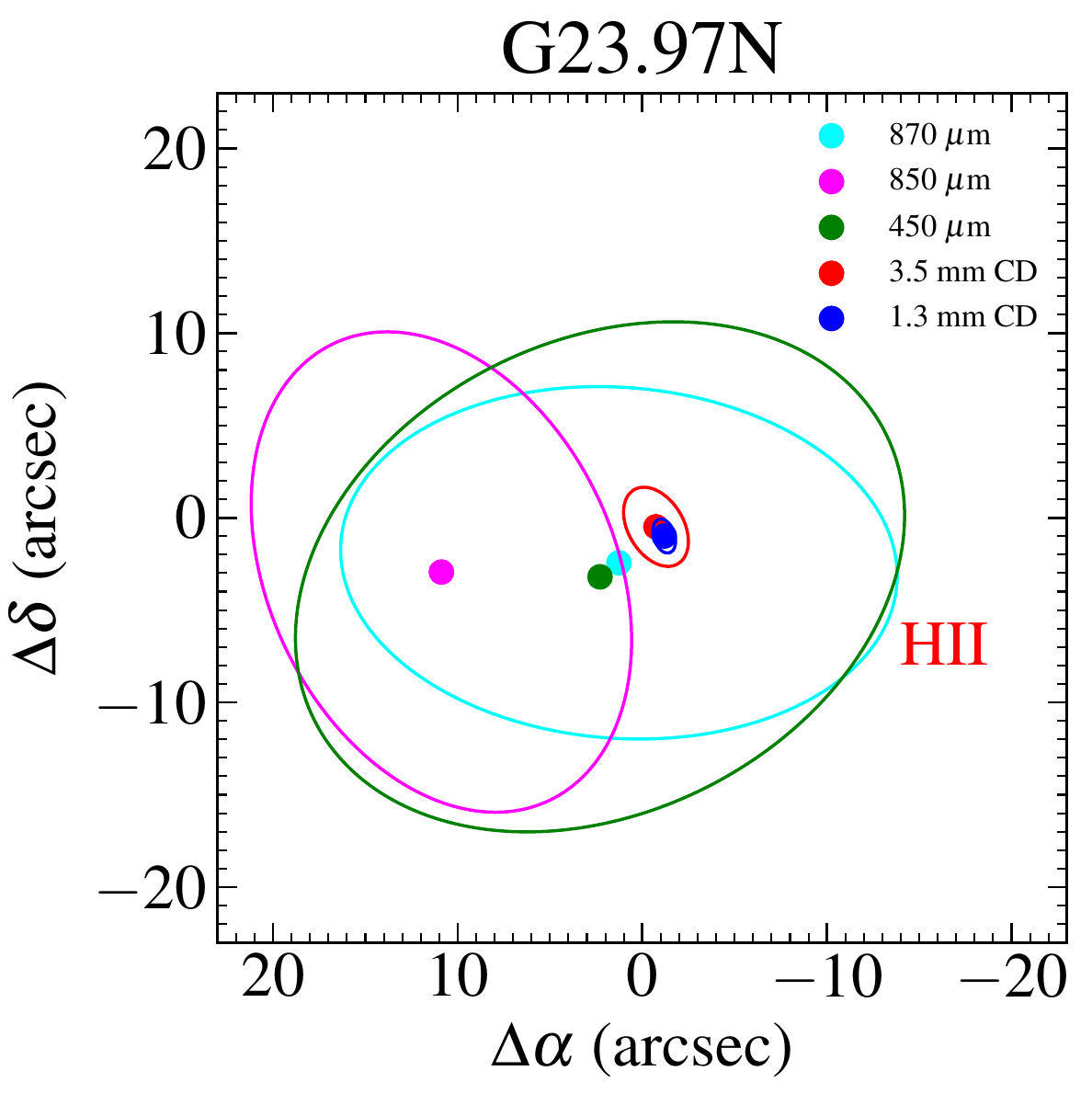}\\
\includegraphics[width=0.245\textwidth, angle=0]{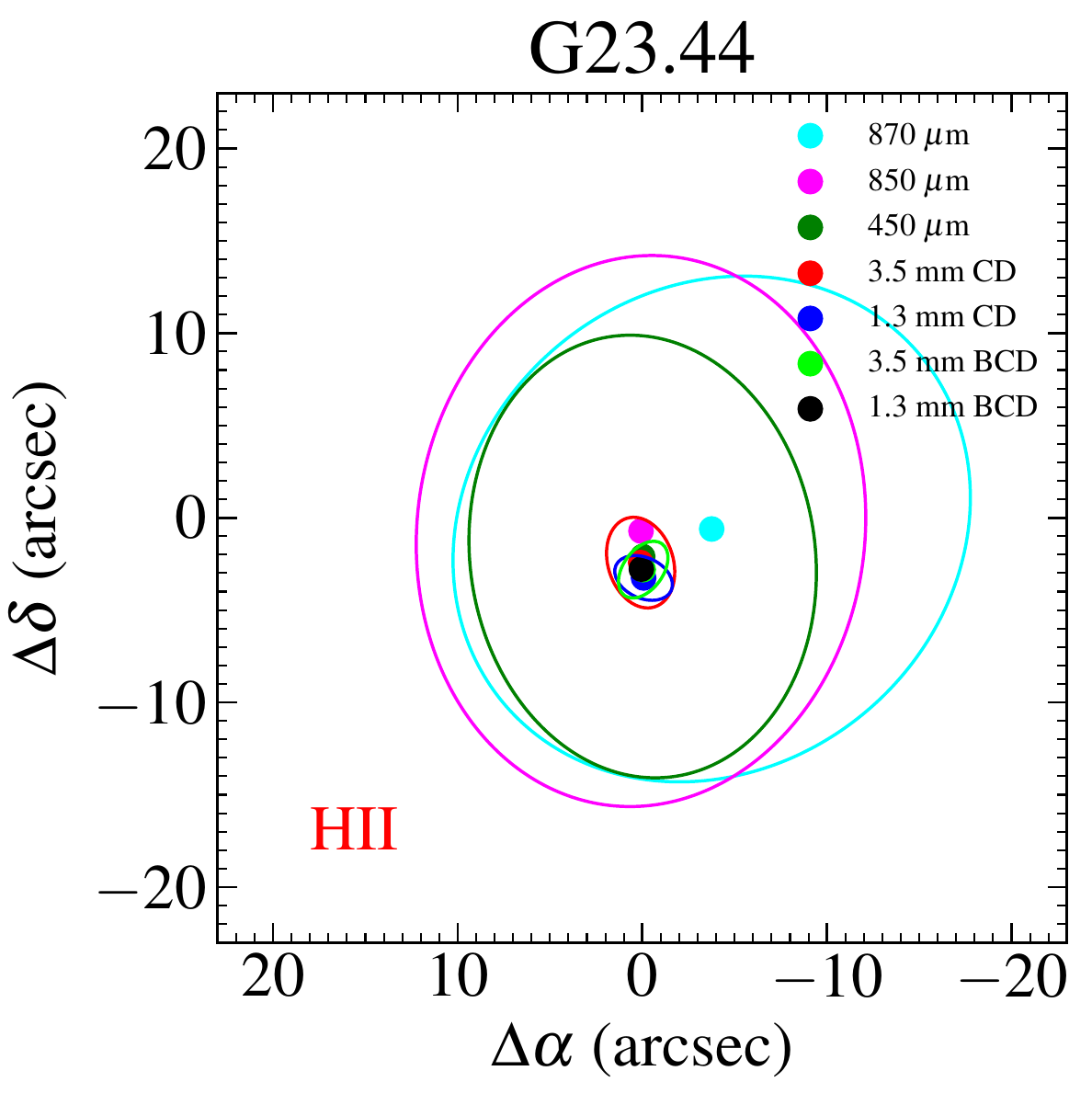}
\includegraphics[width=0.245\textwidth, angle=0]{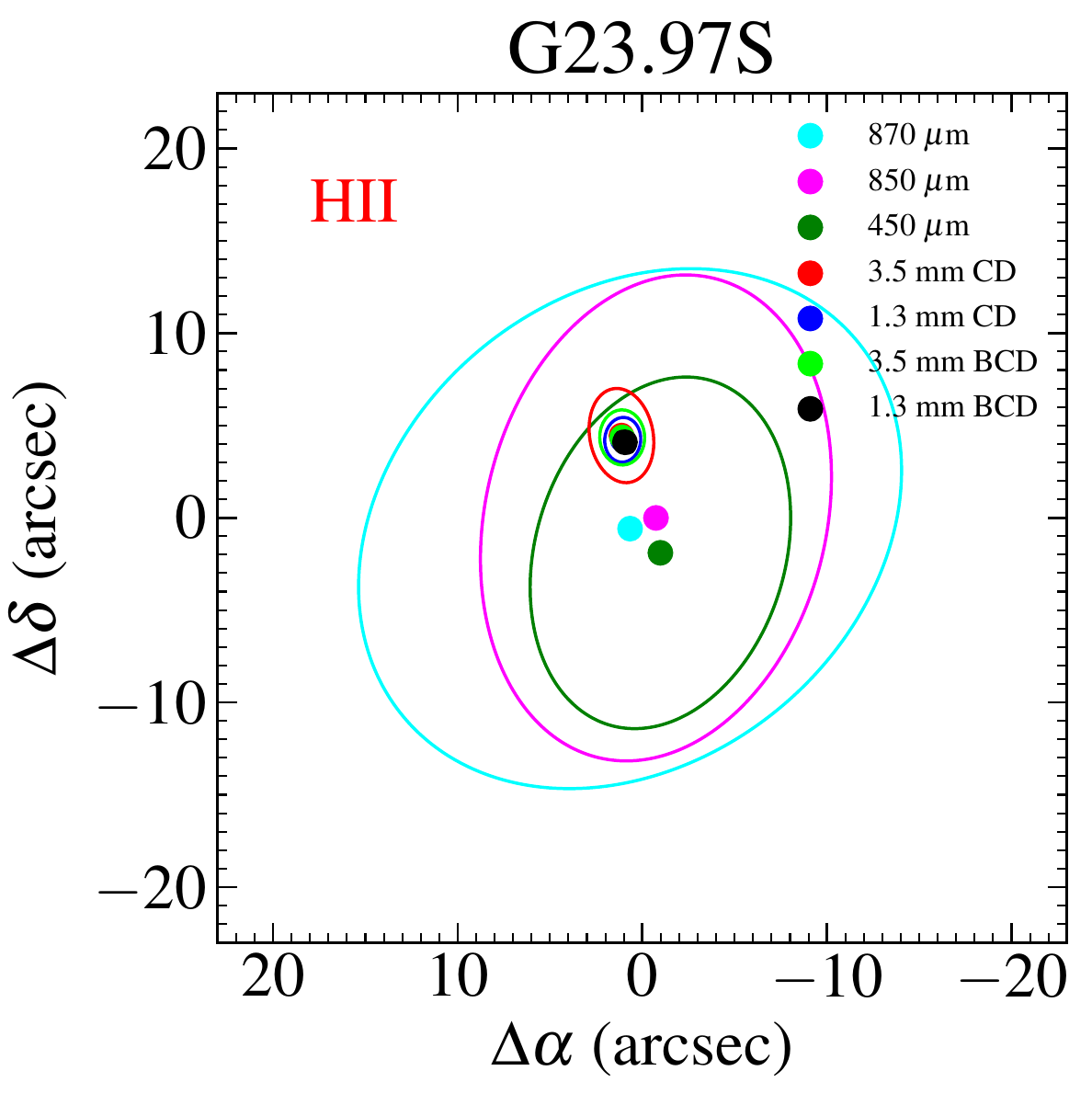}
\includegraphics[width=0.245\textwidth, angle=0]{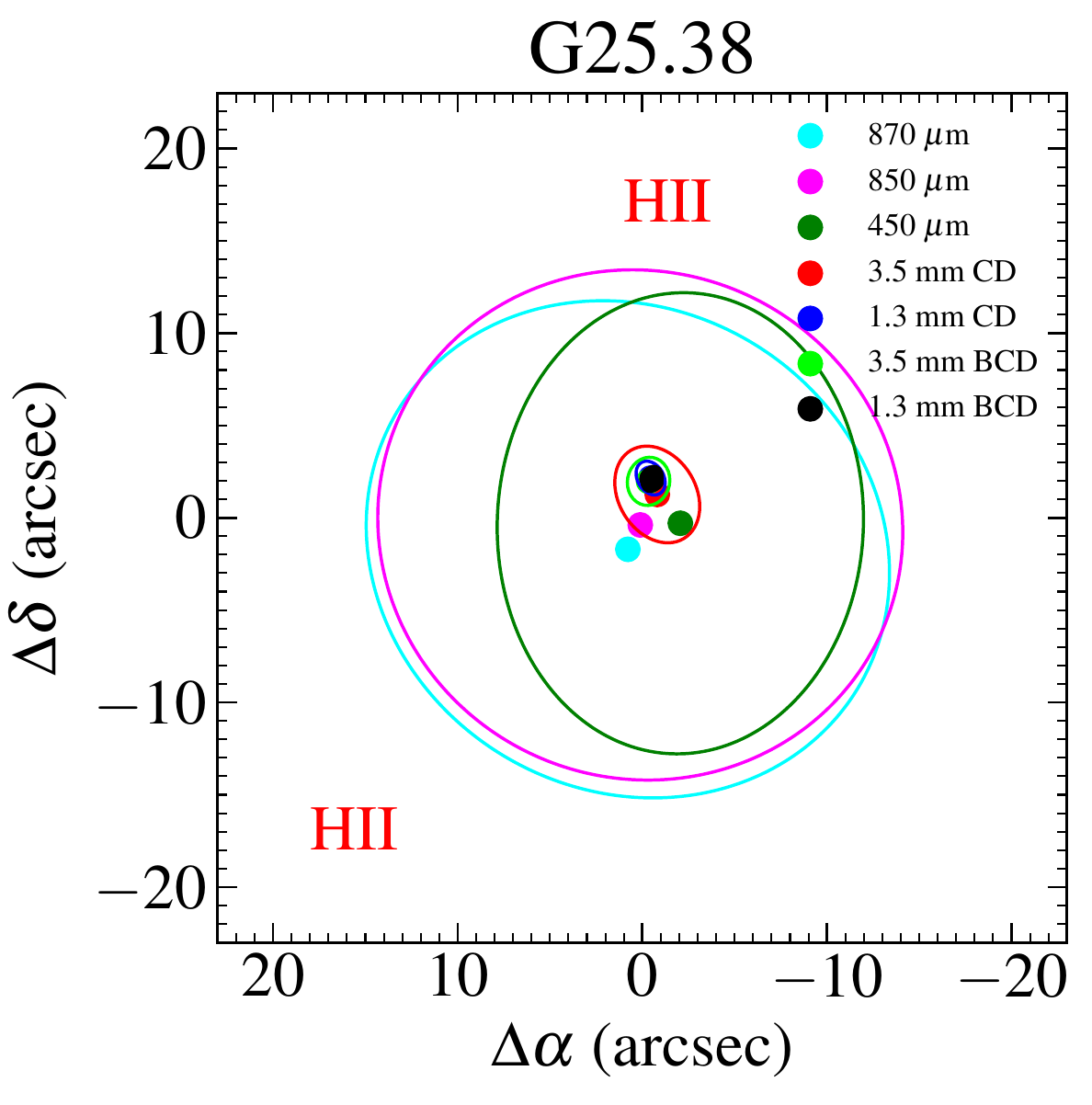}
\includegraphics[width=0.245\textwidth, angle=0]{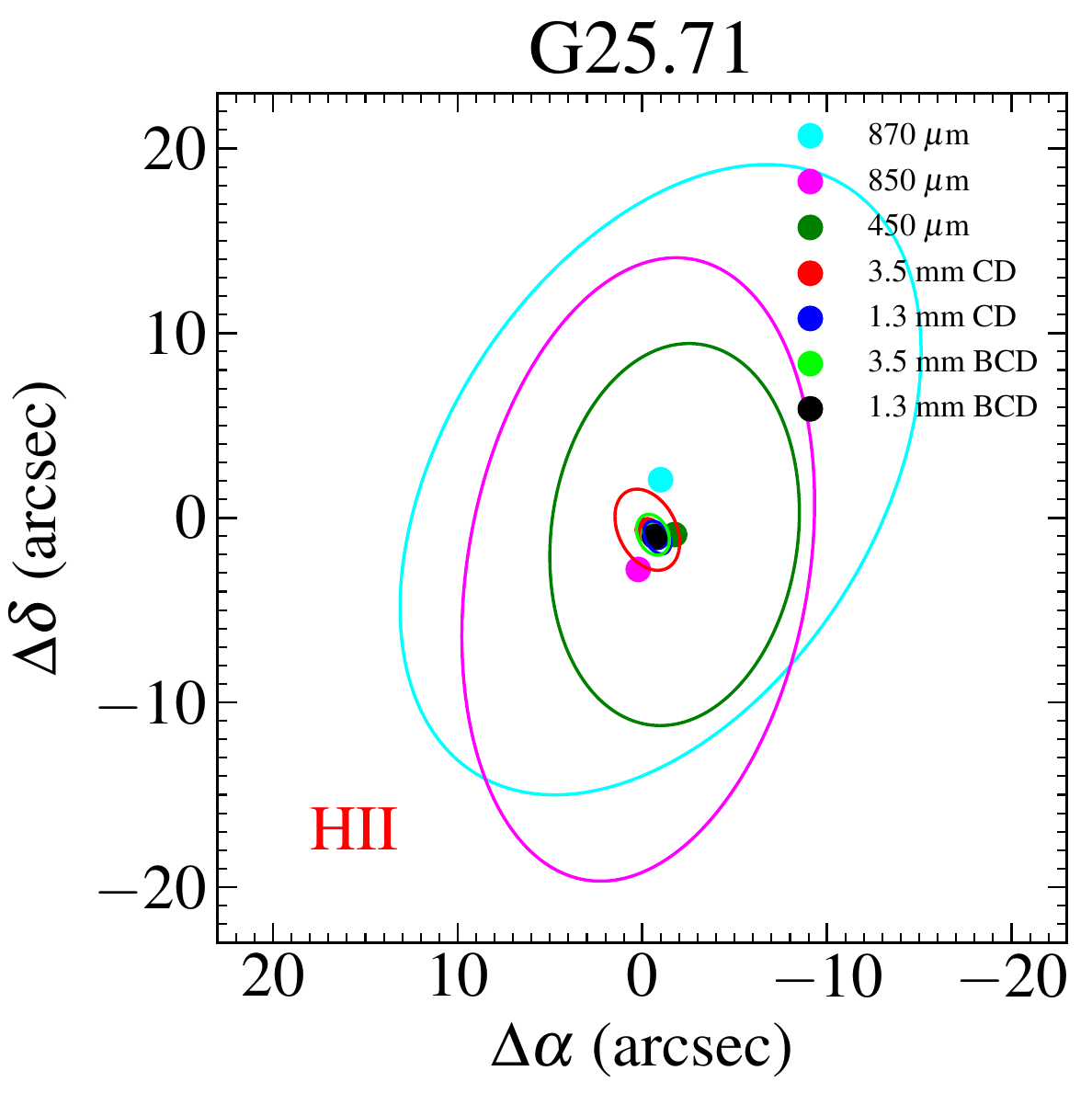}
\caption{The distribution of position offset for the most massive objects (in different color points with associated ellipse) derived from different scale observations. The ellipse shows the Gaussian size of each object. The text ``\HII'' presents the location and direction of nearby \HII regions. The offset (0, 0) is the phase center of the observations.}
\label{Fig_offset_main}
\end{figure*}


Whether, and to what extent, star formation is influenced by triggering remains an open question. Young star-forming regions are often surrounded by evolved OB stars or UC \HII regions \citep{Zhang2017a}. In this study, all sources are close to one or two \HII regions, which have already evolved into an advanced stage. The uniqueness of the sample enables us to study the effect of triggering on the dynamical evolution of gas clumps. The idea is to study the effect of the \HII region on the internal dynamics of the clumps. If the expansion of the \HII region is dynamically important, we expect it to have a noticeable effect on the morphology of the clumps. For example, the density structures of the clump would become spatially asymmetric (e.g., when the high density regions reside close to the trigger source), such that the most massive objects at different scales would align in line directed towards nearby \HII regions (see a schematic illustration in Figure\,\ref{Fig_trigger}). If triggering is dynamically unimportant, we expect to structures at different scales (and thus different densities) to form either spherically-symmetric structure or with a random distribution of asymmetries.

To quantify the spatial asymmetry in the observations, Figure\,\ref{Fig_offset_main} displays the position offset\footnote{In Figure\,\ref{Fig_offset_main}, the pointing accuracy for the LABOCA and SCUBA-2 was, respectively, $\sim$4$''$ \citep{Schuller2009} and 1.5$-$2$''$ (see \url{https://stfc.ukri.org/files/jcmtprospectuspdf/}). The position accuracy in a radio interferometric observation is of the order of 1/10th$-$1/20th of the synthesized beam (see \url{https://www.iram.fr/IRAMFR/IS/IS2002/html\_2/node130.html}).} for the most massive objects derived from observations at each spatial scale. We find that the sources G23.97N, G23.97S, and G23.44 show asymmetric spatial features with high density regions that do not reside at the center of the region and fragmented cores align in a string (see Figure\,\ref{Fig_offset_main}), indicating that the strong stellar wind from the nearby \HII region has affected on the thick clump envelope. In contrast, G25.71 shows a spherically symmetric feature, suggesting that after clump formation, the internal dynamics is self-regulated and independent of outside environment. This is consistent with the picture proposed by \citet{Li2017} that the clump belongs to a regime where the collapse is self-regulated and is dynamically detached from the background fluid. Therefore, our findings imply that although the collapse is self-regulated, external triggering might still be able to create an asymmetric initial condition that might affect further fragmentation. We also note that the current sample is still not large enough, and we expect to be able to constrain this better using this method with a larger dataset. Furthermore, a random numerical simulation will be carried out in a forthcoming work.

\section{Summary}
\label{sect summary}

We present multi-wavelength continuum observations towards a sample (G18.17, G18.21, G23.97N, G23.98, G23.44, G23.97S, G25.38, and G25.71) of high-mass star-forming clumps using radio- and millimeter interferometric data with arcsec resolution taken at 1.3\,cm, 3.5\,mm, 1.3\,mm, and 870\,$\mu$m. Our sample is unique as all the clumps are close to \HII regions. This enables us to understand the effect of an external \HII region on gravitational collapse.

We find that the targeted clumps have a mass range of 228 to 2279\,$\Msun$ with $R_{\rm eff}$ $\sim$ 0.5\,pc and $n_{\rm H_2} \sim 10^5~{\rm cm^{-3}}$. The brightest continuum sources are associated with infrared bright clumps and radio continuum sources. Two of the infrared quiet clumps host compact sources with masses up to $\sim$35\,$\Msun$ and are capable of forming B3-B0 type stars. Two infrared quiet clumps with low surface densitites ($\Sigma\sim0.14$\,g\,cm$^{-2}$) show no compact fragments at 3.5\,mm, indicating that star formation may not have yet started and might not occur in the future. Three sources (G23.97N, G23.98, and G25.38-u) show no indication of star formation despite their considerable mass and surface density and are starless, possibly prestellar clumps. The sources G23.44, G23.97S, and G25.71 may host ultra-compact \HII regions, but they are deeply embedded in a thick and cold dust envelope.

The $mass-size$ relation shows that the fragmentation slope gradually flattens from large scale to small scale and from 870\,$\mu$m to 1.3\,mm, while the \textit{surface density-mass} relation gradually steepens. This indicates that as fragmentation size decrease, the fragment mass remains roughly constant while the surface density increases sharply. The central dense condensations at a scale of around 0.02\,pc can contain most of the mass of the source with little in the surrounding envelope.

Taking advantage of the fact that all clumps have neighboring \HII regions, we test triggered star formation using a novel method where we study the alignment of the centres of mass traced by dust emission at multiple scales with the position of external \HII regions. If triggering is dynamically important, we expect the gas in the clump to be spatially aligned to due to external compression. Our investigations yield no clear patterns, which imply that collapse is self-regulated (regulated by the process of gravitational collapse) and the effect of external triggering is minimized after gravitational collapsed has begun. However, this needs to be confirmed by future studies with larger samples and a random numerical simulation.

\section*{Acknowledgements}
\addcontentsline{toc}{section}{Acknowledgements}

We thank the anonymous referee for prompting many clarifications of this paper. This work is supported by the National Natural Science Foundation of China 11703040 and National Key Basic Research Program of China (973 Program) 2015CB857101. C.-P. Zhang acknowledges support by the MPG-CAS Joint Doctoral Promotion Program (DPP) and China Scholarship Council (CSC) in Germany as a postdoctoral researcher. T. Csengeri and T. Pillai acknowledge support from the \emph{Deut\-sche For\-schungs\-ge\-mein\-schaft, DFG\/} via the SPP (priority programme) 1573 `Physics of the ISM'.









\bibliographystyle{aa}
\bibliography{references}

\appendix

\section{Distinguishing the ionised gas from cold dust}
\label{sect_sed}

\begin{figure*}[ht]
\centering
\includegraphics[width=0.45\textwidth, angle=0]{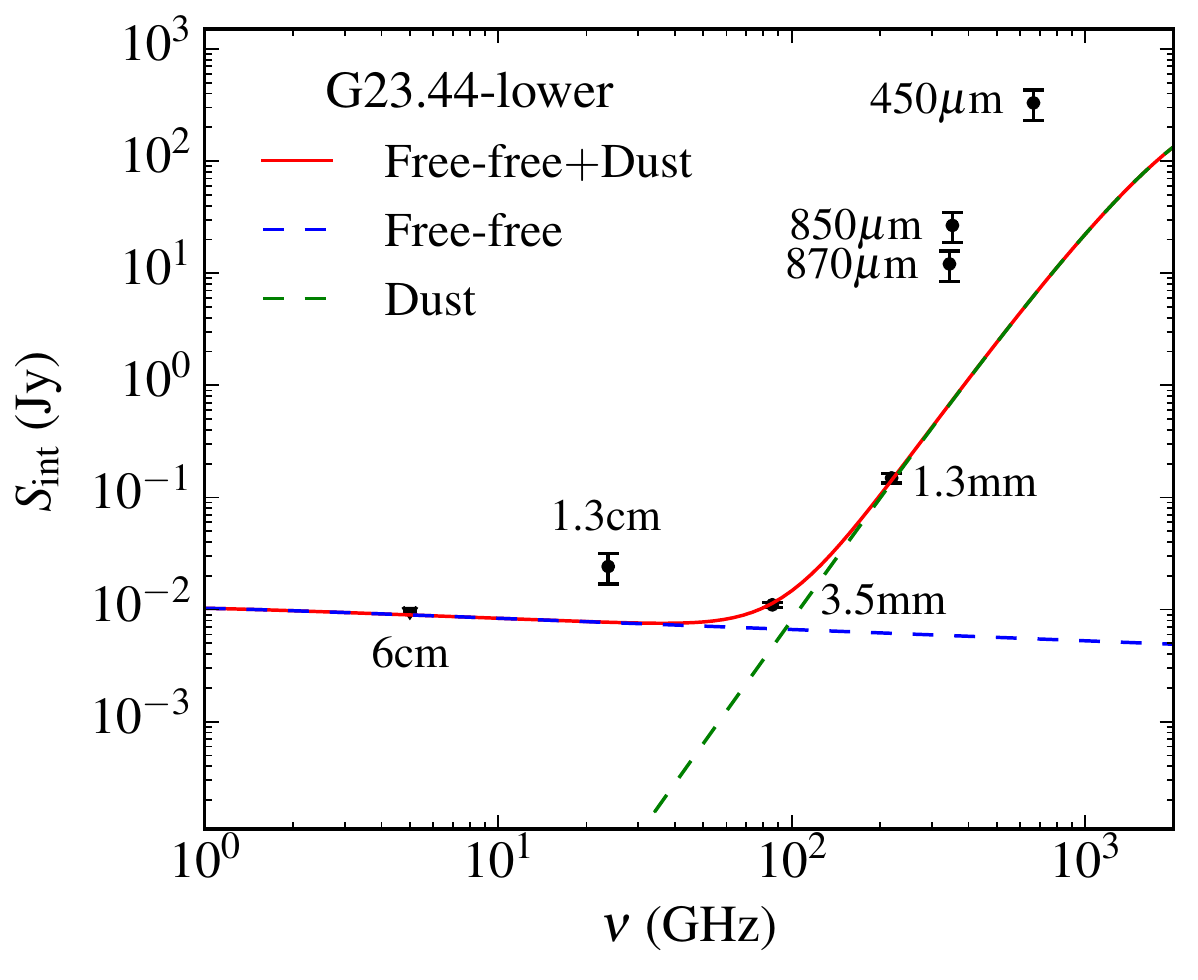}
\includegraphics[width=0.45\textwidth, angle=0]{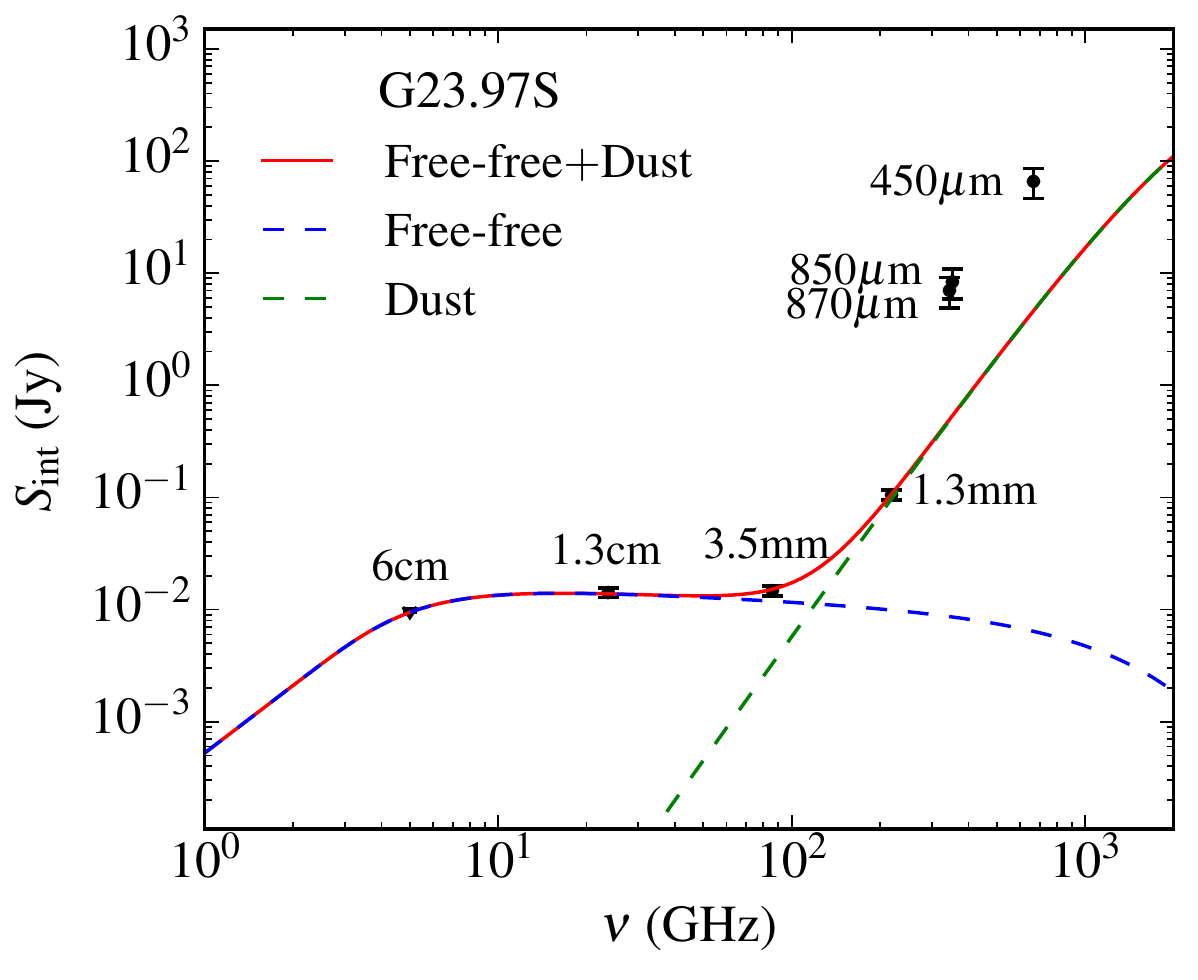}
\includegraphics[width=0.45\textwidth, angle=0]{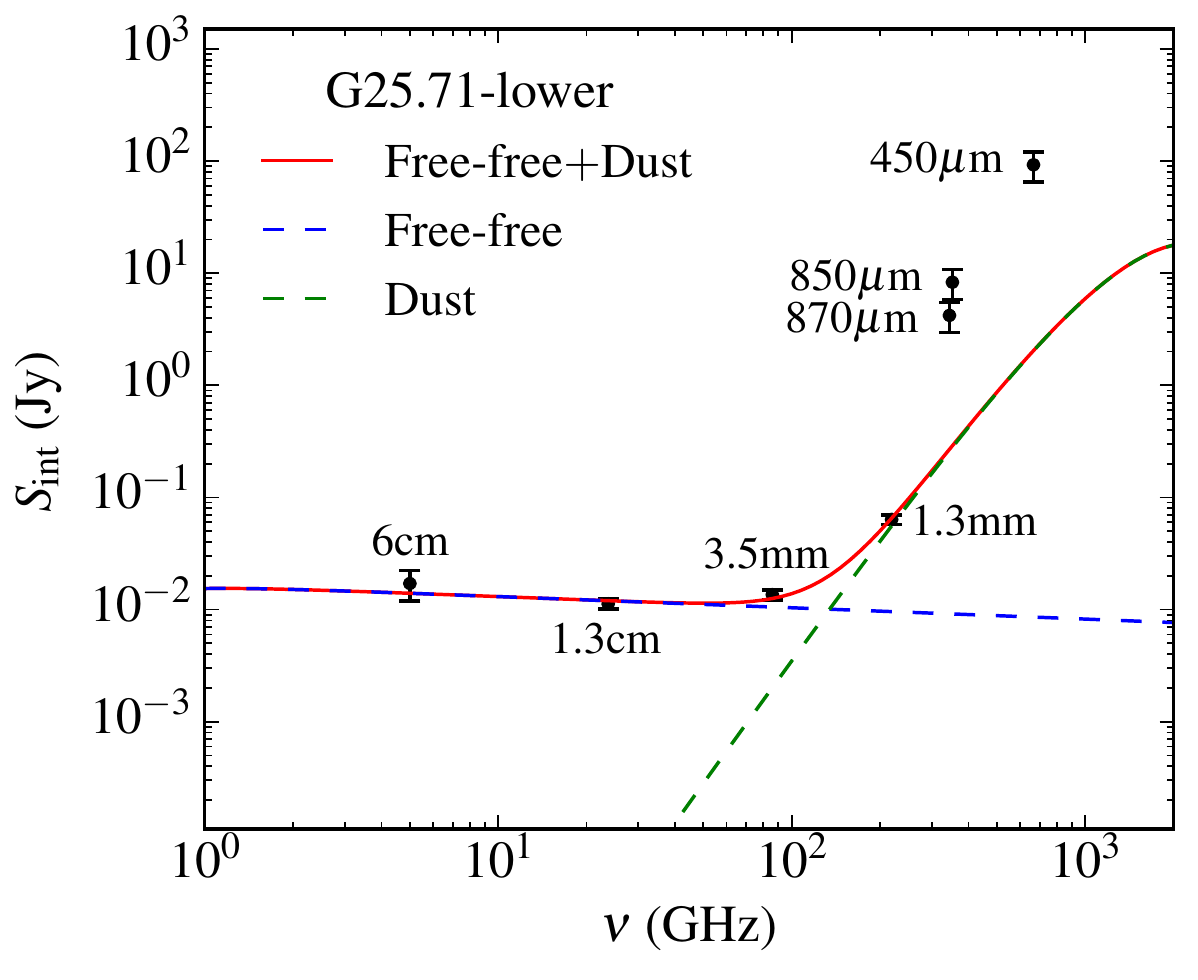}
\includegraphics[width=0.45\textwidth, angle=0]{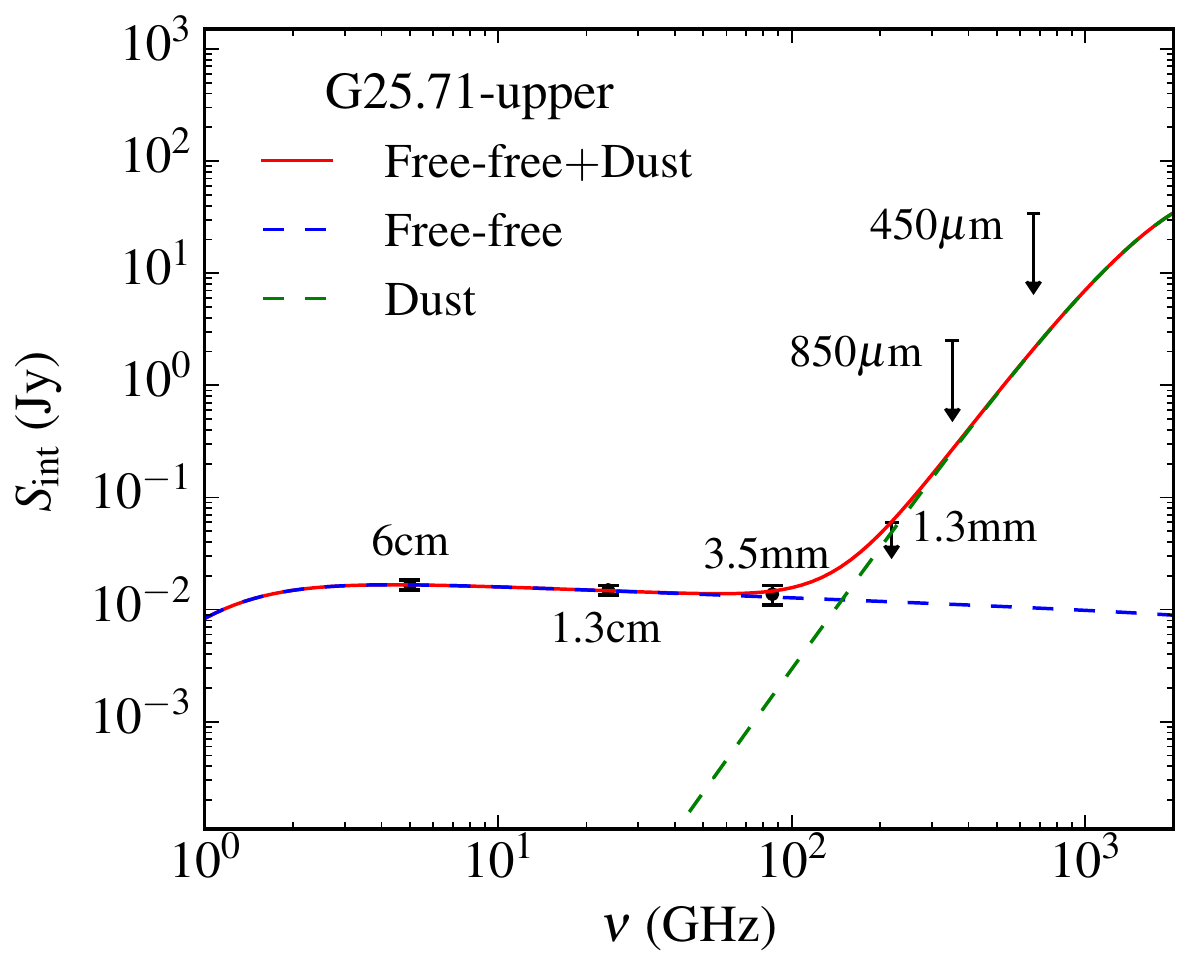}
\caption{SED fittings for cores G23.44-l, G23.97S, G25.71-l, and G25.71-u, combining 6\,cm (CORNISH), 1.3\,cm (VLA), 3.5\,mm (PdBI), 1.3\,mm (PdBI), 870\,$\mu$m (ATLASGAL), 850\,$\mu$m (SCUBA), and 450\,$\mu$m (SCUBA). The spectral index of the dust emissivity is set as $\beta$ = 1.7. The ionised gas and cold dust components were shown in dashed blue and dashed green lines, respectively. For core G25.71-u, there is no observation at 1.3\,mm for its offset away from the primary beam size, so we take the measurements on 850 and 450\,$\mu$m flux as two upper limits.}
\label{Fig_SED}
\end{figure*}

Towards compact \HII regions, the measured continuum fluxes near 3.5\,mm have contributions from both the ionised gas and dust emission \citep{Churchwell2002,Zhang2014}. The free-free emission from ionised gas may strongly boost the flux at 3.5\,mm, when a bright \HII region or protostar is located in the field. For the mass estimations based on the dust, it is, therefore, necessary to distinguish between the contaminating free-free emission and the thermal dust emission. In this work, we used archival continuum data at 6\,cm and 1.3\,cm to look for point sources associated with the dust peaks. The 6\,cm data is from the Co-Ordinated Radio `N' Infrared Survey for High-mass star formation (CORNISH), which has a spatial resolution of 1.5$''$ with a noise level <0.4\,mJy\,beam$^{-1}$. The 1.3\,cm continuum is derived from our VLA observations (see Section\,\ref{sect_vla_obs}) with a noise level of 0.6$-$1.2\,mJy\,beam$^{-1}$. If the 6\,cm or 1.3\,cm emission cannot be detected above 3$\sigma$, we use the 3$\sigma$ upper limit for the fitting of their spectral energy distribution (SED). Towards 27 dust cores, we detect 4 sources (G23.44-1, G23.97S-1, G25.71-1, and G25.71-2) with free-free emission judging from the 1.3\,cm data (see Figure\,\ref{Fig_G23.44}), while only the sources G25.71-1 and G25.71-2 have 6\,cm emission. These detections are consistent with these clumps being classified as infrared bright clumps. We estimate the contamination by free-free emission using SED fitting combining 6\,cm (CORNISH), 1.3\,cm (VLA), 3.5\,mm (PdBI), 1.3\,mm (PdBI), 870\,$\mu$m (ATLASGAL), 850\,$\mu$m (SCUBA), and 450\,$\mu$m (SCUBA).

For SED fitting (see also Figure\,\ref{Fig_SED}), the free-free emission model \citep{mezg1967} is
     \begin{equation}
     \left\{ \begin{aligned}
       S_\nu &= \Omega_s B_\nu(T_{\rm e}) (1-{\rm e}^{-\tau_c}) (\rm Jy) & \\
       \tau_c &= 0.08235~\alpha(\nu, T_{\rm e})
                   \left(\frac{\nu}{{\rm GHz}}\right)^{-2.1}
                   \left(\frac{T_{\rm e}}{{\rm K}}\right)^{-1.35}
                   \left(\frac{EM}{{\rm pc~cm^{-6}}}\right) & \\
     \end{aligned}, \right.
     \end{equation}
where $S_v$ is the integration flux at frequency $\nu$, $\Omega_s$ is the solid angle, $B_\nu(T_{\rm e})$ is the Planck function with variable electron temperature $T_{\rm e}$, $\alpha(\nu, T_{\rm e})$ $\sim$ 1, and $EM$ is the emission measure. The dust emission enters \citep{Woody1989} as
   \begin{eqnarray}
   \left(\frac{S_\nu}{\rm Jy}\right) = 52.36
                \left(\frac{M}{\Msun}\right)
                \left(\frac{0.25}{{\lambda_{\rm mm}}}\right)^{\beta+3}
   \left[{\rm e}^{\left(\frac{14.4}{{\lambda_{\rm mm}T_{\rm dust}}}\right)}-1\right]^{-1}
   \left(\frac{D}{{\rm kpc}}\right)^{-2}, &
     \end{eqnarray}
where $M$ is the mass of dusty gas, $\beta$ is the spectral index of the dust emissivity, $T_{\rm dust}$ is the dust temperature (approximately equal to kinetic temperature; see Section\,\ref{sect_sed}) in Kelvin, $\lambda_{\rm mm}$ is the wavelength in mm, and $D$ is the distance. In the models, we let $M$, $T_{\rm e}$, and $EM$ vary as free parameters, set $\beta$ to 1.7 \citep{Battersby2011,PlanckXIV2014}, and use the least square method to fit the data. For the 6\,cm data, we use its 3$\sigma$ upper limit; for other wavelength data, we use \texttt{Gaussclumps} to extract their fluxes as listed in Tables\,\ref{tab_atlasgal}, \ref{tab_3.5mm}, and \ref{tab_1.3mm}. The 870, 850, and 450\,$\mu$m data are from single dish observations and have much lower resolutions than other interferometer data in this work. However, the single dish observations can be used to measure total flux of each source. To estimate missing masses at different scales between interferometer and single dish observations, the wavelengths at 870, 850, and 450\,$\mu$m don't participate in the SED fitting.  Additionally, we assume that the turnover occurs shortward of 3.5\,mm at wavelengths where the free-free is already negligible compared to the dust emission.

From the SED fitting, we find that the contribution from the dust emission to the measured total continuum flux is significant, e.g., 40.8\% for G23.44 (No.1 for CD and BCD configuration observations) and 21.7\% for G23.97S (No.1 for CD and BCD configuration observations) at 3.5\,mm. These results are indicated in Table\,\ref{tab_3.5mm}. The 1.3\,mm flux is also a key parameter in distinguishing the dust component from free-free emission with SED fitting. If the cores are not detected at 6\,cm and 1.3\,cm emission, we will assume that the contribution is 100\% from dust emission.

\section{Missing flux}
\label{sect_missing}

Interferometric observations filter out emission from large scales due to missing information on short spacings for the 1.3\,mm observations. According to the interferometer baseline range of 16.0\,m (11.7\,k$\lambda$) to 452.1\,m (331.1\,k$\lambda$), we obtain a rough estimate of about 21.6$''$ above which flux is filtered out. This means that we recover emission from the compact sources only up to these scales, which impacts the mass estimates. As Figure\,\ref{Fig_SED} shows, the fitted fluxes at 870\,$\mu$m are much lower than the measured fluxes. The difference between the measured and fitted fluxes at 870\,$\mu$m suggests that between 92.0\% to 94.6\% of the flux is filtered out by the interferometer. Assuming that the gas emitting and 1.3\,mm and at 870\,$\mu$m has the same origin, we can extrapolate that typically a large fraction of the flux at 1.3\,mm is lost by filtering from the clump to the condensation scale. We find two clumps where all the flux seems to be filtered by the interferometer suggesting that there are no bright compact objects formed yet within these clumps.

In practice, the high resolution interferometric observations typically resolve the clumps into several fragments. We derived their masses based on our SED fitting with which we disentangled contribution of the thermal dust and the free-free emission, and therefore could estimate the mass of each fragment on 5000\,AU size scales. If we take into account that the 870\,$\mu$m flux is not entirely emitted from the same region as the 1.3\,mm condensation, to more accurately derive missing flux we should estimate the mass ratio between the most massive condensation and the clustered condensations of each clump. In the 1.3\,mm CD configuration observations, the most massive condensation within each cluster takes between 40.6\% and 66.3\% of the total condensation mass, which means that about 50\% of the clump mass had been resolved out into other fragments at higher resolutions. Based on above, there may exist about 46.0\% ($50\%\times92.0\%$) to 47.3\% ($50\%\times94.6\%$) missing flux above a typical scale of 21.6$''$, due to missing short spacings.

\section{Individual clumps}
\label{sect_individual}

\subsection{G18.17}
\label{sect_G18.17}
\begin{figure*}
\centering
\subfigure[]{\includegraphics[width=0.65\textwidth, angle=0]{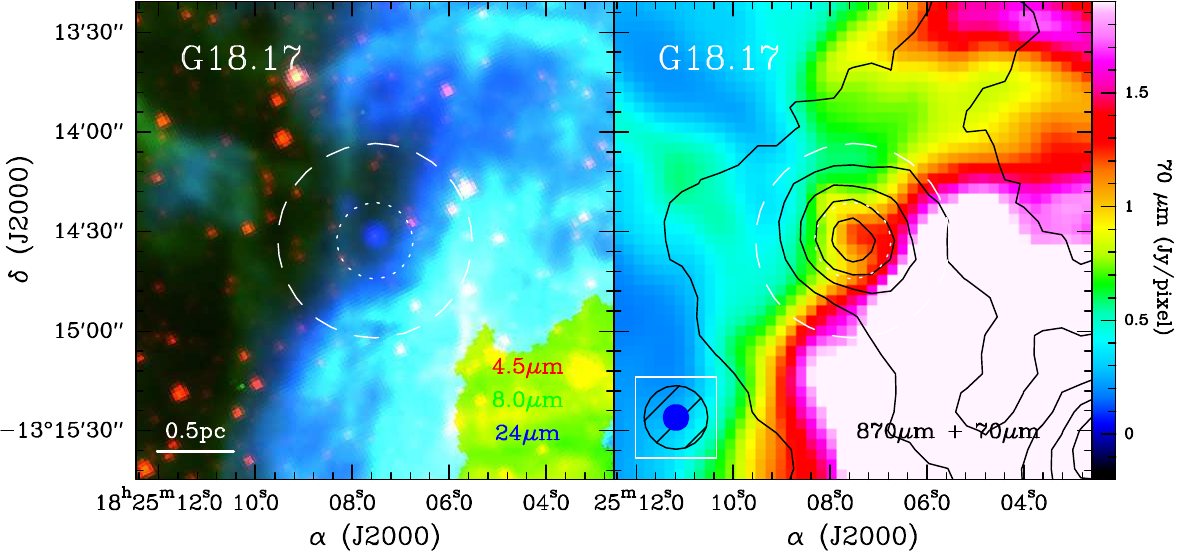}}
\subfigure[]{\includegraphics[width=0.33\textwidth, angle=0]{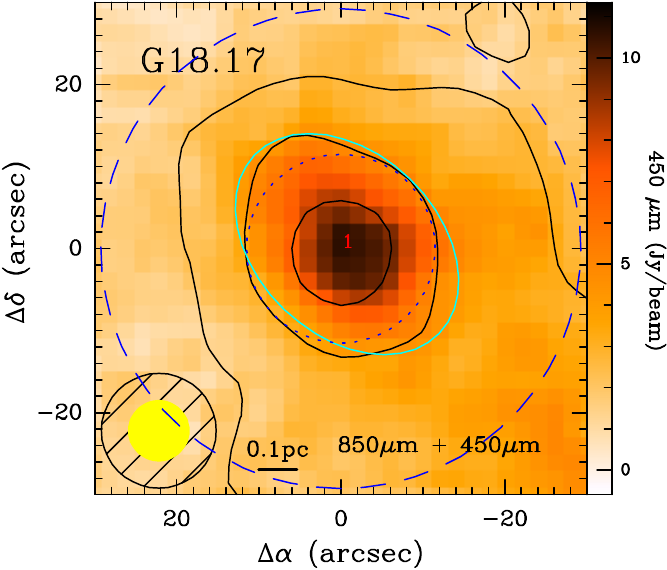}}
\subfigure[]{\includegraphics[width=0.33\textwidth, angle=0]{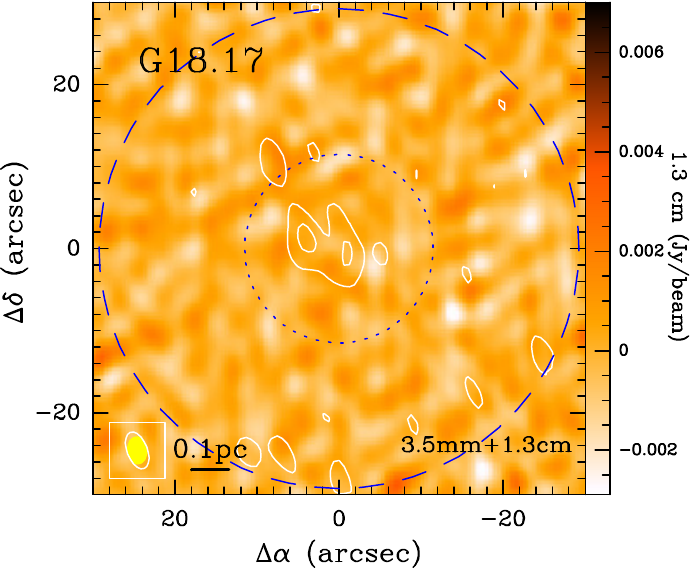}}
\subfigure[]{\includegraphics[width=0.33\textwidth, angle=0]{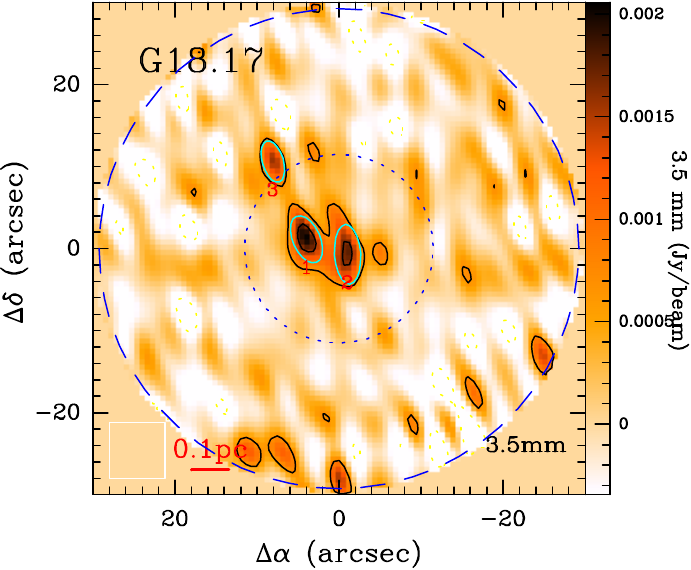}}
\subfigure[]{\includegraphics[width=0.33\textwidth, angle=0]{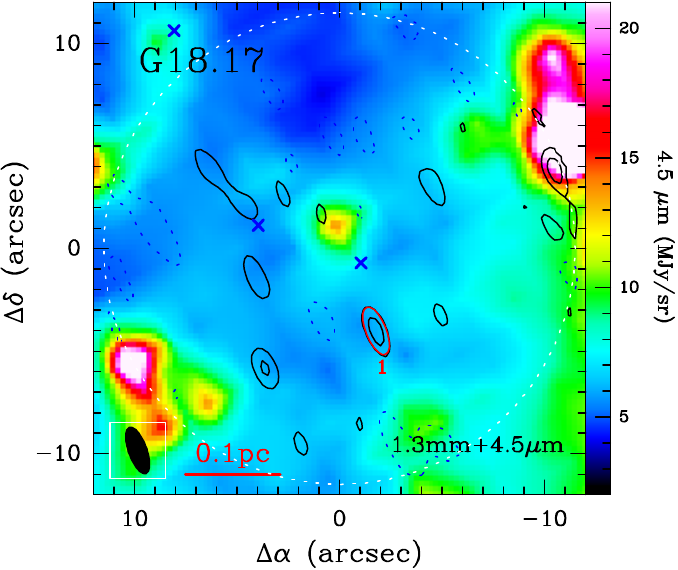}}
\caption{G18.17: {\it (a)} (left) Three-color image of 4.5\,$\mu$m (blue), 8.0\,$\mu$m (green), and 24\,$\mu$m (red); (right) 870\,$\mu$m contours overlaid on a 70\,$\mu$m color scale. The 870\,$\mu$m contour levels start at 6$\sigma$ in steps of 8$\sigma$ ($\sigma$ = 54\,$\mjyb$). {\it (b)} 850\,$\mu$m contours overlaid on a 450\,$\mu$m color scale. The 850\,$\mu$m contour levels start at 6$\sigma$ in steps of 8$\sigma$ ($\sigma$ = 83\,$\mjyb$). {\it (c)} 3.5\,mm contours overlaid on a 1.3\,cm color scale. For CD configuration observations, {\it (d)} the 3.5\,mm contour levels start at -3$\sigma$, 3$\sigma$ in steps of 4$\sigma$ ($\sigma$ = 0.23\,$\mjyb$), and {\it (e)} the 1.3\,mm contour levels start at -3$\sigma$, 3$\sigma$ in steps of 2$\sigma$ ($\sigma$ = 1.67\,$\mjyb$) superimposed on a 24\,$\mu$m color scale. The ellipses with numbers indicate the positions of extracted sources. The crosses in the 1.3\,mm CD configuration observations indicate the positions of the corresponding 3.5\,mm CD configuration observations. The dashed and dotted circles in each subfigure indicate the primary beam scales of 3.5 and 1.3\,mm PdBI tracks, respectively.}
\label{Fig_G18.17}
\end{figure*}

This clump is infrared quiet with flux $S_{\rm 24\,\mu m}$ = 214.9\,mJy at 24\,$\mu$m (see Table\,\ref{tab_infrared}), and has a neighboring UC \HII region \citep[see Figure\,\ref{Fig_G18.17}(a) and also the work in][]{Zhang2017a}. It is located at a near kinematic distance of 3.73\,kpc \citep{Wienen2012,Reid2009}. Given the strong infrared bright background, it is difficult to present our relatively weak target in Figure\,\ref{Fig_G18.17}(a). At 8.0\,$\mu$m there is one apparently bow-shaped border separating the clump G18.17 from the UC \HII region, indicating radiation feedback from the UC \HII region nearby \citep{Zhang2017a}. At 70\,$\mu$m, both of them are probably merging together, but we can still identify the clump G18.17 with a convex morphology.

There is no detected emission at 6 and 1.3\,cm (see Figure\,\ref{Fig_G18.17}(c)), so G18.17 is not a candidate UC \HII region. There is strong dust emission at 870, 850, and 450\,$\mu$m, which is highly consistent with the weak 70\,$\mu$m emission. The derived mass is about 387.3\,$\Msun$ with an effective radius $R_{\rm eff}$ = 0.30\,pc at 870\,$\mu$m. From the physical parameters, we believe that the clump G18.17 is a very early high-mass star-forming candidate. This clump has fragmented into three cores each with a mass above 17.6\,$\Msun$ at 3.5\,mm, but there is almost no corresponding object at 1.3\,mm above 5$\sigma$ (see Figures\,\ref{Fig_G18.17}(d)(e)).

\subsection{G18.21}
\label{sect_G18.21}
\begin{figure*}
\centering
\subfigure[]{\includegraphics[width=0.65\textwidth, angle=0]{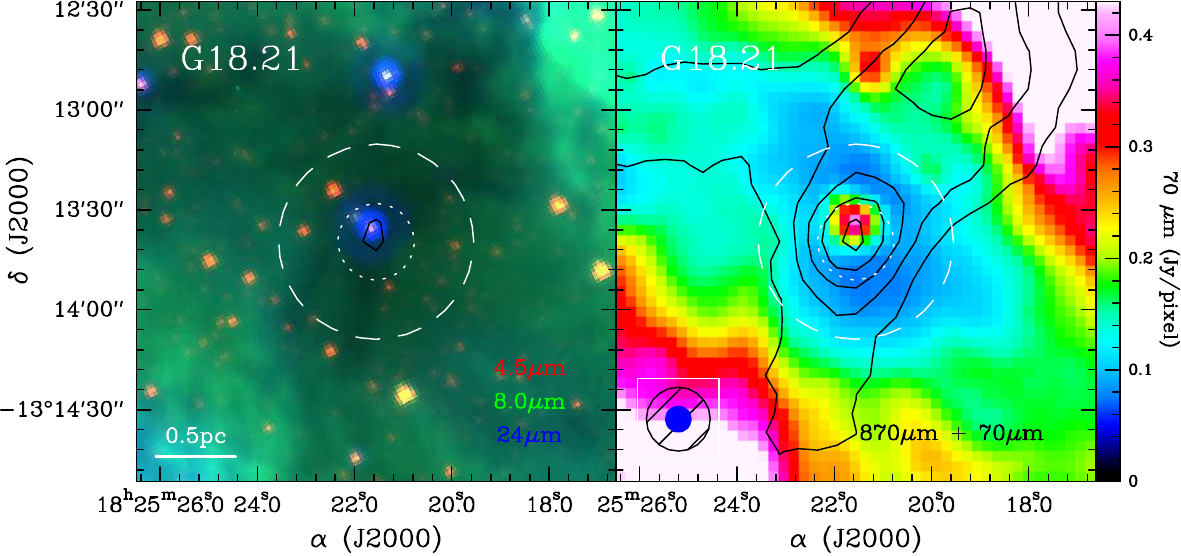}}
\subfigure[]{\includegraphics[width=0.33\textwidth, angle=0]{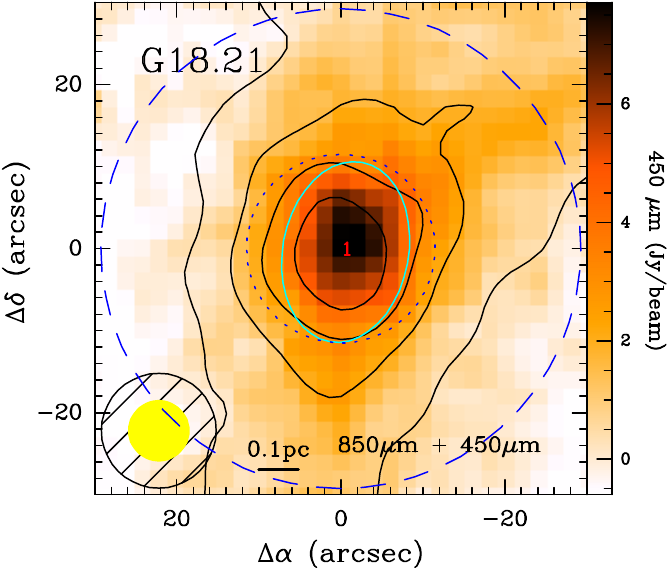}}
\subfigure[]{\includegraphics[width=0.33\textwidth, angle=0]{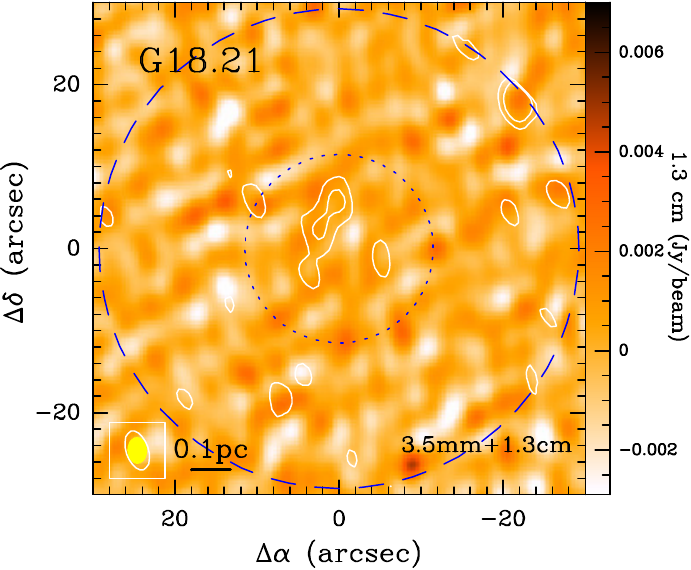}}
\subfigure[]{\includegraphics[width=0.33\textwidth, angle=0]{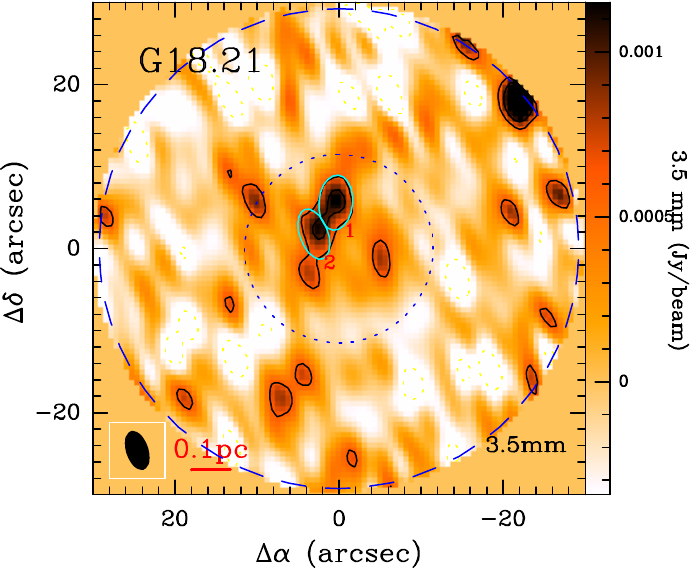}}
\subfigure[]{\includegraphics[width=0.33\textwidth, angle=0]{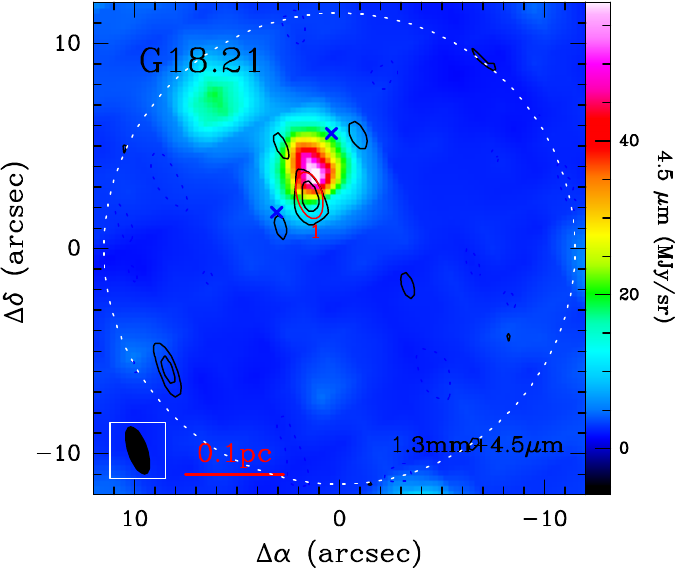}}
\caption{G18.21: {\it (a)} (left) Three-color image of 4.5\,$\mu$m (blue), 8.0\,$\mu$m (green), and 24\,$\mu$m (red); (right) 870\,$\mu$m contours overlaid on a 70\,$\mu$m color scale. The 870\,$\mu$m contour levels start at 6$\sigma$ in steps of 7$\sigma$ ($\sigma$ = 54\,$\mjyb$). {\it (b)} 850\,$\mu$m contours overlaid on a 450\,$\mu$m color scale. The 850\,$\mu$m contour levels start at 1.2$\sigma$ in steps of 4.8$\sigma$ ($\sigma$ = 83\,$\mjyb$). {\it (c)} 3.5\,mm contours overlaid on a 1.3\,cm color scale. For CD configuration observations, {\it (d)} the 3.5\,mm contour levels start at -3$\sigma$, 3$\sigma$ in steps of 2$\sigma$ ($\sigma$ = 0.20\,$\mjyb$), and {\it (e)} the 1.3\,mm contour levels start at -3$\sigma$, 3$\sigma$ in steps of 2$\sigma$ ($\sigma$ = 1.81\,$\mjyb$) superimposed on a 24\,$\mu$m color scale. The ellipses with numbers indicate the positions of extracted sources. The crosses in the 1.3\,mm CD configuration observations indicate the positions of the corresponding 3.5\,mm CD configuration observations. The dashed and dotted circles in each subfigure indicate the primary beam scales of 3.5 and 1.3\,mm PdBI tracks, respectively.}
\label{Fig_G18.21}
\end{figure*}

From Figure\,\ref{Fig_G18.21}(a), the clump G18.21 has two bright UC \HII regions nearby \citep[see also the work in][]{Zhang2017a}. The photometric flux of the clump at 24\,$\mu$m is 773.3\,mJy, and belongs to infrared quiet source (see Table\,\ref{tab_infrared}), The emission at 70\,$\mu$m is very weak at only 4.6\,Jy. It is located at a near kinematic distance of 3.60\,kpc \citep{Wienen2012,Reid2009}. The clump G18.21 is embedded in an infrared dark cloud. There is no detected emission at 6 or 1.3\,cm. The bright emission at 870, 850, and 450\,$\mu$m shows that this clump is a cold and massive star-forming object with a mass of 501.8\,$\Msun$ at 870\,$\mu$m.

The clump has fragmented into two cores each with mass above 15.3\,$\Msun$ ($R_{\rm eff}$ = 0.045$-$0.052 pc) at 3.5\,mm. The condensation with a mass of 5.9\,$\Msun$ ($R_{\rm eff}$ =  0.016\,pc) identified at 1.3\,mm is located at the middle position of the two cores at 3.5\,mm. There is an obvious offset between the 3.5, 1.3 and 4.5\,$\mu$m peaks.

\subsection{G23.97N}
\label{sect_G23.97N}
\begin{figure*}
\centering
\subfigure[]{\includegraphics[width=0.65\textwidth, angle=0]{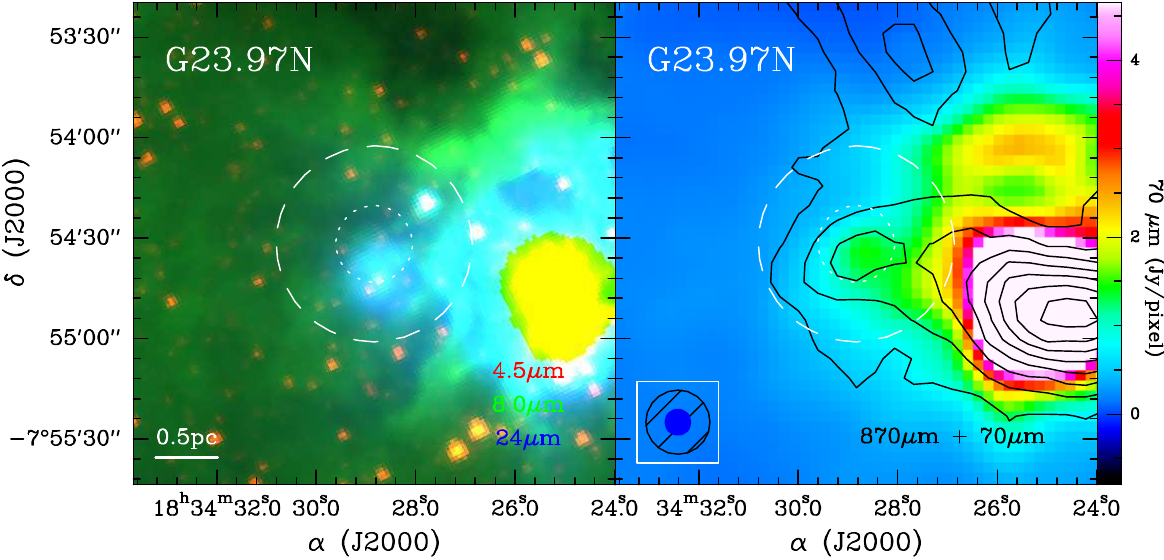}}
\subfigure[]{\includegraphics[width=0.33\textwidth, angle=0]{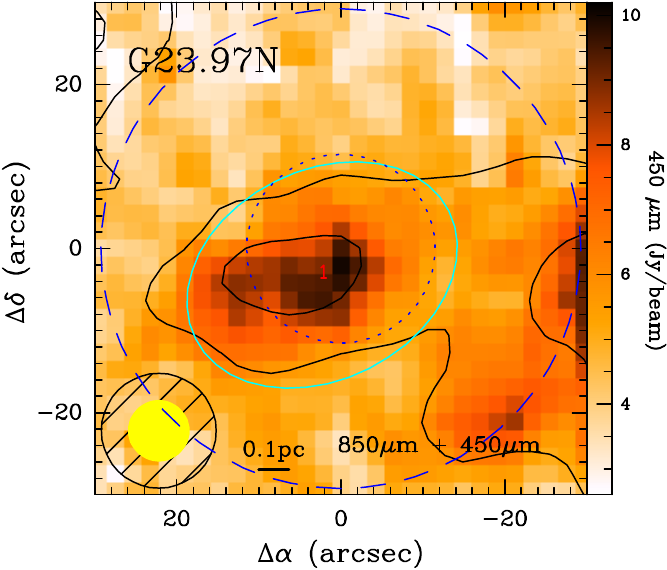}}
\subfigure[]{\includegraphics[width=0.33\textwidth, angle=0]{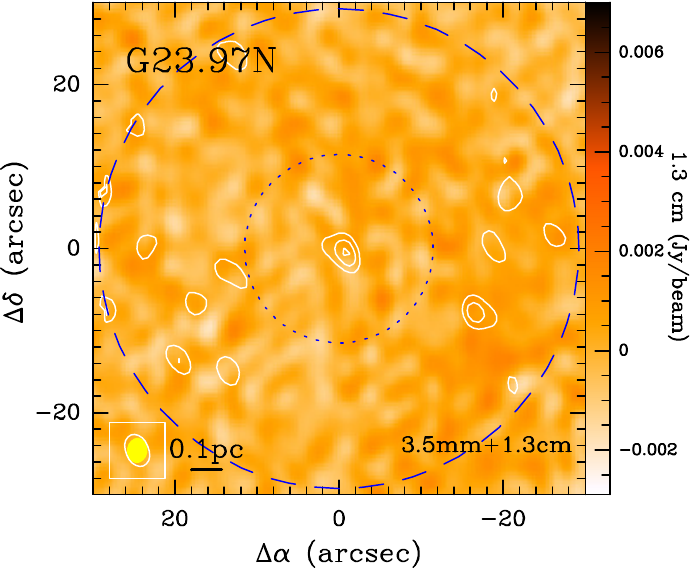}}
\subfigure[]{\includegraphics[width=0.33\textwidth, angle=0]{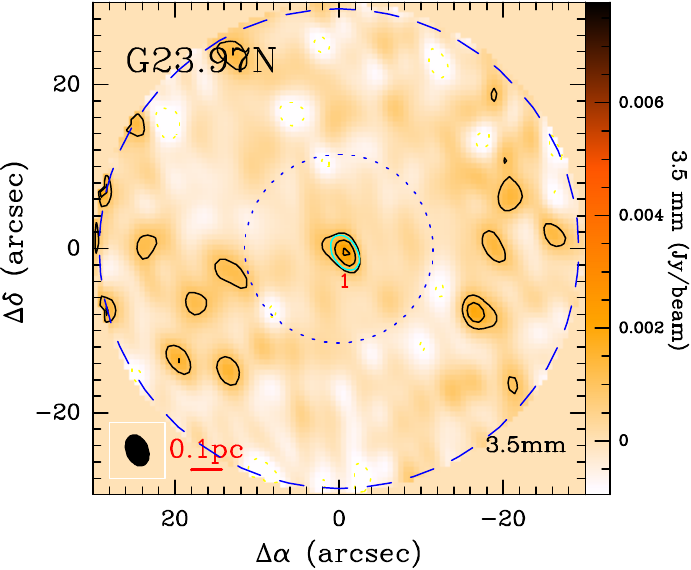}}
\subfigure[]{\includegraphics[width=0.33\textwidth, angle=0]{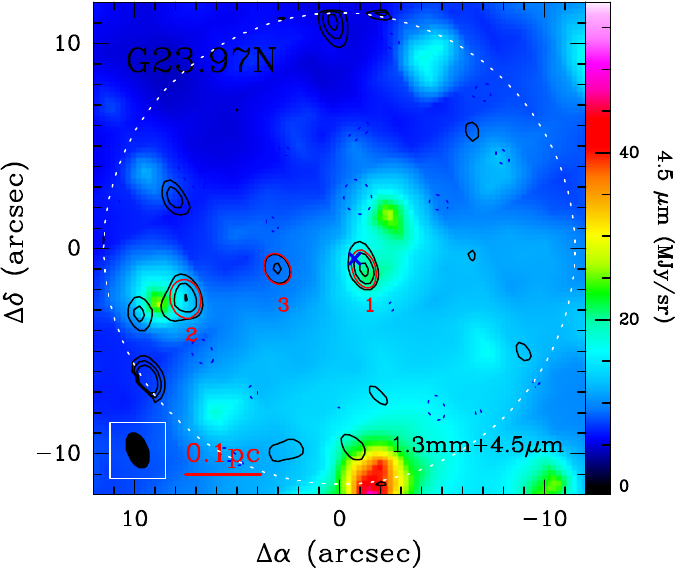}}
\caption{G23.97N: {\it (a)} (left) Three-color image of 4.5\,$\mu$m (blue), 8.0\,$\mu$m (green), and 24\,$\mu$m (red); (right) 870\,$\mu$m contours overlaid on a 70\,$\mu$m color scale. The 870\,$\mu$m contour levels start at 6$\sigma$ in steps of 6$\sigma$ ($\sigma$ = 54\,$\mjyb$). {\it (b)} 850\,$\mu$m contours overlaid on a 450\,$\mu$m color scale. The 850\,$\mu$m contour levels start at 3.6$\sigma$ in steps of 4.8$\sigma$ ($\sigma$ = 83\,$\mjyb$). {\it (c)} 3.5\,mm contours overlaid on a 1.3\,cm color scale. For CD configuration observations, {\it (d)} the 3.5\,mm contour levels start at -3$\sigma$, 3$\sigma$ in steps of 1.9$\sigma$ ($\sigma$ = 0.32\,$\mjyb$), and {\it (e)} the 1.3\,mm contour levels start at -3$\sigma$, 3$\sigma$ in steps of 2$\sigma$ ($\sigma$ = 1.62\,$\mjyb$) superimposed on a 24\,$\mu$m color scale. The ellipses with numbers indicate the positions of extracted sources. The crosses in the 1.3\,mm CD configuration observations indicate the positions of the corresponding 3.5\,mm CD configuration observations. The dashed and dotted circles in each subfigure indicate the primary beam scales of 3.5 and 1.3\,mm PdBI tracks, respectively.}
\label{Fig_G23.97N}
\end{figure*}

The clump G23.97N is in a similar condition to G18.17. It is located at a near kinematic distance of 4.68\,kpc \citep{Wienen2012,Reid2009}. In Figure\,\ref{Fig_G23.97N}(e), it is clear that the 4.5\,$\mu$m point source is located far away from the core position at 1.3\,mm in Figure\,\ref{Fig_G23.97N}(e), but is consistent with the positions of the 8.0 and 24\,$\mu$m point sources. The bright infrared point source nearby may be a background or foreground object corresponding to the clump G23.97N. Based on the above, we suggest that G23.97N, with very weak 70\,$\mu$m emission, may be a prestellar core.

There is no detected free-free emission in Figure\,\ref{Fig_G23.97N}(c). The emission at 870, 850, and 450\,$\mu$m is very weak, relative to the other clumps in this work. The clump mass is 227.5\,$\Msun$ ($R_{\rm eff}$ = 0.33\,pc) at 870\,$\mu$m; the core mass is 38.7\,$\Msun$ ($R_{\rm eff}$ = 0.049\,pc) at 3.5\,mm, while the corresponding condensation mass is 8.1\,$\Msun$ ($R_{\rm eff}$ = 0.019\,pc) at 1.3\,mm. Actually, we detect three intermediate-mass condensations in a line above 5$\sigma$ at 1.3\,mm. Their distributions are correlative to the filament structure at 850\,$\mu$m in Figure\,\ref{Fig_G23.97N}(b).

\subsection{G23.98}
\label{sect_G23.98}
\begin{figure*}
\centering
\subfigure[]{\includegraphics[width=0.65\textwidth, angle=0]{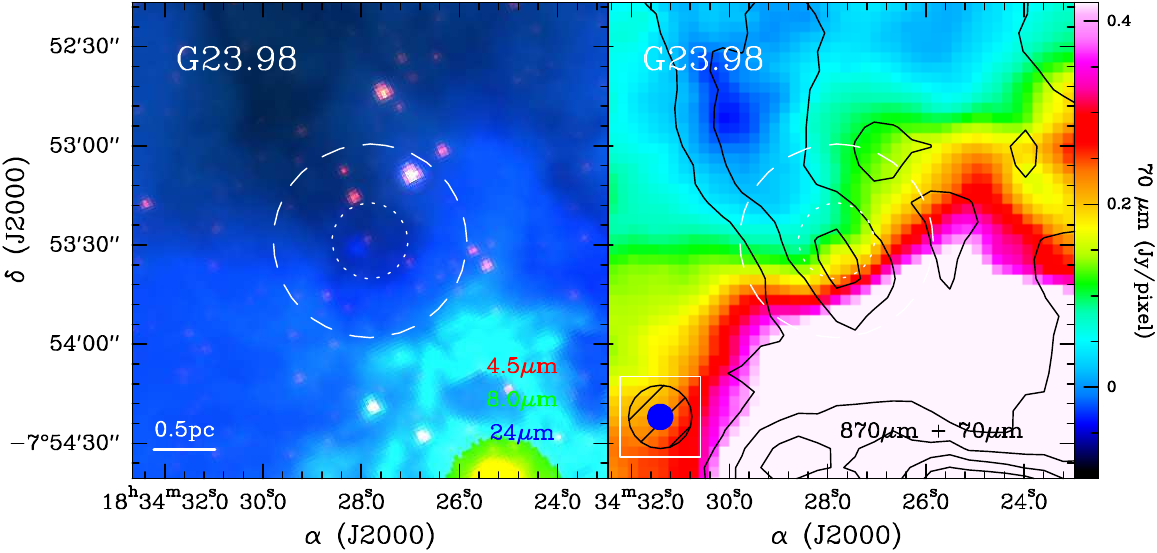}}
\subfigure[]{\includegraphics[width=0.33\textwidth, angle=0]{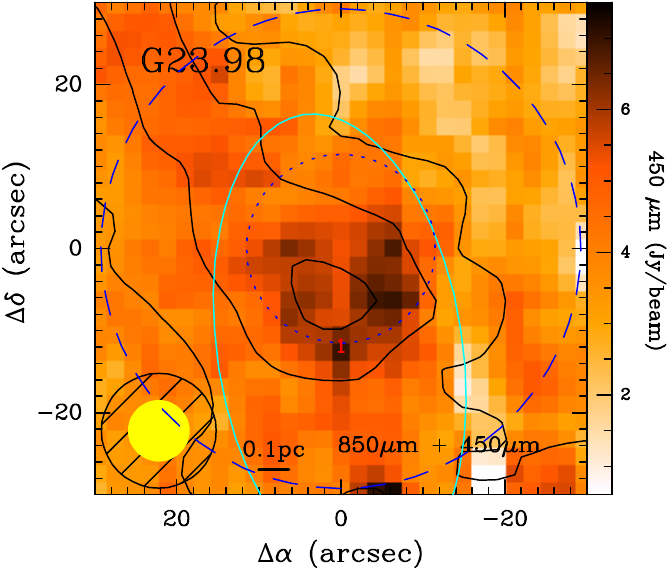}}
\subfigure[]{\includegraphics[width=0.33\textwidth, angle=0]{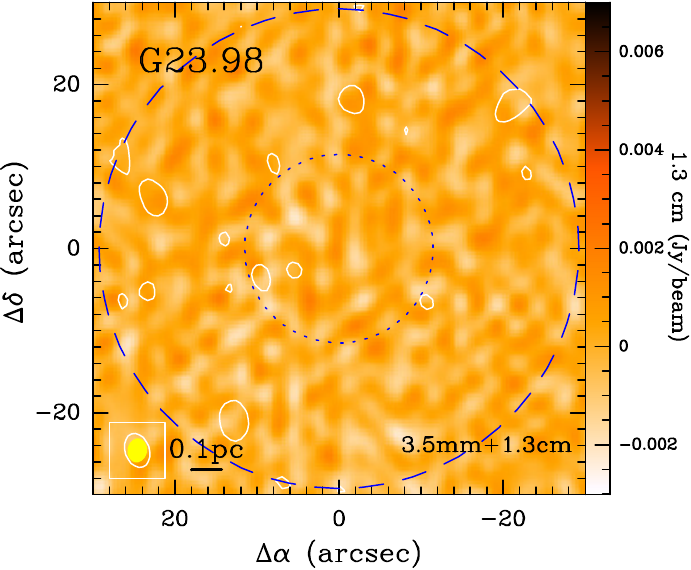}}
\subfigure[]{\includegraphics[width=0.33\textwidth, angle=0]{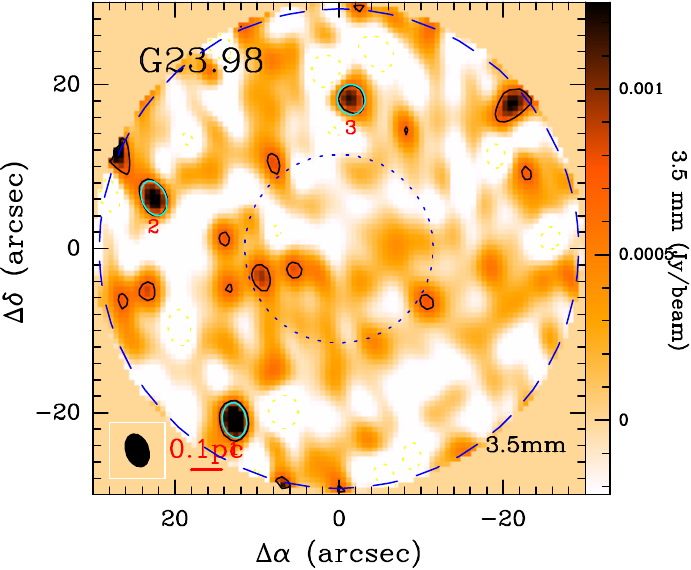}}
\subfigure[]{\includegraphics[width=0.33\textwidth, angle=0]{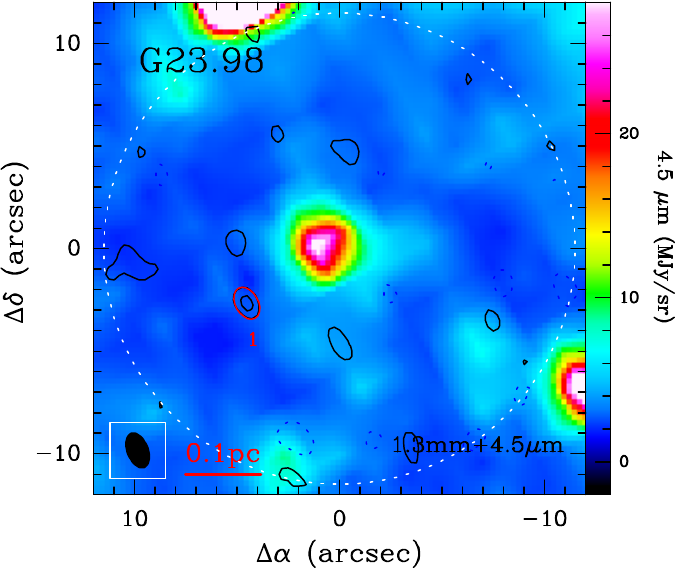}}
\caption{G23.98: {\it (a)} (left) Three-color image of 4.5\,$\mu$m (blue), 8.0\,$\mu$m (green), and 24\,$\mu$m (red); (right) 870\,$\mu$m contours overlaid on a 70\,$\mu$m color scale. The 870\,$\mu$m contour levels start at 6$\sigma$ in steps of 6$\sigma$ ($\sigma$ = 54\,$\mjyb$). {\it (b)} 850\,$\mu$m contours overlaid on a 450\,$\mu$m color scale. The 850\,$\mu$m contour levels start at 3.6$\sigma$ in steps of 4.8$\sigma$ ($\sigma$ = 83\,$\mjyb$). {\it (c)} 3.5\,mm contours overlaid on a 1.3\,cm color scale. For CD configuration observations, {\it (d)} the 3.5\,mm contour levels start at -3$\sigma$, 3$\sigma$ in steps of 4$\sigma$ ($\sigma$ = 0.22\,$\mjyb$), and {\it (e)} the 1.3\,mm contour levels start at -3$\sigma$, 3$\sigma$ in steps of 2$\sigma$ ($\sigma$ = 1.65\,$\mjyb$) superimposed on a 24\,$\mu$m color scale. The ellipses with numbers indicate the positions of extracted sources. The crosses in the 1.3\,mm CD configuration observations indicate the positions of the corresponding 3.5\,mm CD configuration observations. The dashed and dotted circles in each subfigure indicate the primary beam scales of 3.5 and 1.3\,mm PdBI tracks, respectively.}
\label{Fig_G23.98}
\end{figure*}

The clump G23.98 is infrared quiet with a 24\,$\mu$m flux of only 8.7\,mJy and may be a prestellar core (Figure\,\ref{Fig_G23.98}). It is located at a near kinematic distance of 4.68\,kpc \citep{Wienen2012,Reid2009}. Possibly due to contamination from the bright UC \HII region nearby, it is difficult to show this weak source clearly at infrared wavelengths. This core may be optically thick condition or in a very early evolutionary stage.

The emission at 870, 850, and 450\,$\mu$m is also very weak. The mass is 258.1\,$\Msun$ ($R_{\rm eff}$ = 0.35\,pc) at 870\,$\mu$m. There is no free-free emission at 6 or 1.3\,cm, and no dust emission at 3.5\,mm within its primary beam size. However, we detected only one condensation with a mass of 6.7\,$\Msun$ ($R_{\rm eff}$ = 0.018\,pc) at 1.3\,mm, which is consistent with the clump G23.98 at 850\,$\mu$m.

\subsection{G23.44}
\label{sect_G23.44}
The clump G23.44 has a trigonometric-parallax distance of 5.88\,kpc \citep{Brunthaler2009}. In Figure\,\ref{Fig_G23.44}(a), the three-color image shows one vague point source within 1.3\,mm primary beam. The 70\,$\mu$m emission shows two adjacent clumps, designated lower (G23.44-l) and upper (G23.44-u), respectively. Their 24\,$\mu$m fluxes are 300.1 and 1683.5\,mJy, respectively. The clumps G23.44-l and -u are thus classified into infrared bright and quiet sources, respectively. In Figure\,\ref{Fig_G23.44}(c), we find that there is strong free-free emission (24.20\,$\mjyb$) with extended morphology at 1.3\,cm for G23.44-l, but there is not any emission detected at 6\,cm. The integrated flux at 3.5\,mm is 11.01\,mJy for core No.1 in Figure\,\ref{Fig_G23.44}(d). The integrated flux at 1.3\,cm is more than twice that at 3.5\,mm, which means that the measured integrated flux at 3.5\,mm is purely from free-free emission. That is very interesting and not convincing. We suggest that the derived integrated flux at 1.3\,cm has a high uncertainty. We have to adopt an upper limit at 6\,cm for SED fitting, in order to estimate the core mass at 3.5\,mm (see Section\,\ref{sect_sed}).

The emission at 870 and 850\,$\mu$m shows one very compact and bright clump. The mass is 1498.6\,$\Msun$ ($R_{\rm eff}$ = 0.47\,pc) at 870\,$\mu$m. The emission at 450\,$\mu$m has two resolved components. In the 3.5\,mm CD configuration observations in Figure\,\ref{Fig_G23.44}(d), the clump fragments into 6 cores or two clusters (a lower cluster surrounding core No.1 and upper cluster surrounding core No.2). The two clusters are further resolved in the 3.5\,mm BCD configuration observations in Figure\,\ref{Fig_G23.44}(f). The masses of No.1 and No.2 are 27.6\,$\Msun$ ($R_{\rm eff}$ = 0.044\,pc) and 55.2\,$\Msun$ ($R_{\rm eff}$ = 0.038\,pc), respectively. In the 1.3\,mm CD configuration observations in Figure\,\ref{Fig_G23.44}(e), there are three massive and compact cores. In the 1.3\,mm BCD configuration observations in Figure\,\ref{Fig_G23.44}(g), there are 16 condensations in total. The most massive condensations No.1 and No.2 have a mass of 7.1\,$\Msun$ ($R_{\rm eff}$ = 0.016\,pc) and 5.6\,$\Msun$ ($R_{\rm eff}$ = 0.013\,pc), respectively. The upper clusters are close to the boundary of the primary beam at 1.3\,mm, so their masses have large uncertainty.

\subsection{G23.97S}
\label{sect_G23.97S}
\begin{figure*}
\centering
\subfigure[]{\includegraphics[width=0.65\textwidth, angle=0]{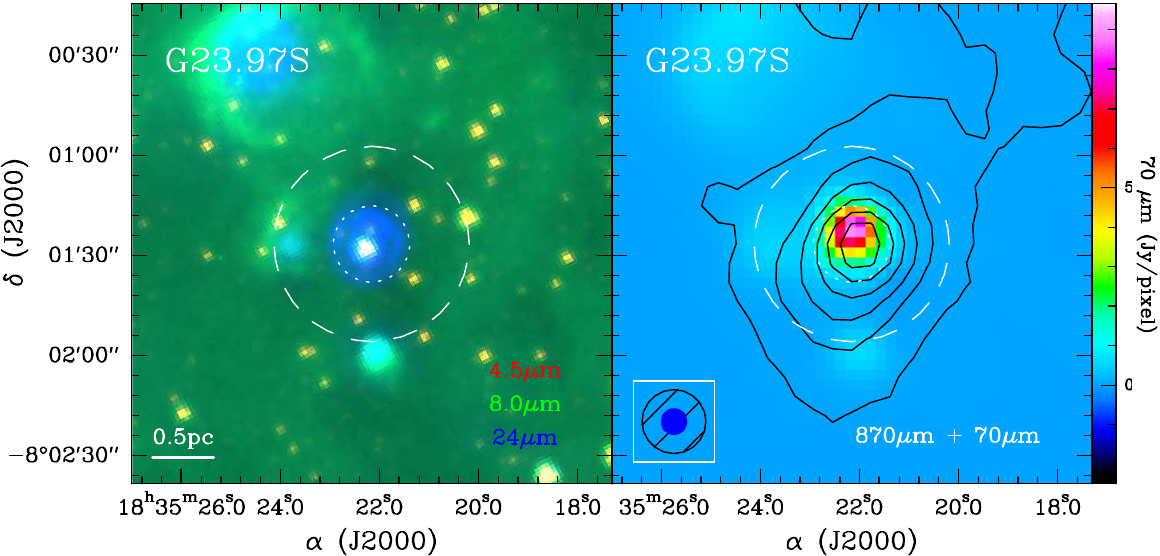}}
\subfigure[]{\includegraphics[width=0.33\textwidth, angle=0]{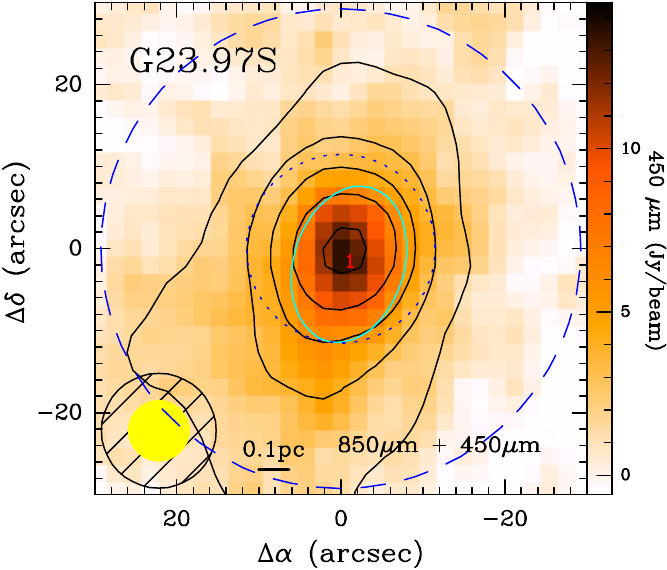}}
\subfigure[]{\includegraphics[width=0.33\textwidth, angle=0]{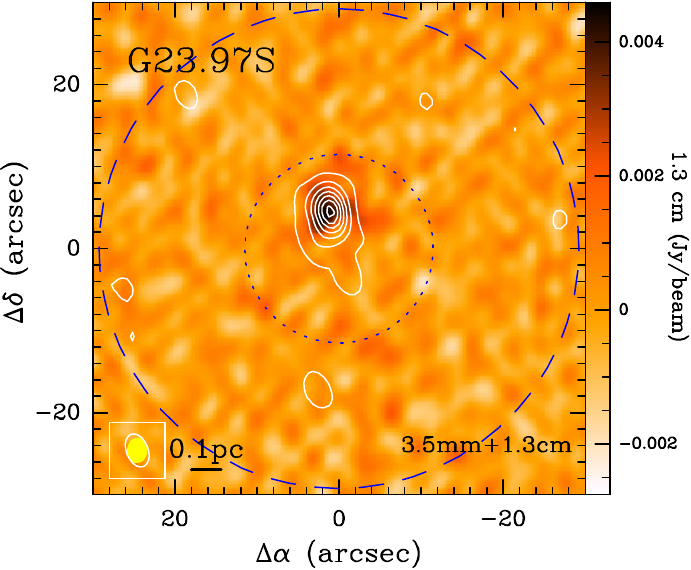}}
\subfigure[]{\includegraphics[width=0.33\textwidth, angle=0]{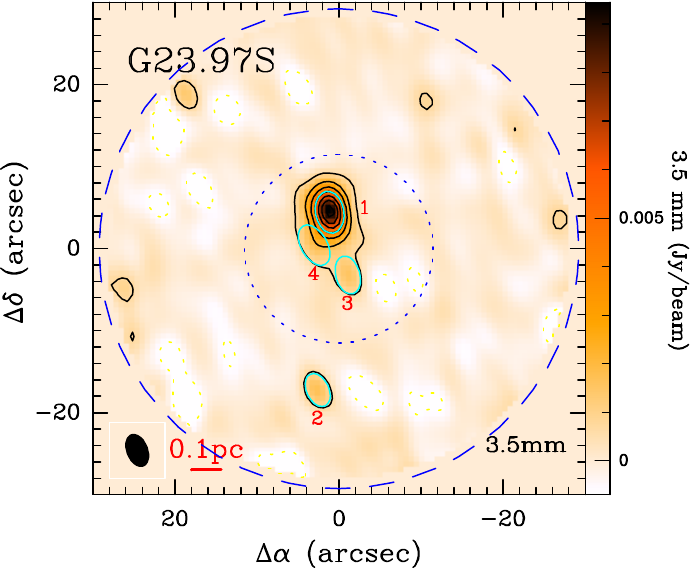}}
\subfigure[]{\includegraphics[width=0.33\textwidth, angle=0]{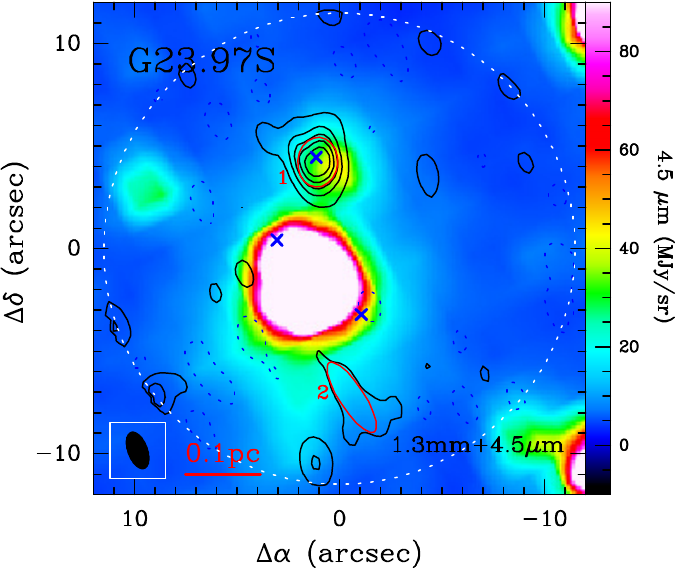}}
\subfigure[]{\includegraphics[width=0.33\textwidth, angle=0]{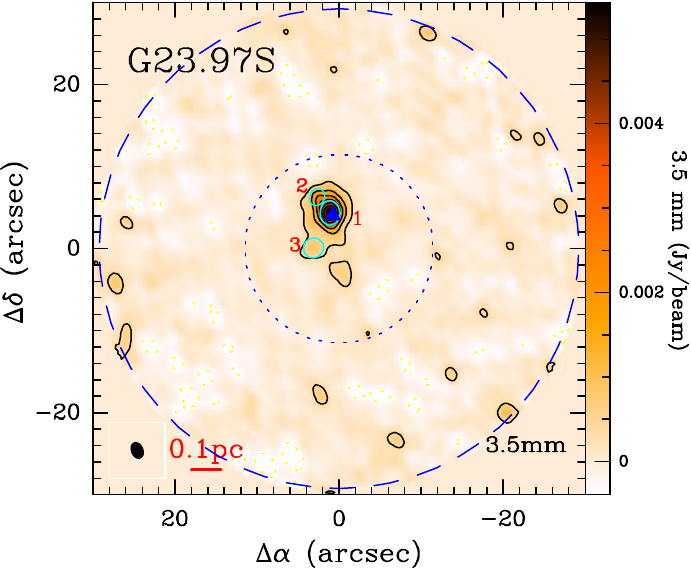}}
\subfigure[]{\includegraphics[width=0.33\textwidth, angle=0]{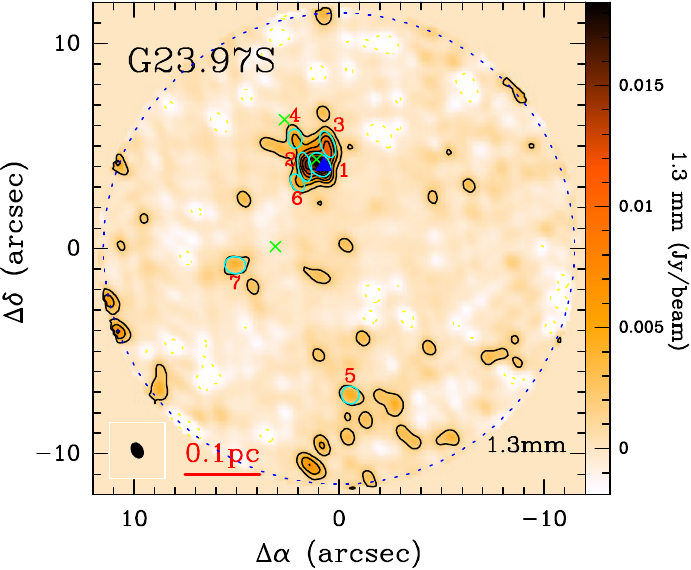}}
\caption{G23.97S: {\it (a)} (left) Three-color image of 4.5\,$\mu$m (blue), 8.0\,$\mu$m (green), and 24\,$\mu$m (red); (right) 870\,$\mu$m contours overlaid on a 70\,$\mu$m color scale. The 870\,$\mu$m contour levels start at 6$\sigma$ in steps of 10$\sigma$ ($\sigma$ = 54\,$\mjyb$). {\it (b)} 850\,$\mu$m contours overlaid on a 450\,$\mu$m color scale. The 850\,$\mu$m contour levels start at 6$\sigma$ in steps of 8$\sigma$ ($\sigma$ = 83\,$\mjyb$). {\it (c)} 3.5\,mm contours overlaid on a 1.3\,cm color scale. For CD configuration observations, {\it (d)} the 3.5\,mm contour levels start at -3$\sigma$, 3$\sigma$ in steps of 6$\sigma$ ($\sigma$ = 0.23\,$\mjyb$), and {\it (e)} the 1.3\,mm contour levels start at -3$\sigma$, 3$\sigma$ in steps of 5$\sigma$ ($\sigma$ = 1.43\,$\mjyb$). For BCD configuration observations, {\it (f)} the 3.5\,mm contour levels start at -3$\sigma$, 3$\sigma$ in steps of 6$\sigma$ ($\sigma$ = 0.13\,$\mjyb$), and {\it (g)} the 1.3\,mm contour levels start at -3$\sigma$, 3$\sigma$ in steps of 4$\sigma$ ($\sigma$ = 0.64\,$\mjyb$) superimposed on a 24\,$\mu$m color scale. The ellipses with numbers indicate the positions of extracted sources. The crosses in the 1.3\,mm (B)CD configuration observations indicate the peak positions of the corresponding cores and condensations in the 3.5\,mm (B)CD configuration observations. The dashed and dotted circles in each subfigure indicate the primary beam scales of 3.5 and 1.3\,mm PdBI tracks, respectively. The blue triangle ``$\blacktriangle$'' indicates the position of 6668 MHz methanol masers.}
\label{Fig_G23.97S}
\end{figure*}

The clump G23.97S is located at a near kinematic distance of 4.70\,kpc \citep{Wienen2012,Reid2009}. In Figure\,\ref{Fig_G23.97S}(a), there are two connected infrared point sources within the 1.3\,mm primary beam of the PdBI. Comparing with other wavelengths, we argue that the upper infrared source is associated with our target. The clump at infrared wavelengths is bright compared to the background. However, the 24 and 70\,$\mu$m fluxes are just 709.6\,mJy and 179.8\,Jy. This source is therefore infrared quiet, but there is detectable free-free emission at 1.3\,cm, so this source has a corresponding UC \HII region.

This clump has strong emission at 870, 850, and 450\,$\mu$m. The derived mass is 708.3\,$\Msun$ within $R_{\rm eff}$ = 0.39\,pc at 870\,$\mu$m. In the 3.5\,mm and 1.3\,mm CD configuration observations (Figures\,\ref{Fig_G23.97S}(d)(e)), we find that only the single core No.1 dominates the clusters. The core mass is 17.5\,$\Msun$ with $R_{\rm eff}$ = 0.055\,pc at 3.5\,mm, and 25.3\,$\Msun$ with $R_{\rm eff}$ = 0.029\,pc at 1.3\,mm. The clump fragments into a cluster of condensations with BCD configuration observations, especially in the 1.3\,mm BCD configuration observations in Figure\,\ref{Fig_G23.97S}(g). The most massive condensation No.1 is located in the south-west of the cluster. The condensation No.1 has a mass of 12.3\,$\Msun$ with $R_{\rm eff}$ = 0.015\,pc. The condensations No.2 and No.5 are intermediately massive (above 4.5\,$\Msun$).

\subsection{G25.38}
\label{sect_G25.38}
\begin{figure*}
\centering
\subfigure[]{\includegraphics[width=0.65\textwidth, angle=0]{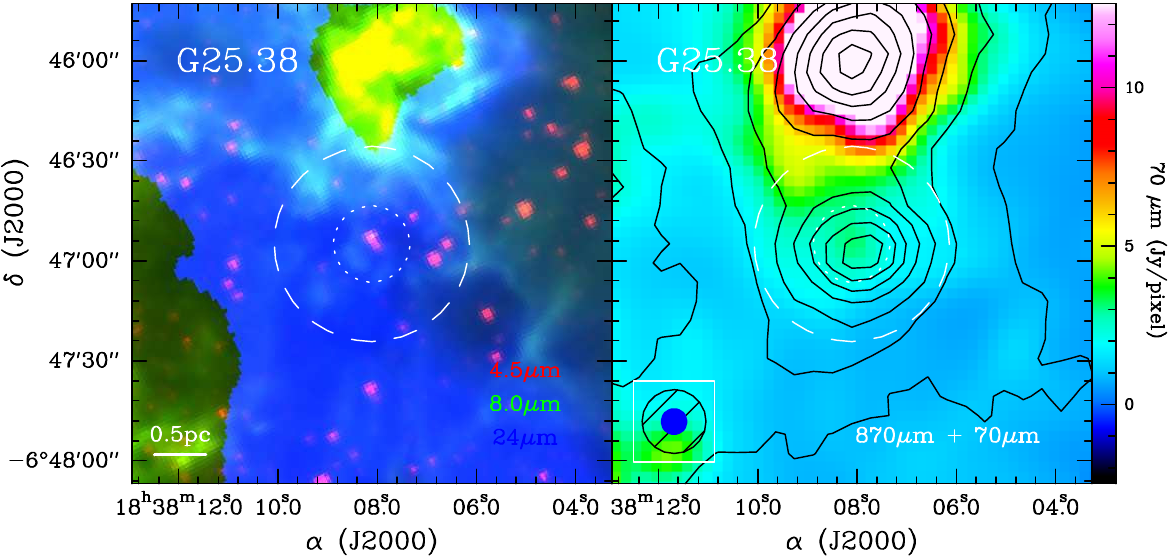}}
\subfigure[]{\includegraphics[width=0.33\textwidth, angle=0]{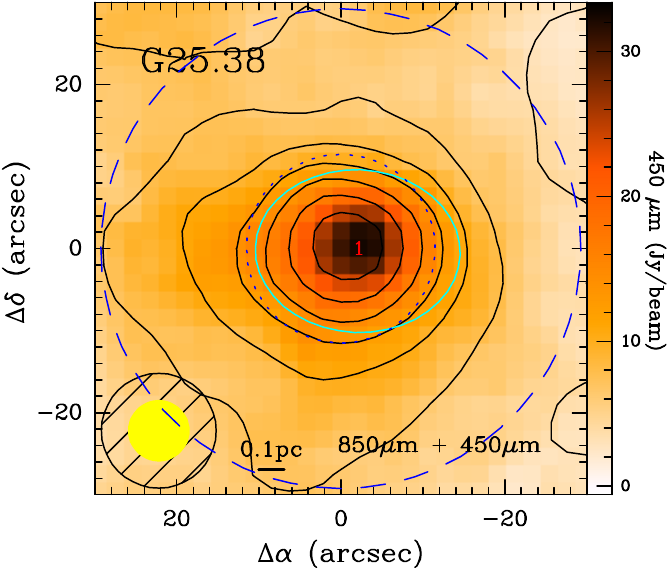}}
\subfigure[]{\includegraphics[width=0.33\textwidth, angle=0]{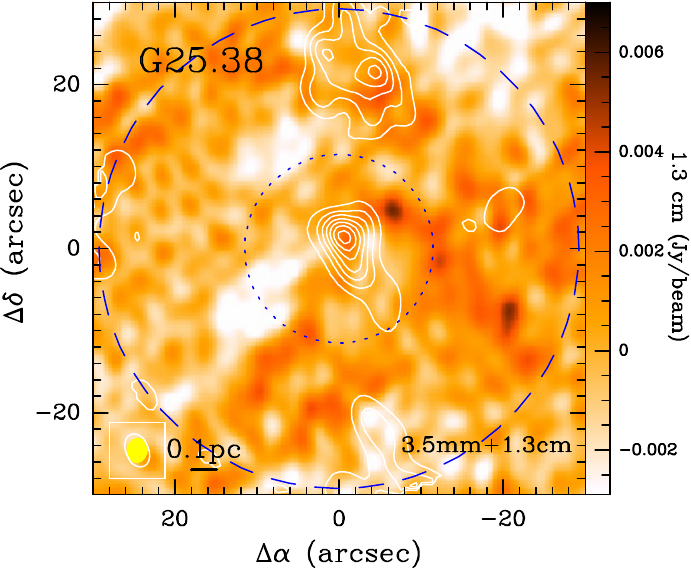}}
\subfigure[]{\includegraphics[width=0.33\textwidth, angle=0]{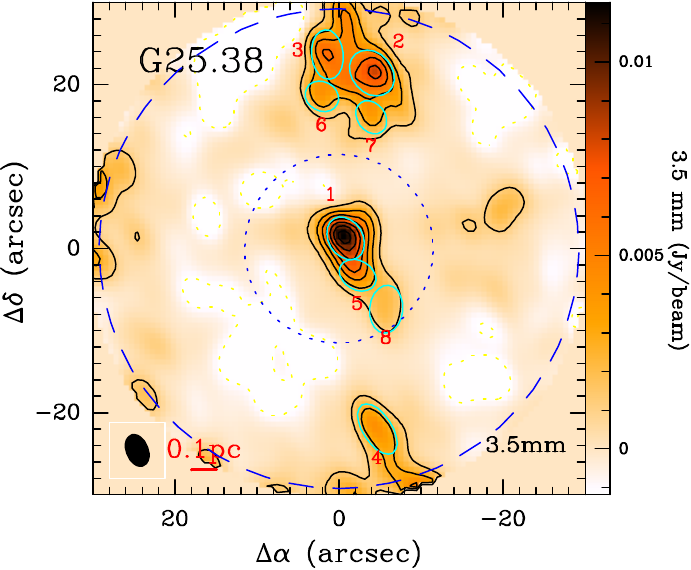}}
\subfigure[]{\includegraphics[width=0.33\textwidth, angle=0]{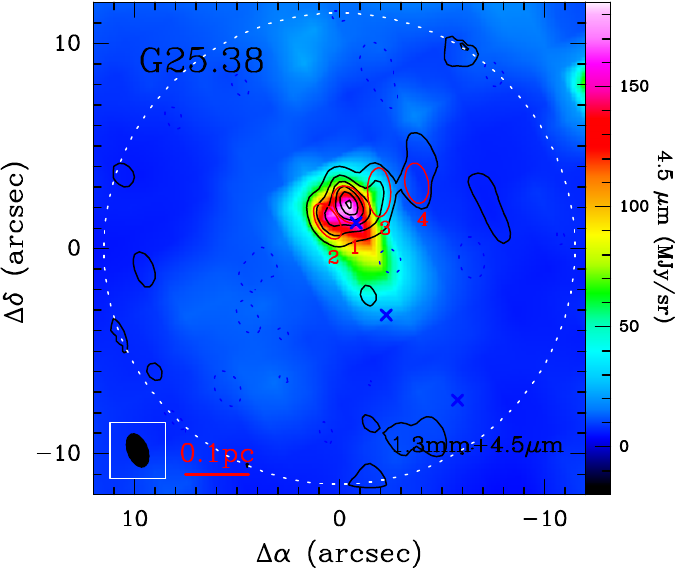}}
\subfigure[]{\includegraphics[width=0.33\textwidth, angle=0]{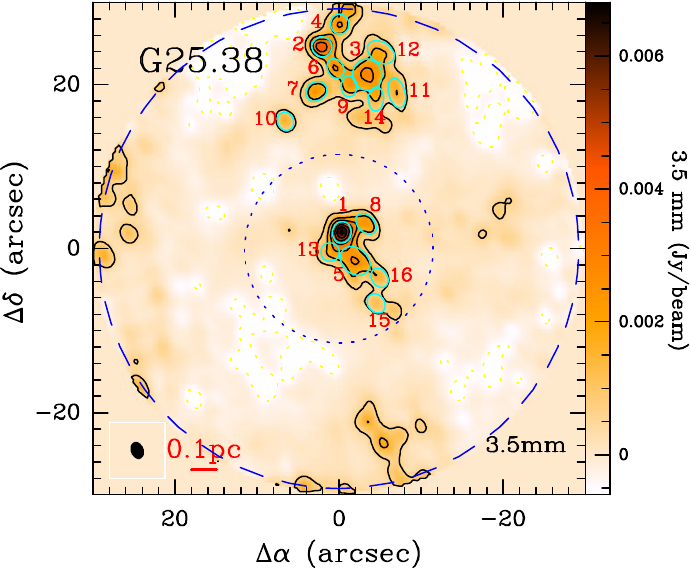}}
\subfigure[]{\includegraphics[width=0.33\textwidth, angle=0]{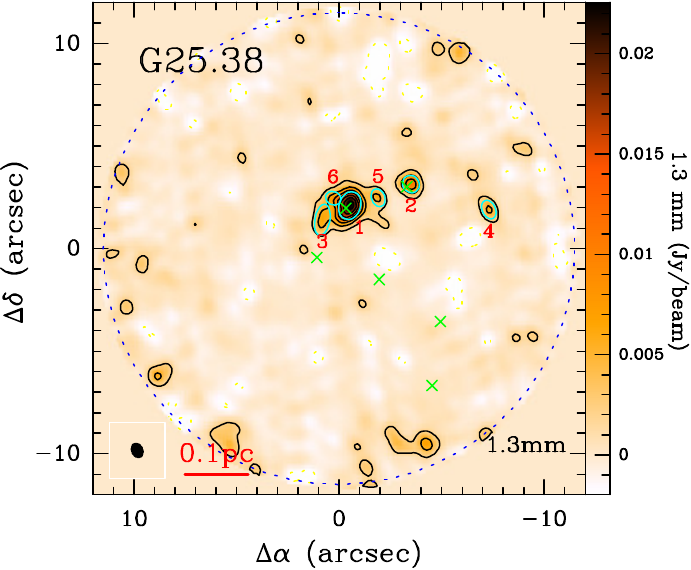}}
\caption{G25.38: {\it (a)} (left) Three-color image of 4.5\,$\mu$m (blue), 8.0\,$\mu$m (green), and 24\,$\mu$m (red); (right) 870\,$\mu$m contours overlaid on a 70\,$\mu$m color scale. The 870\,$\mu$m contour levels start at 5.3$\sigma$ in steps of 8.8$\sigma$ ($\sigma$ = 61\,$\mjyb$). {\it (b)} 850\,$\mu$m contours overlaid on a 450\,$\mu$m color scale. The 850\,$\mu$m contour levels start at 6$\sigma$ in steps of 8$\sigma$ ($\sigma$ = 83\,$\mjyb$). {\it (c)} 3.5\,mm contours overlaid on a 1.3\,cm color scale. For CD configuration observations, {\it (d)} the 3.5\,mm contour levels start at -3$\sigma$, 3$\sigma$ in steps of 4$\sigma$ ($\sigma$ = 0.39\,$\mjyb$), and {\it (e)} the 1.3\,mm contour levels start at -3$\sigma$, 3$\sigma$ in steps of 6$\sigma$ ($\sigma$ = 1.34\,$\mjyb$). For BCD configuration observations, {\it (f)} the 3.5\,mm contour levels start at -3$\sigma$, 3$\sigma$ in steps of 5$\sigma$ ($\sigma$ = 0.20\,$\mjyb$), and {\it (g)} the 1.3\,mm contour levels start at -3$\sigma$, 3$\sigma$ in steps of 5$\sigma$ ($\sigma$ = 0.66\,$\mjyb$) superimposed on a 24\,$\mu$m color scale. The ellipses with numbers indicate the positions of extracted sources. The crosses in the 1.3\,mm (B)CD configuration observations indicate the peak positions of the corresponding cores and condensations in the 3.5\,mm (B)CD configuration observations. The dashed and dotted circles in each subfigure indicate the primary beam scales of 3.5 and 1.3\,mm PdBI tracks, respectively.}
\label{Fig_G25.38}
\end{figure*}

The clump G25.38 is located at a near kinematic distance of 5.60\,kpc \citep{ande2009,Ai2013,Urquhart2013}. In Figure\,\ref{Fig_G25.38}(a)(c), there is a very vague infrared point source within the 1.3\,mm primary beam of the PdBI. At 3.5\,mm, the clump G25.38 fragments into two clusters (Figure\,\ref{Fig_G25.38}(d)). In Table\,\ref{tab_infrared}, the derived 24\,$\mu$m fluxes are 252.3\,mJy for G25.38-l, and no detection for G25.38-u, so they are infrared quiet sources. The infrared luminosity of G25.38-u has been contaminated by the extended UC \HII region nearby. At 1.3\,cm in Figure\,\ref{Fig_G25.38}(c), there is just extremely weak emission (about 3$\sigma$) located at the peak position of the infrared point source.

At 870, 850, and 450\,$\mu$m, the clump shows a compact and bright structure. The clump mass is 1080.6\,$\Msun$ with $R_{\rm eff}$ = 0.45\,pc at 870\,$\mu$m. The G25.38-u is located at the boundary of the 3.5\,mm primary beam of the PdBI, which will introduce high noise. We believed that the G25.38-u is an independent clump, rather than a piece of the UC \HII region nearby. However, it is likely that feedback from the UC \HII region nearby has affected G25.38-u and G25.38-l. In Figure\,\ref{Fig_G25.38}(f), the clump has fragmented into 16 cores. The most massive object No.1 within 1.3\,mm primary beam has a mass of 52.2\,$\Msun$ with $R_{\rm eff}$ = 0.040\,pc in the 3.5\,mm BCD configuration observations. In Figure\,\ref{Fig_G25.38}(e), there are 4 cores in a filament morphology perpendicular to its parent clump (No.1, No.5, and No.8 are also in a filament structure, see Figure\,\ref{Fig_G25.38}(d)), which may be a dynamical effect. This case is  similar to sources G28.34-P1 \citep{Wang2011} and G11.11-P6 \citep{Wang2014}. They explained that it was due to cylindrical fragmentation. In Figure\,\ref{Fig_G25.38}(g), we can see one more compact structure than that in Figure\,\ref{Fig_G25.38}(e). The condensation No.1 has a mass of 25.9\,$\Msun$ with $R_{\rm eff}$ = 0.019\,pc in the 1.3\,mm BCD configuration observations.

\subsection{G25.71}
\label{sect_G25.71}
\begin{figure*}
\centering
\subfigure[]{\includegraphics[width=0.65\textwidth, angle=0]{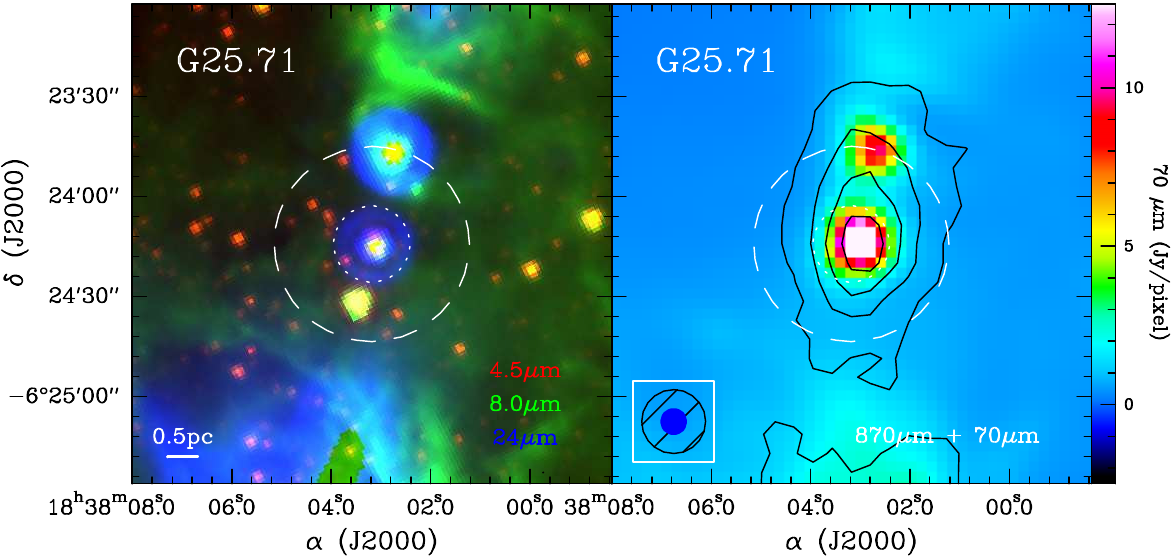}}
\subfigure[]{\includegraphics[width=0.33\textwidth, angle=0]{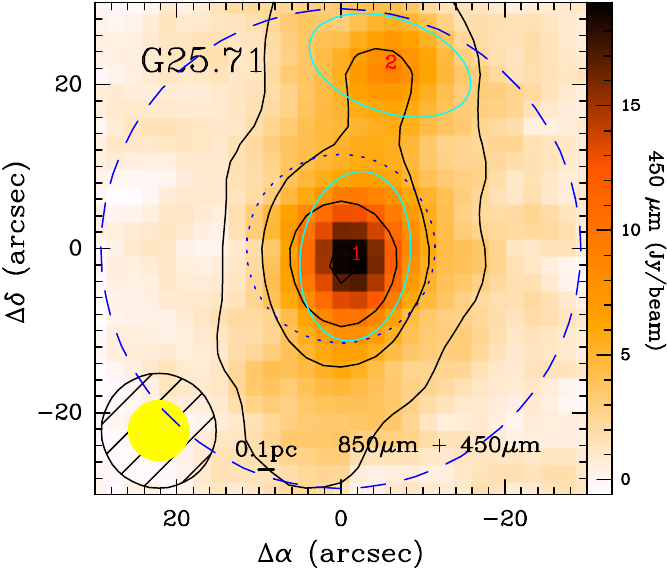}}
\subfigure[]{\includegraphics[width=0.33\textwidth, angle=0]{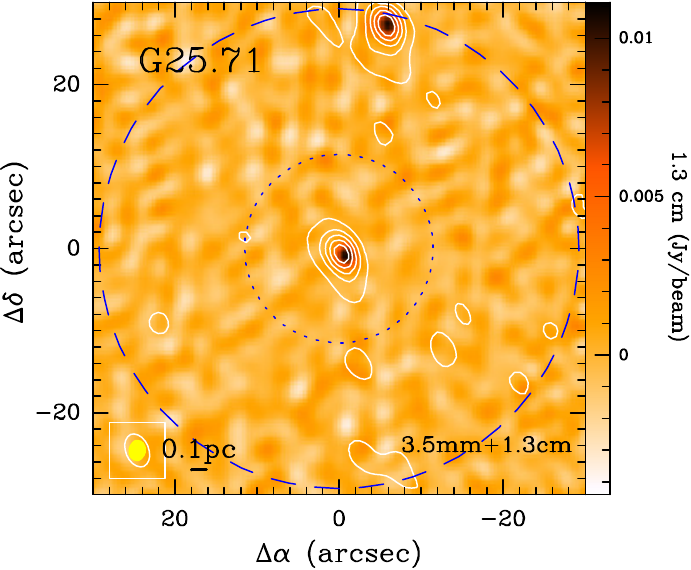}}
\subfigure[]{\includegraphics[width=0.33\textwidth, angle=0]{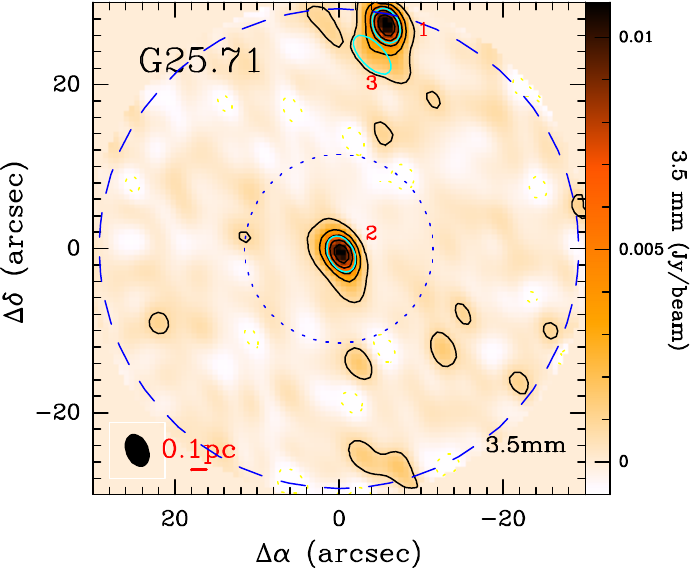}}
\subfigure[]{\includegraphics[width=0.33\textwidth, angle=0]{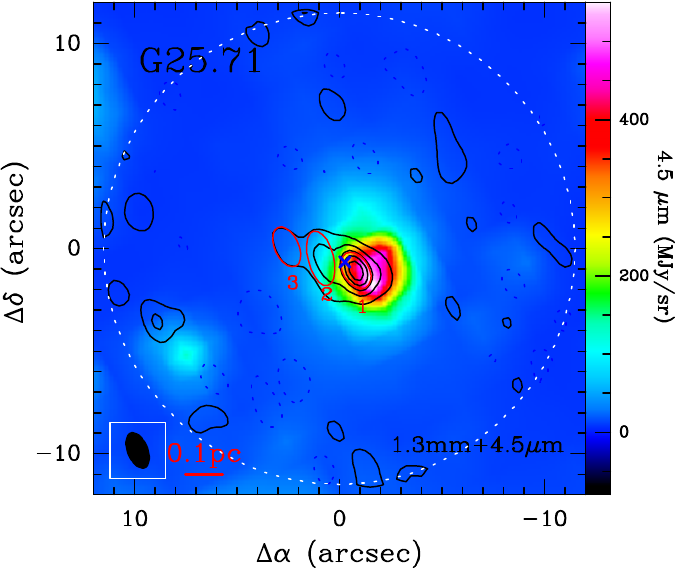}}
\subfigure[]{\includegraphics[width=0.33\textwidth, angle=0]{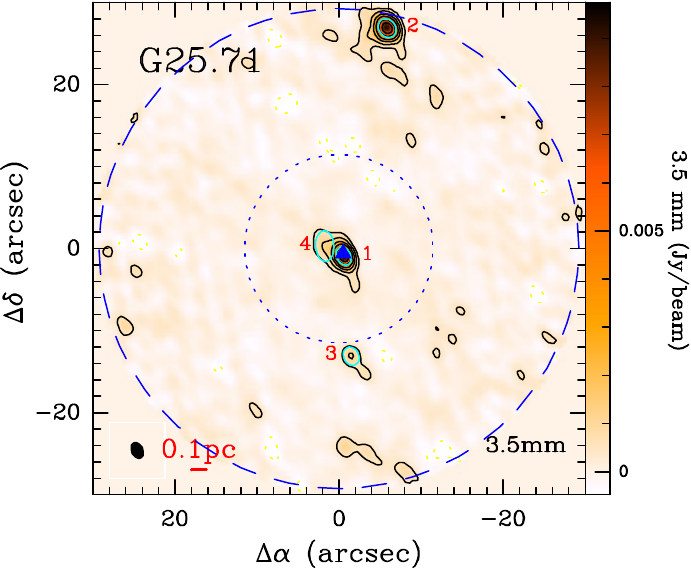}}
\subfigure[]{\includegraphics[width=0.33\textwidth, angle=0]{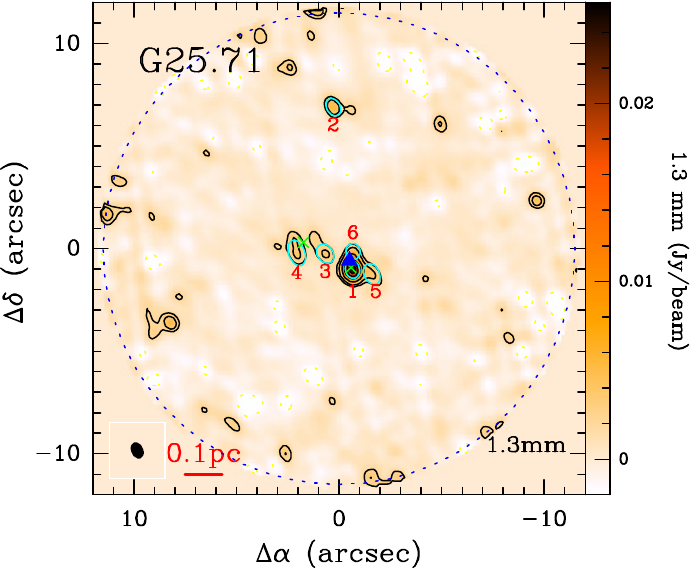}}
\caption{G25.71: {\it (a)} (left) Three-color image of 4.5\,$\mu$m (blue), 8.0\,$\mu$m (green), and 24\,$\mu$m (red); (right) 870\,$\mu$m contours overlaid on a 70\,$\mu$m color scale. The 870\,$\mu$m contour levels start at 5.3$\sigma$ in steps of 7$\sigma$ ($\sigma$ = 61\,$\mjyb$). {\it (b)} 850\,$\mu$m contours overlaid on a 450\,$\mu$m color scale. The 850\,$\mu$m contour levels start at 6$\sigma$ in steps of 8$\sigma$ ($\sigma$ = 83\,$\mjyb$). {\it (c)} 3.5\,mm contours overlaid on a 1.3\,cm color scale. For CD configuration observations, {\it (d)} the 3.5\,mm contour levels start at -3$\sigma$, 3$\sigma$ in steps of 8$\sigma$ ($\sigma$ = 0.26\,$\mjyb$), and {\it (e)} the 1.3\,mm contour levels start at -3$\sigma$, 3$\sigma$ in steps of 4$\sigma$ ($\sigma$ = 1.34\,$\mjyb$). For BCD configuration observations, {\it (f)} the 3.5\,mm contour levels are -3$\sigma$, 3$\sigma$, 7$\sigma$, 13$\sigma$, 20$\sigma$, and to maximum in steps of 13$\sigma$ ($\sigma$ = 0.16\,$\mjyb$), and {\it (g)} the 1.3\,mm contour levels are -3$\sigma$, 3$\sigma$, 5$\sigma$, 10$\sigma$ and to maximum in steps of 8$\sigma$ ($\sigma$ = 0.66\,$\mjyb$) superimposed on a 24\,$\mu$m color scale. The ellipses with numbers indicate the positions of extracted sources. The crosses in the 1.3\,mm (B)CD configuration observations indicate the peak positions of the corresponding cores and condensations in the 3.5\,mm (B)CD configuration observations. The dashed and dotted circles in each subfigure indicate the primary beam scales of 3.5 and 1.3\,mm PdBI tracks, respectively. The blue triangle ``$\blacktriangle$'' indicates the position of 6668\,MHz methanol masers.}
\label{Fig_G25.71}
\end{figure*}

The clump G25.71 is located at a near kinematic distance of 9.50\,kpc \citep{Urquhart2013,Lockman1989}. In Figure\,\ref{Fig_G25.71}(a), the extended UC \HII region is located to the south-east of the clump G25.71. There are two fairly bright infrared point sources G25.71-l ($S_{\rm 24\, \mu m}$ = 6197.1\,mJy) and G25.71-u ($S_{\rm 24\, \mu m}$ = 16109.0\,mJy) within the 3.5\,mm primary beam (see in Table\,\ref{tab_infrared}). Based the criteria in Section\,\ref{sect_infrared}, both the G25.71-l and G25.71-u are infrared bright. The emission at 870, 850, and 450\,$\mu$m shows an extended structure in a north-south direction. For the free-free emission at 1.3\,cm, two compact points are well located at the peak positions of G25.71-u and G25.71-l, which are likely two compact \HII region candidates.

In Figure\,\ref{Fig_G25.71}(d)(f), there are similar structures in the CD configuration observations and BCD configuration observations at 3.5\,mm, but the core No.1 in the CD configuration observations fragments into three condensations in the BCD configuration observations. The mass of condensation No.1 is 71.2\,$\Msun$ with $R_{\rm eff}$ = 0.052\,pc in the 3.5\,mm BCD configuration observations. In Figure\,\ref{Fig_G25.71}(e)(g), the cores show an elongated structure in an east-west direction, which may be from dynamical feedback or cylindrical fragmentation, similar to clump G25.38 (see analysis in Section\,\ref{sect_G25.38}). The core No.1 in the 1.3\,mm CD configuration observations fragmented into three condensations in the 1.3\,mm BCD configuration observations. The mass of condensation No.1 is 52.5\,$\Msun$ with $R_{\rm eff}$ = 0.021\,pc in the 1.3\,mm BCD configuration observations.

\section{Other figures and tables}

\begin{figure*}
\centering
\subfigure[]{\includegraphics[width=0.33\textwidth,angle=0]{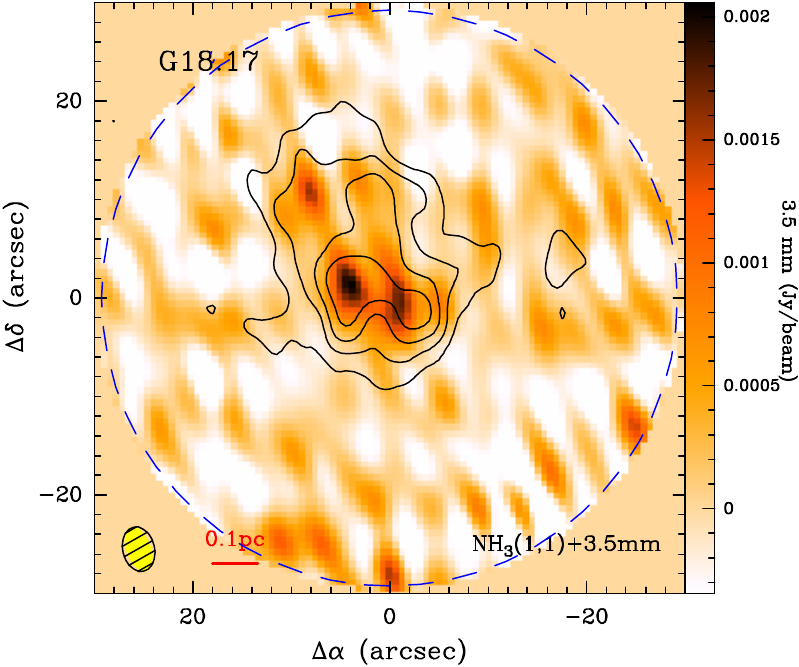}}
\subfigure[]{\includegraphics[width=0.33\textwidth,angle=0]{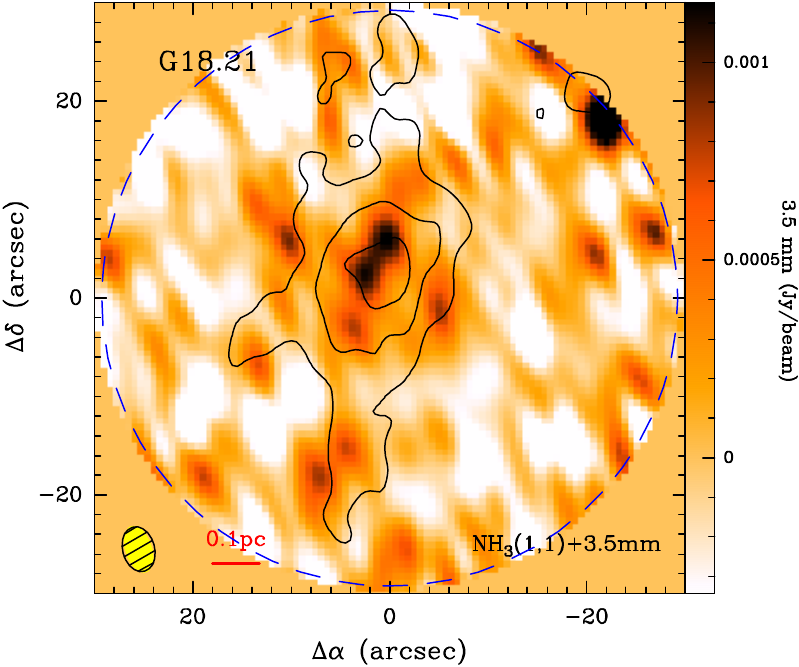}}
\subfigure[]{\includegraphics[width=0.33\textwidth,angle=0]{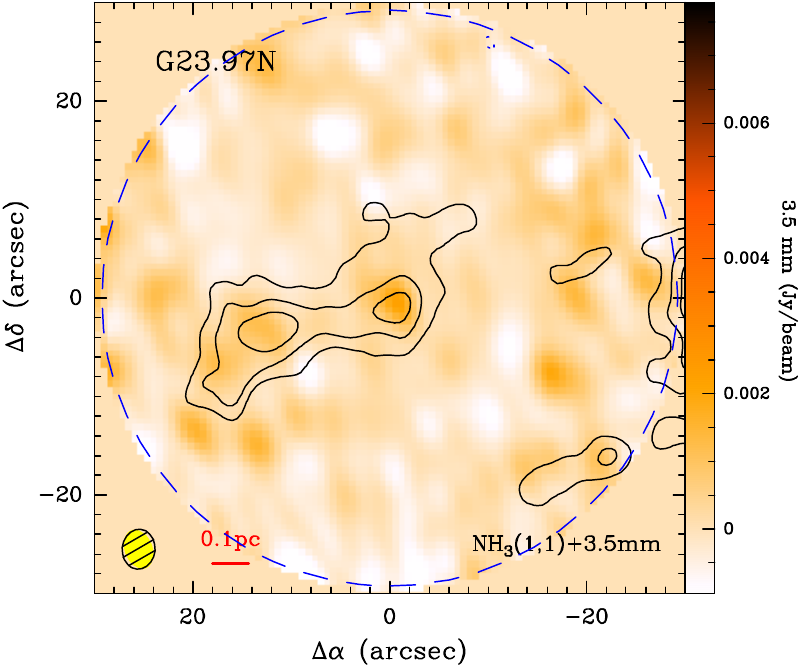}}
\subfigure[]{\includegraphics[width=0.33\textwidth,angle=0]{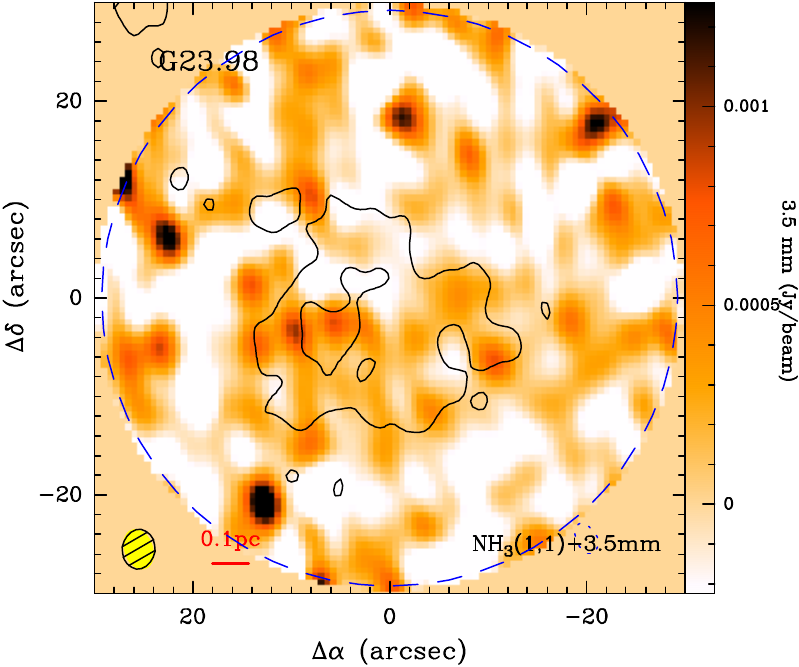}}
\subfigure[]{\includegraphics[width=0.33\textwidth,angle=0]{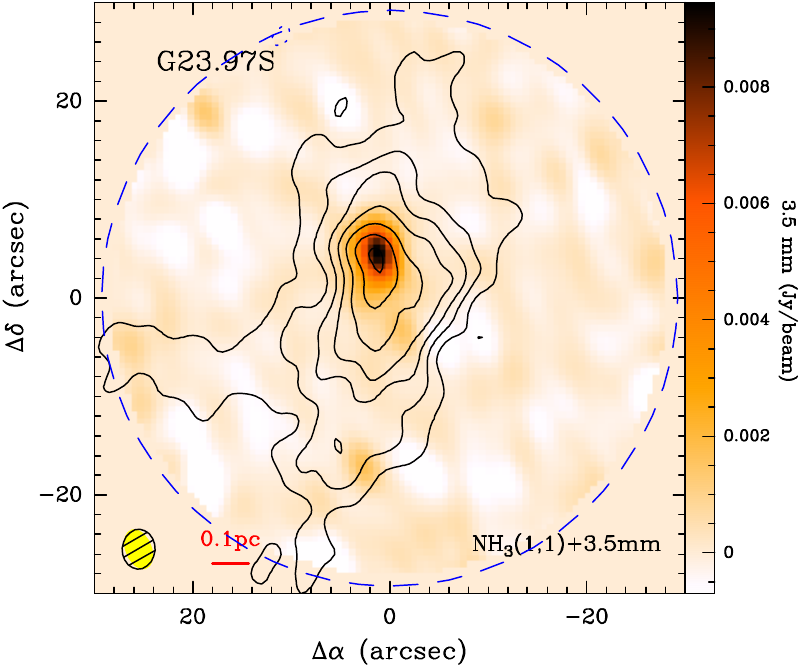}}
\subfigure[]{\includegraphics[width=0.33\textwidth,angle=0]{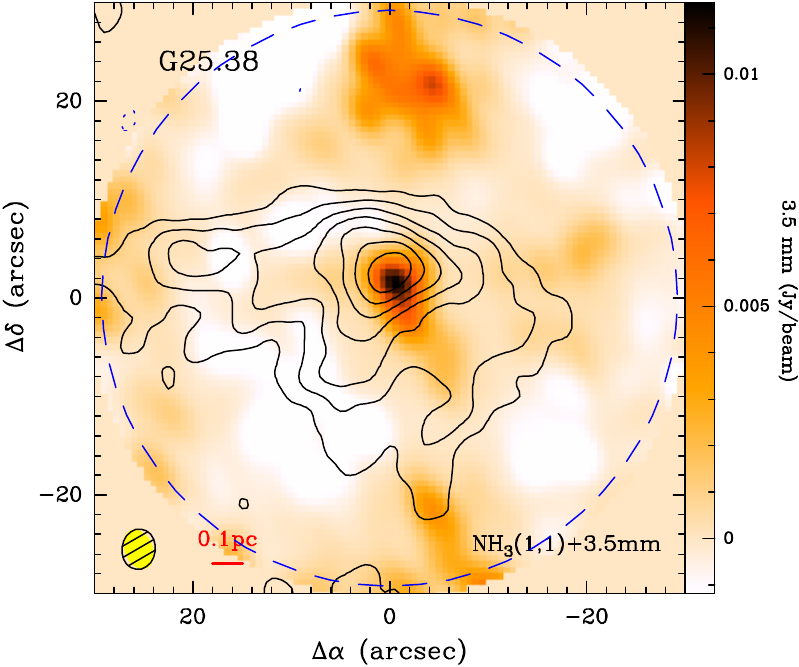}}
\subfigure[]{\includegraphics[width=0.33\textwidth,angle=0]{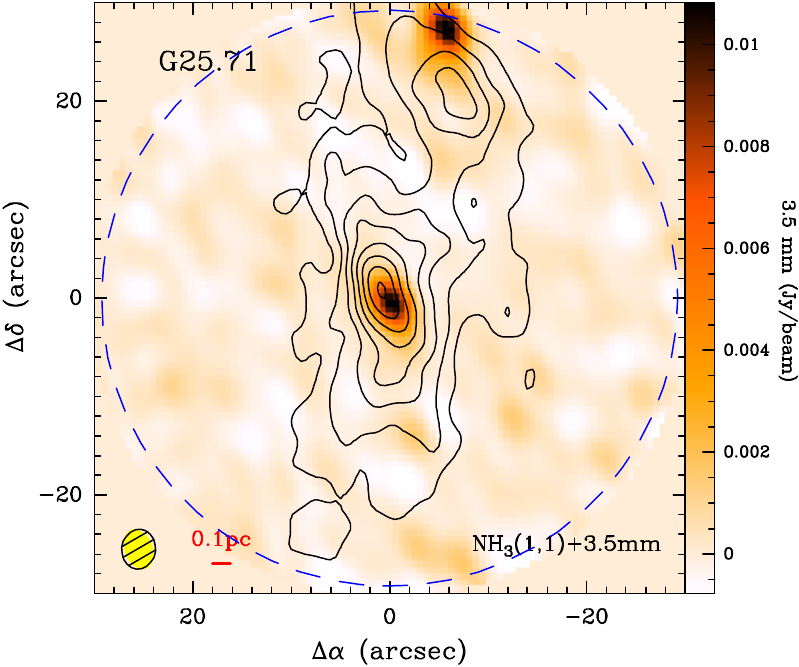}}
\caption{NH$_3$ (1,\,1) integrated-intensity contours overlaid on a 3.5\,mm continuum with velocity range covering only the main line. The contour levels start at -3$\sigma$ in steps of 3$\sigma$ for NH$_3$ (1,\,1) with $\sigma_{\rm (a)-(g)} =$ 12.5, 20.2, 9.1, 9.8, 12.6, 12.9, 10.2 $\mjybkms$. The synthesized beam size of each subfigure is indicated at the bottom-left corner.}
\label{Fig_nh3-11_appendix}
\end{figure*}

\begin{figure*}
\centering
\subfigure[]{\includegraphics[width=0.33\textwidth,angle=0]{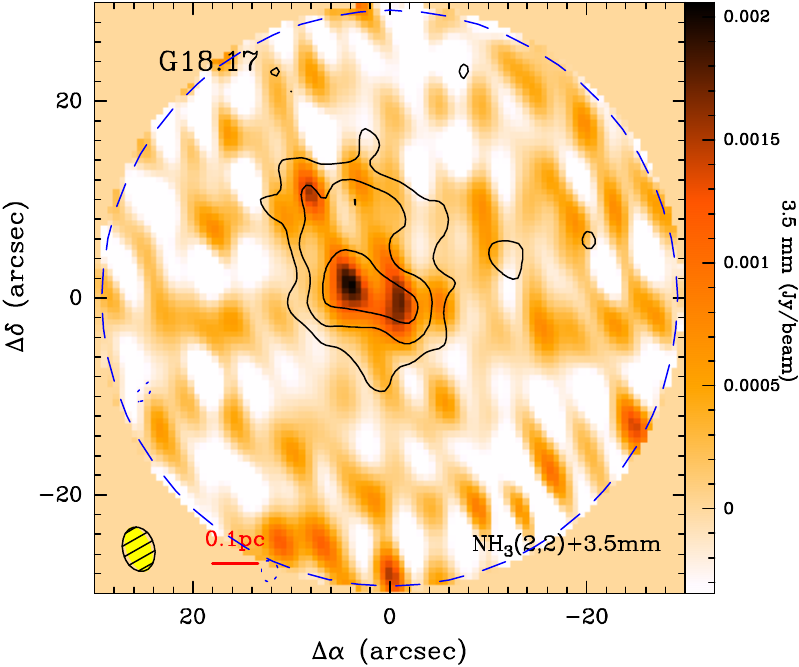}}
\subfigure[]{\includegraphics[width=0.33\textwidth,angle=0]{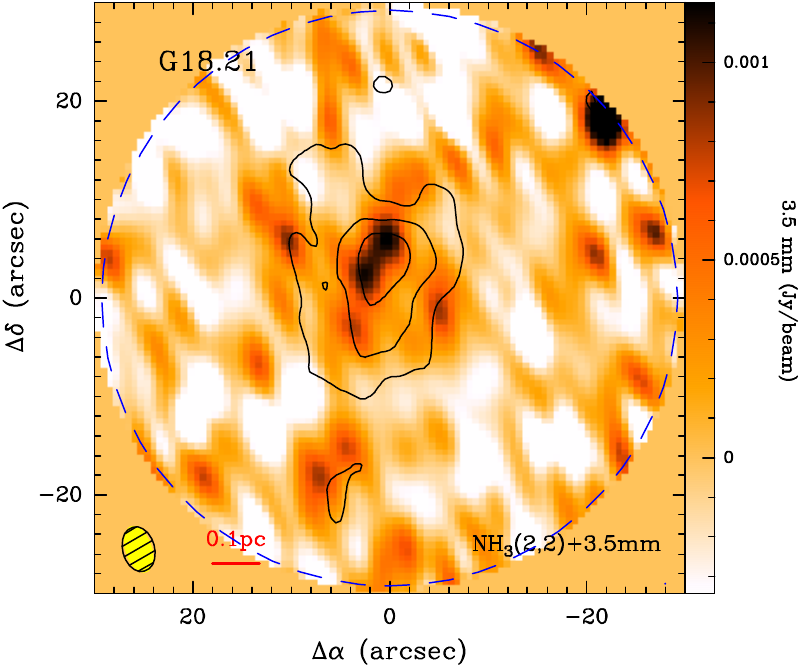}}
\subfigure[]{\includegraphics[width=0.33\textwidth,angle=0]{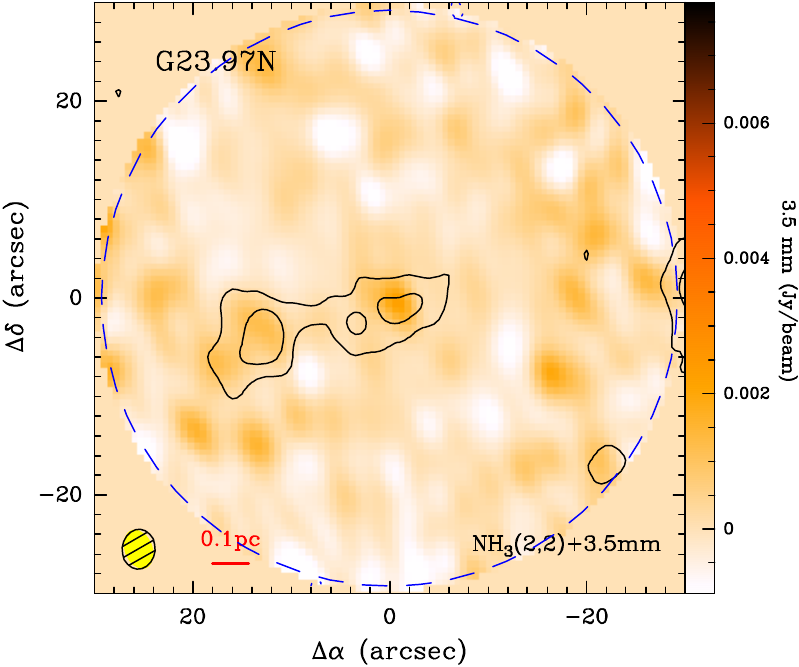}}
\subfigure[]{\includegraphics[width=0.33\textwidth,angle=0]{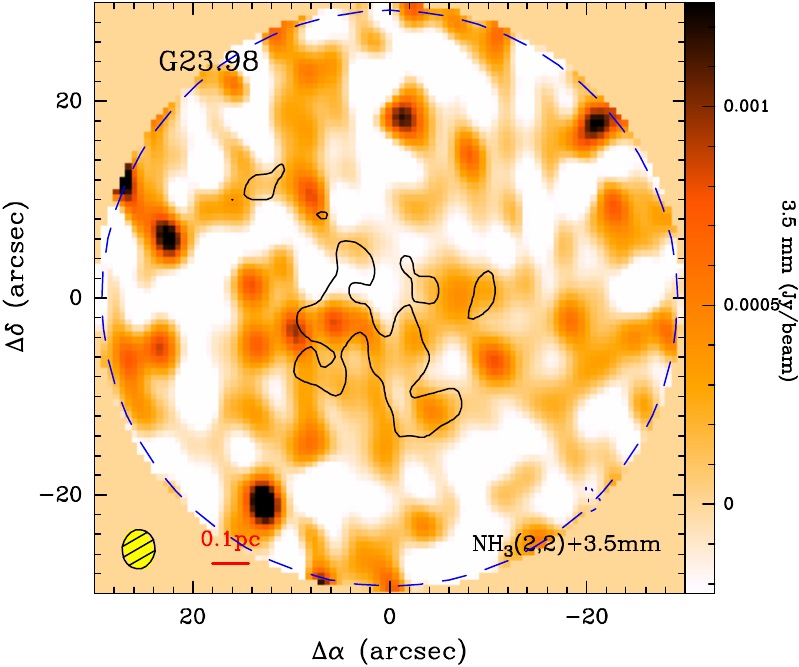}}
\subfigure[]{\includegraphics[width=0.33\textwidth,angle=0]{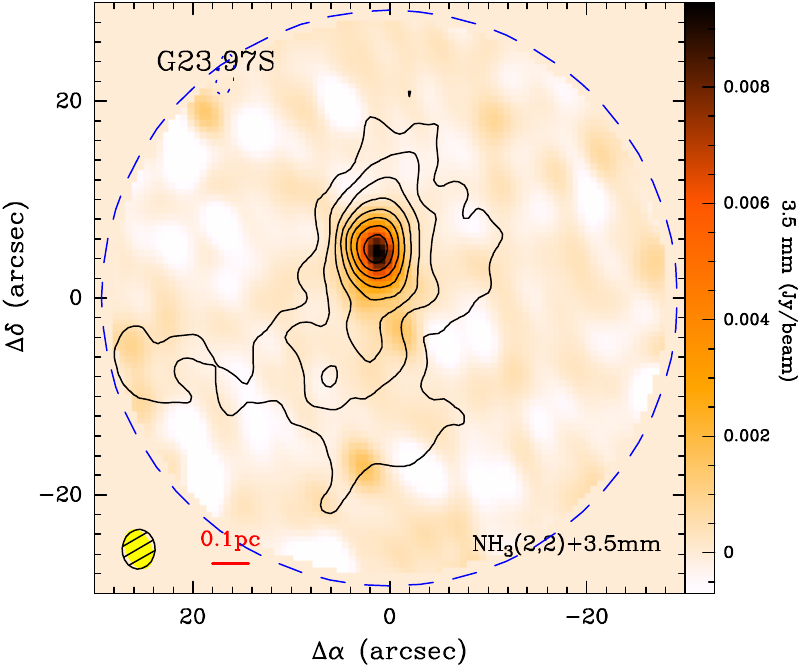}}
\subfigure[]{\includegraphics[width=0.33\textwidth,angle=0]{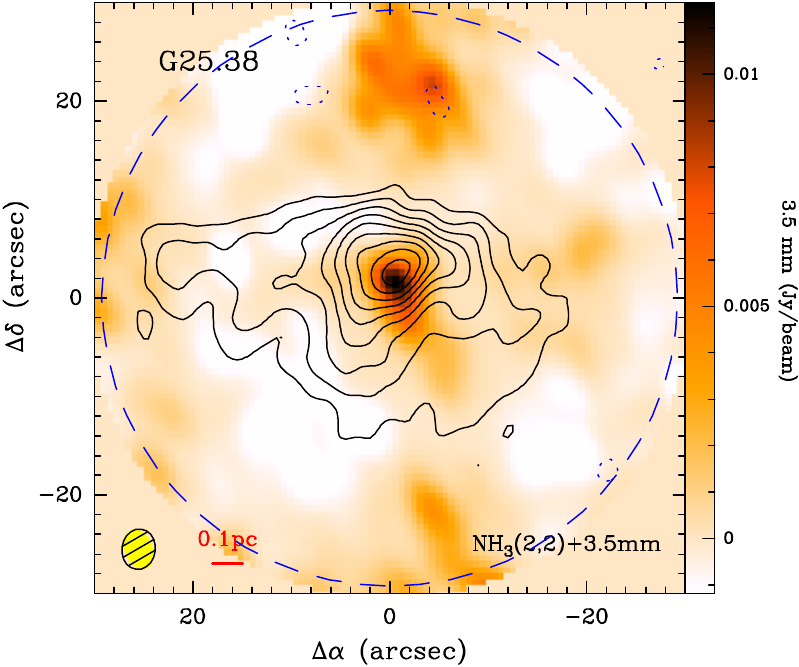}}
\subfigure[]{\includegraphics[width=0.33\textwidth,angle=0]{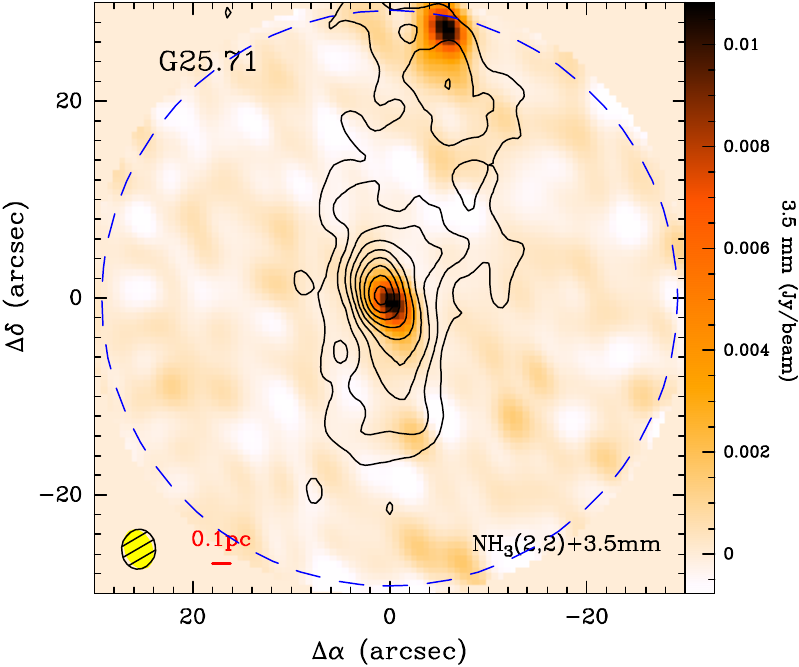}}
\caption{NH$_3$ (2,\,2) integrated-intensity contours overlaid on a 3.5\,mm continuum with velocity range covering only the main line. The contour levels start at -3$\sigma$ in steps of 3$\sigma$ for NH$_3$ (2,\,2) with $\sigma_{\rm (a)-(g)} =$ 10.8, 13.5, 7.9, 7.2, 8.2, 9.7, 8.3 $\mjybkms$. The synthesized beam size of each subfigure is indicated at the bottom-left corner.}
\label{Fig_nh3-22_appendix}
\end{figure*}

\begin{figure*}
\centering
\subfigure[]{\includegraphics[width=0.33\textwidth,angle=0]{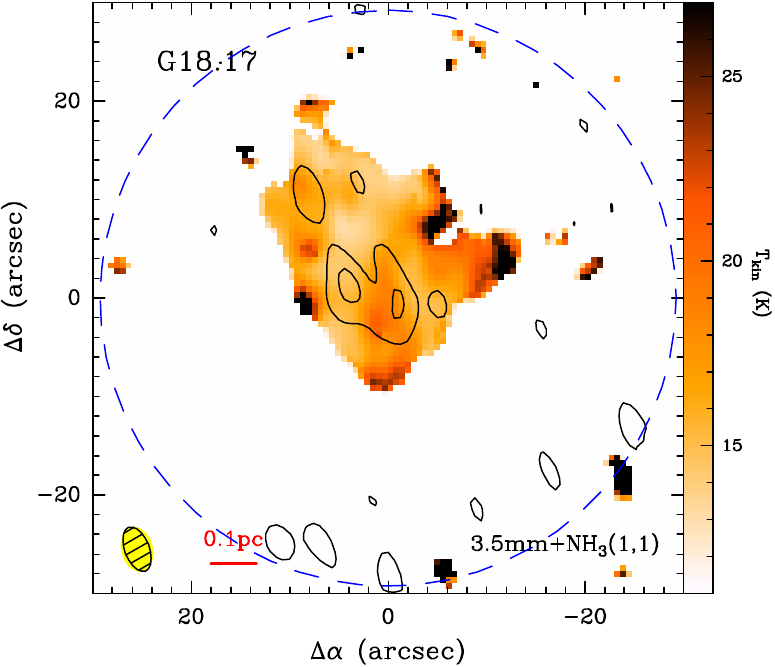}}
\subfigure[]{\includegraphics[width=0.33\textwidth,angle=0]{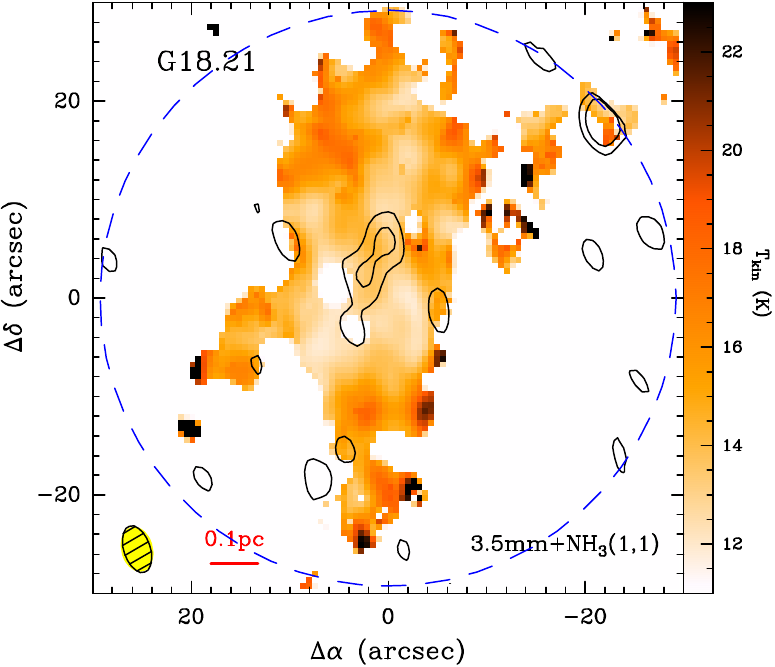}}
\subfigure[]{\includegraphics[width=0.33\textwidth,angle=0]{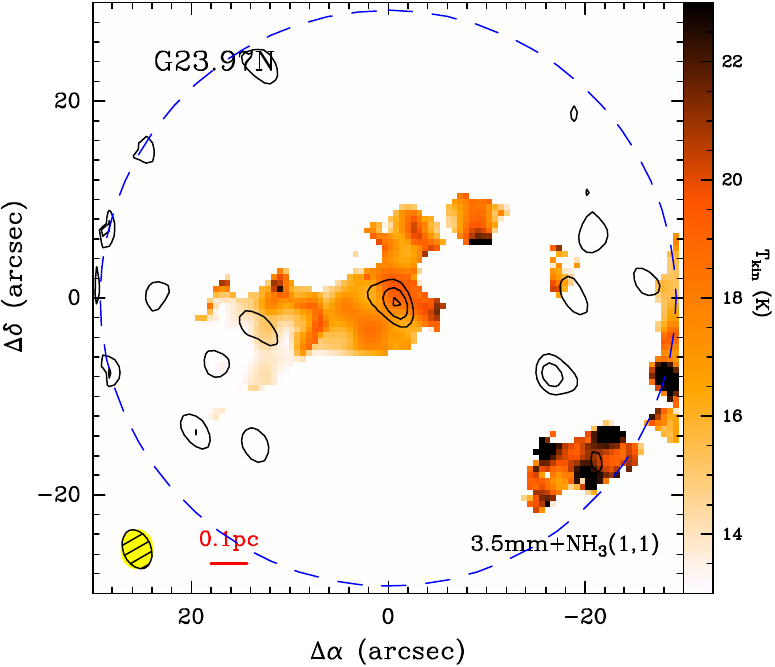}}
\subfigure[]{\includegraphics[width=0.33\textwidth,angle=0]{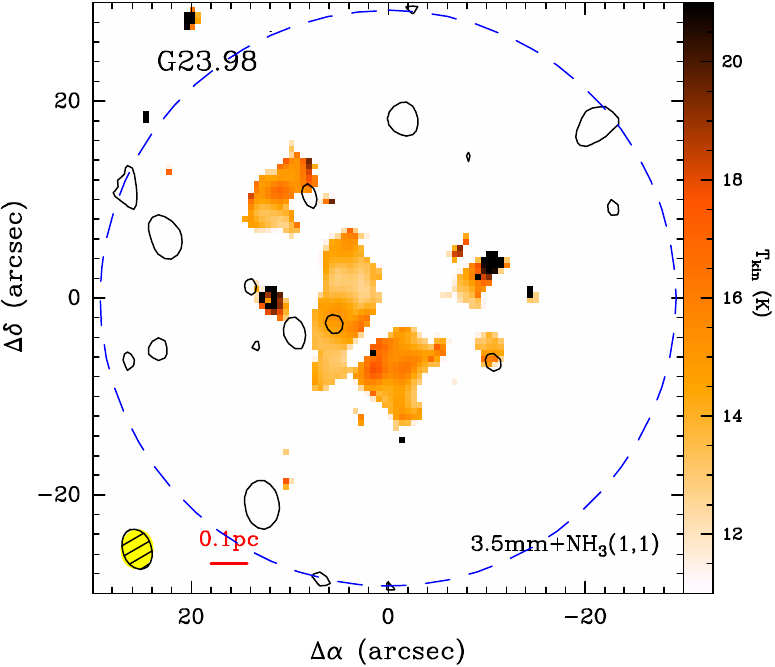}}
\subfigure[]{\includegraphics[width=0.33\textwidth,angle=0]{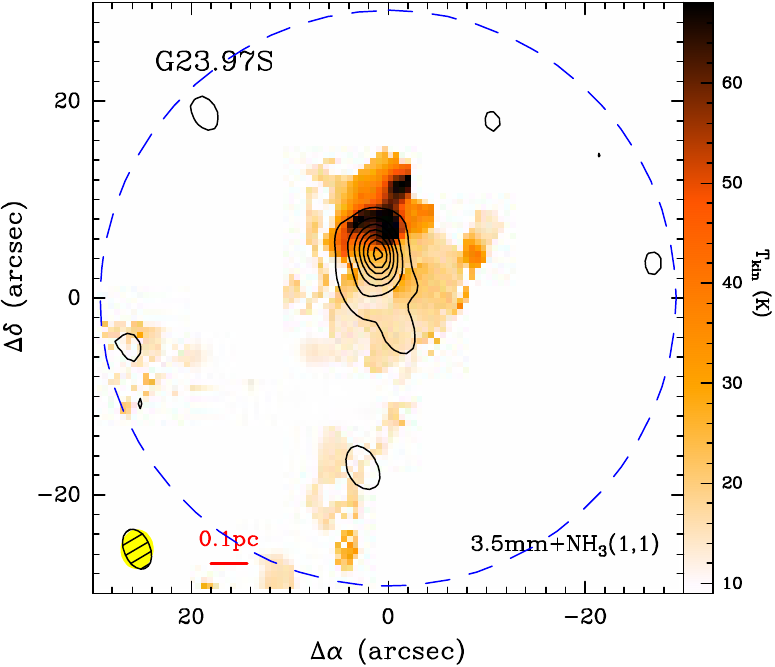}}
\subfigure[]{\includegraphics[width=0.33\textwidth,angle=0]{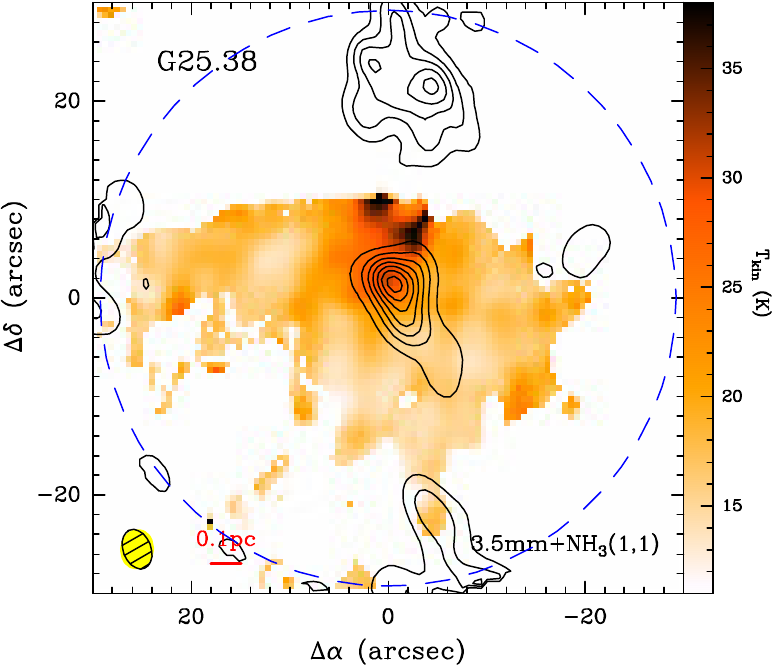}}
\subfigure[]{\includegraphics[width=0.33\textwidth,angle=0]{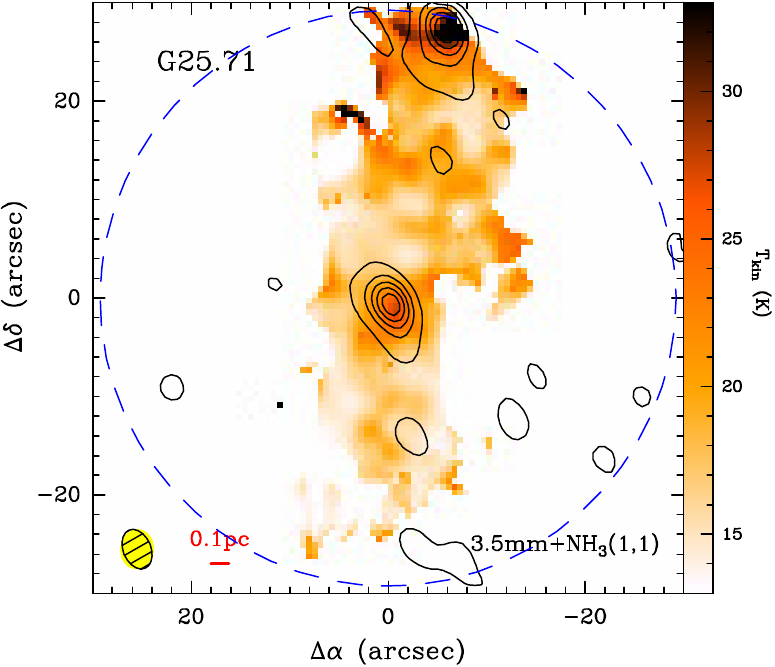}}
\caption{The kinetic temperature $T_{\rm kin}$ maps overlaid by 3.5\,mm continuum contours. The synthesized beam size of each subfigure is indicated at the bottom-left corner.}
\label{Fig_temperature_appendix}
\end{figure*}

\begin{figure*}
\centering
\subfigure[]{\includegraphics[width=0.33\textwidth,angle=0]{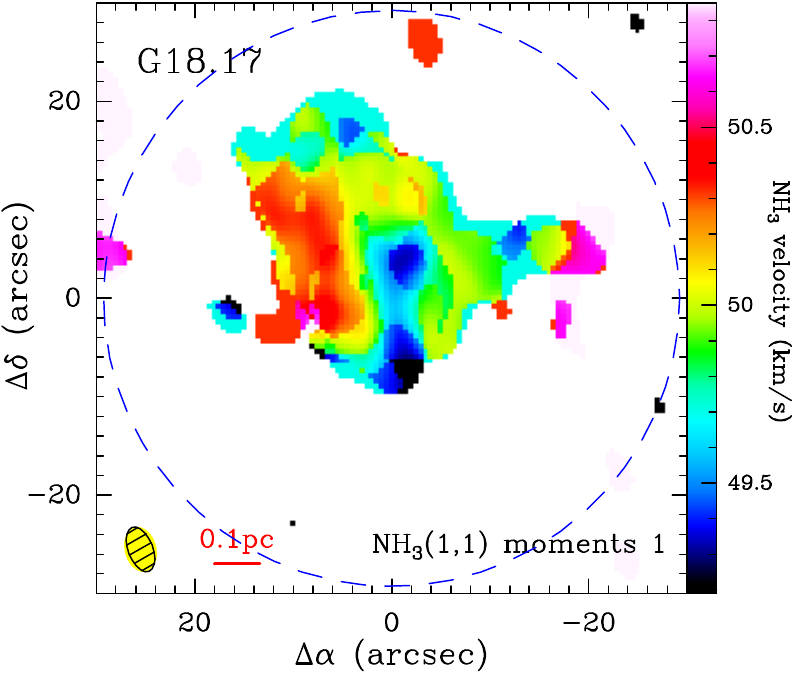}}
\subfigure[]{\includegraphics[width=0.33\textwidth,angle=0]{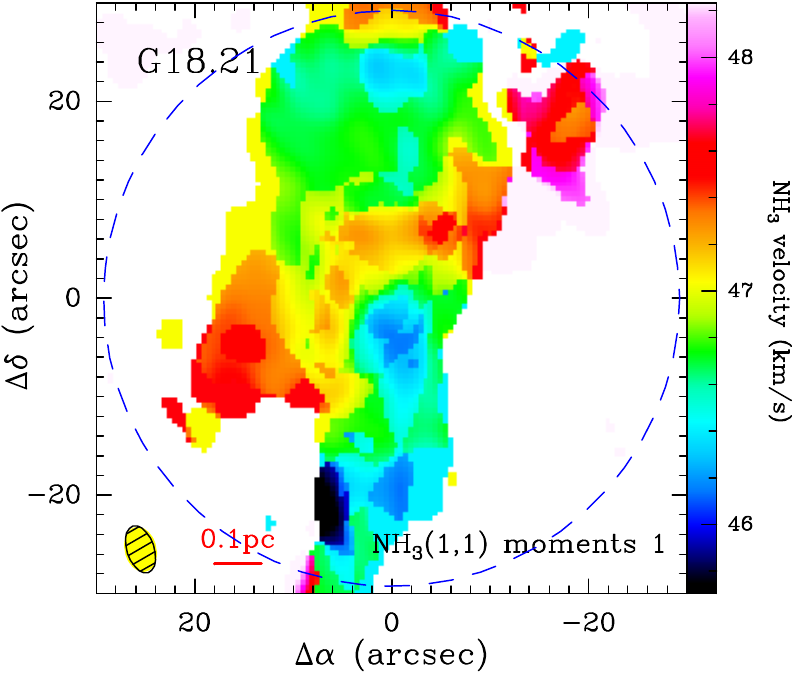}}
\subfigure[]{\includegraphics[width=0.33\textwidth,angle=0]{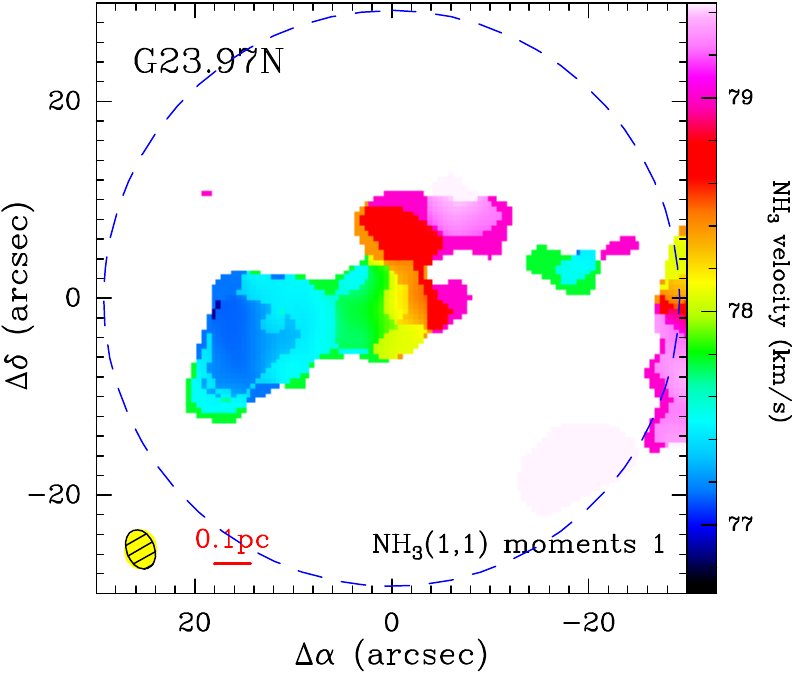}}
\subfigure[]{\includegraphics[width=0.33\textwidth,angle=0]{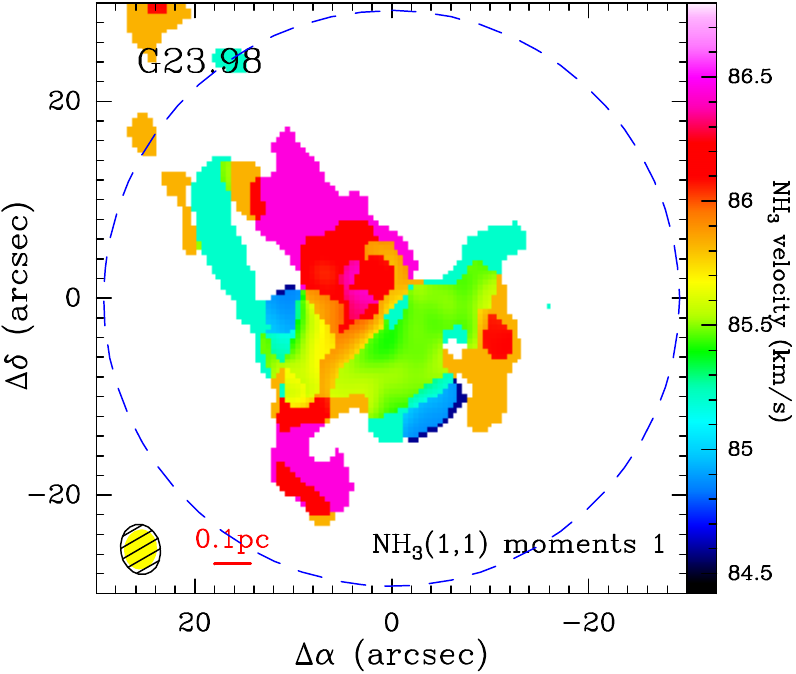}}
\subfigure[]{\includegraphics[width=0.33\textwidth,angle=0]{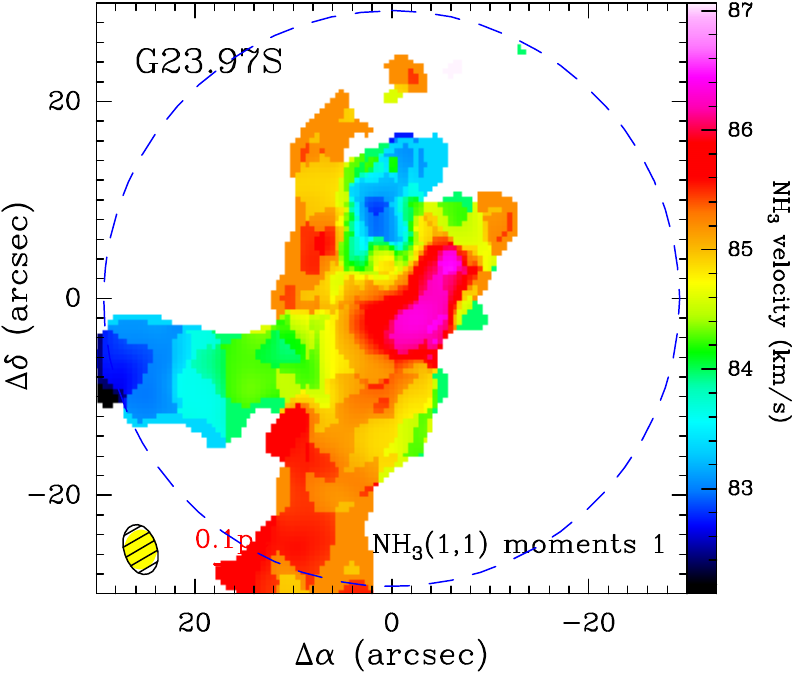}}
\subfigure[]{\includegraphics[width=0.33\textwidth,angle=0]{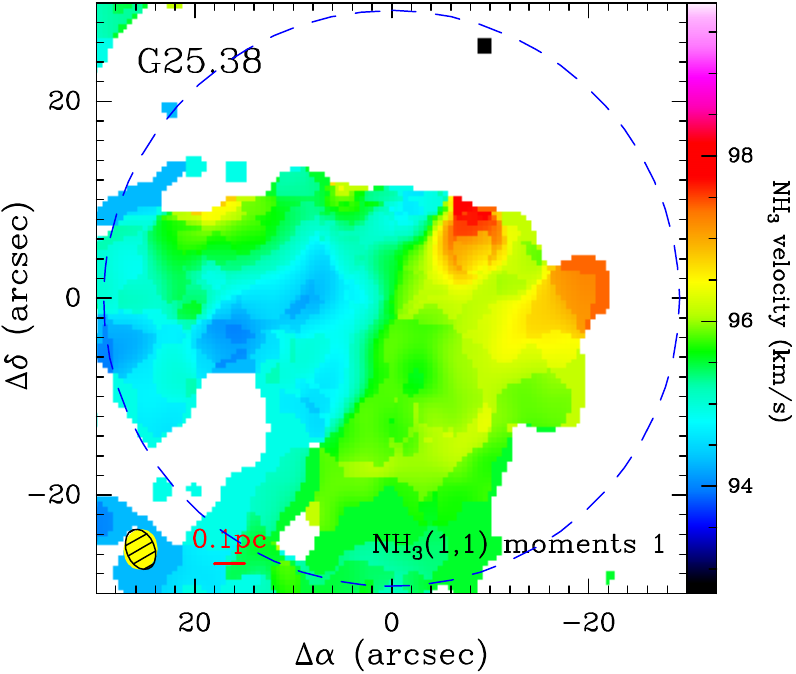}}
\subfigure[]{\includegraphics[width=0.33\textwidth,angle=0]{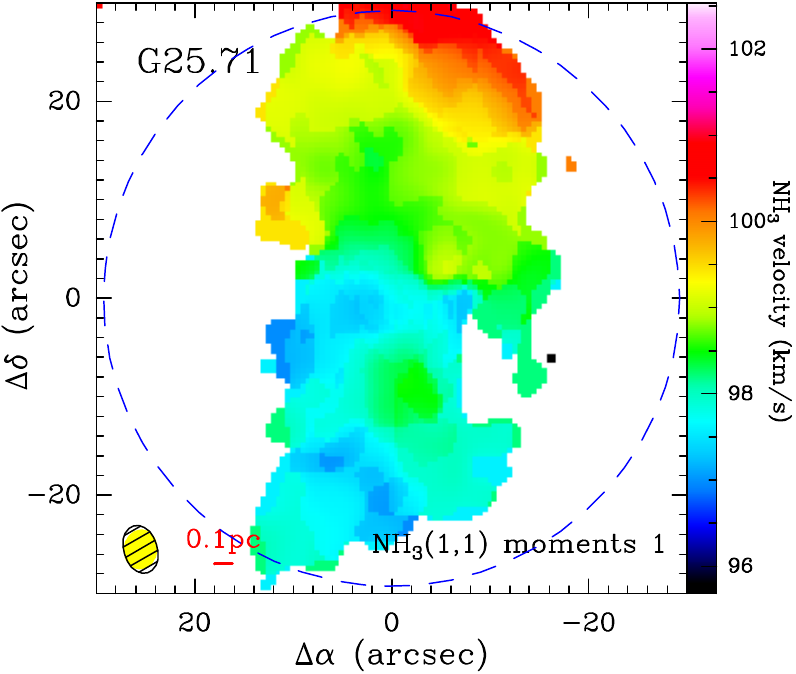}}
\caption{Velocity distribution (moment\,1) of NH$_3$ (1,\,1) line. The synthesized beam size of each subfigure is indicated at the bottom-left corner.}
\label{Fig_mom1_appendix}
\end{figure*}

\begin{figure*}
\centering
\subfigure[]{\includegraphics[width=0.33\textwidth,angle=0]{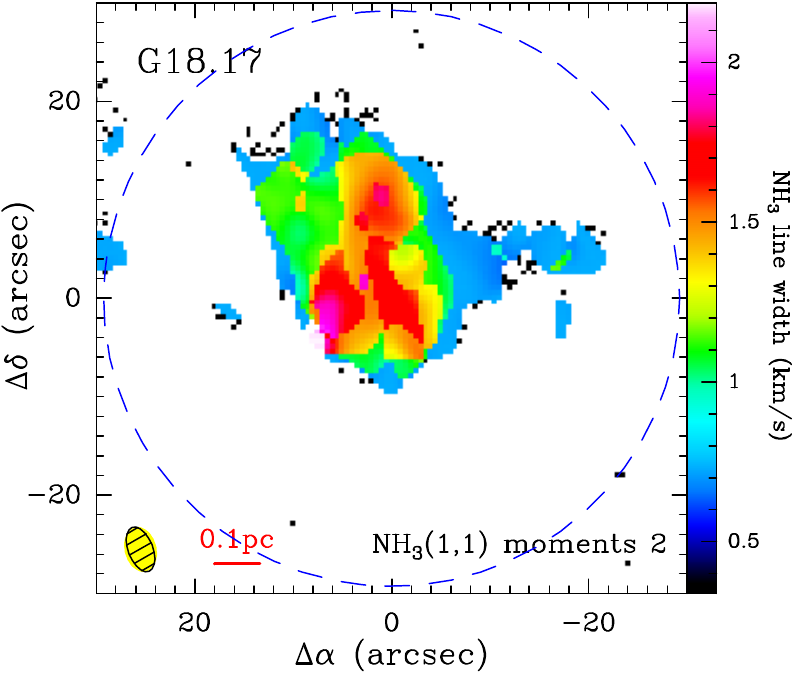}}
\subfigure[]{\includegraphics[width=0.33\textwidth,angle=0]{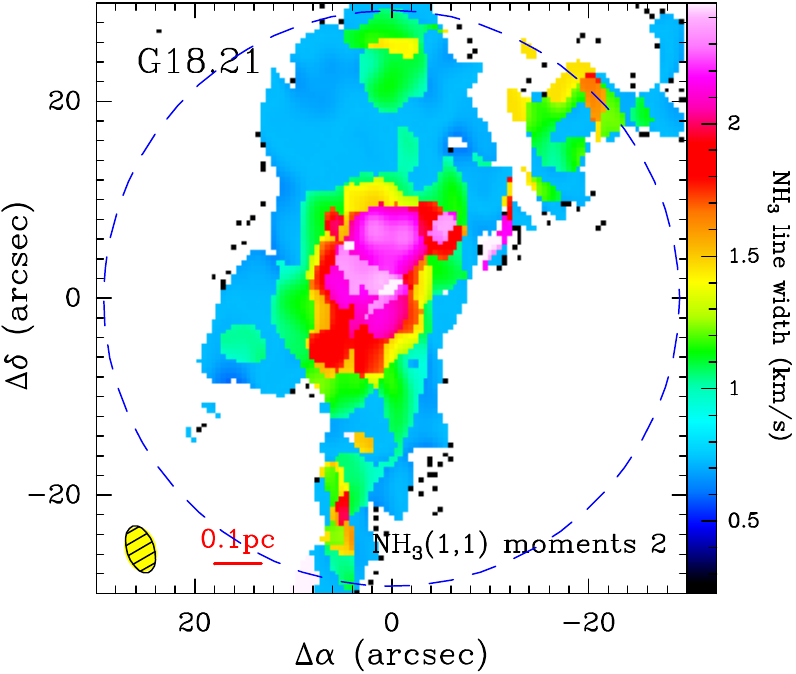}}
\subfigure[]{\includegraphics[width=0.33\textwidth,angle=0]{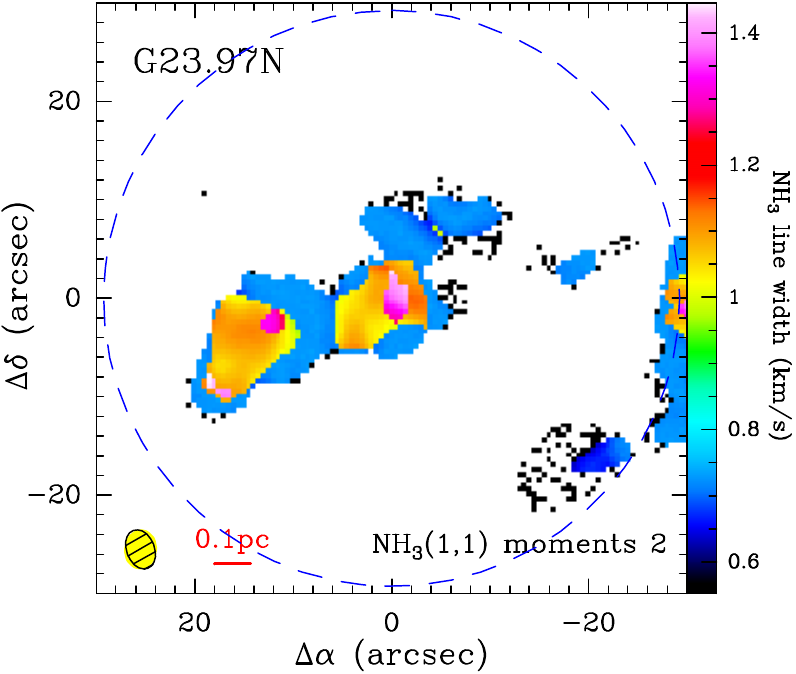}}
\subfigure[]{\includegraphics[width=0.33\textwidth,angle=0]{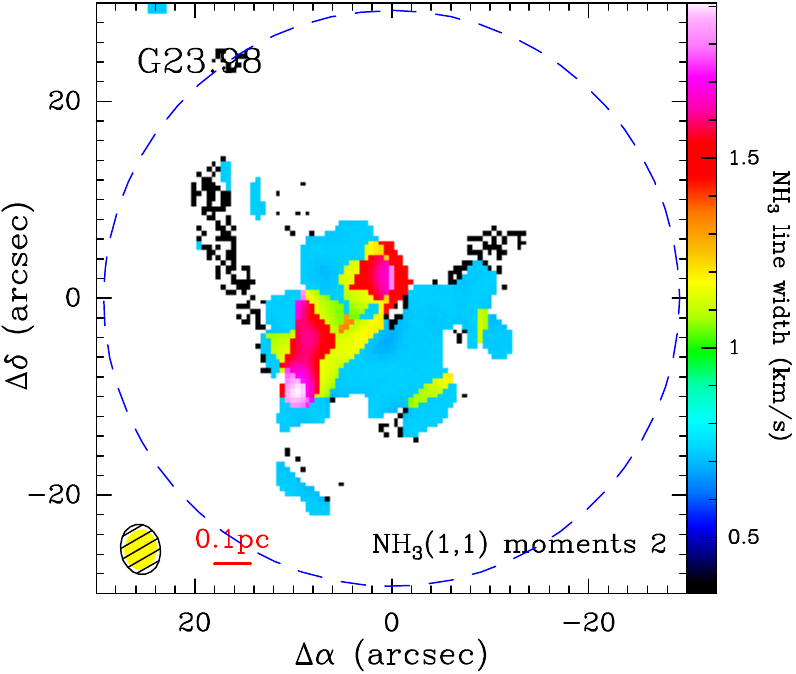}}
\subfigure[]{\includegraphics[width=0.33\textwidth,angle=0]{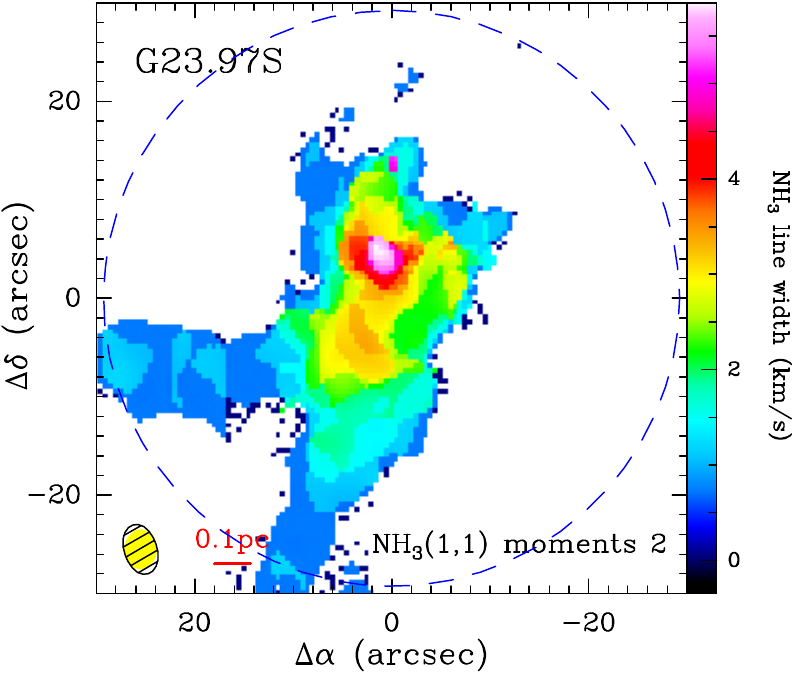}}
\subfigure[]{\includegraphics[width=0.33\textwidth,angle=0]{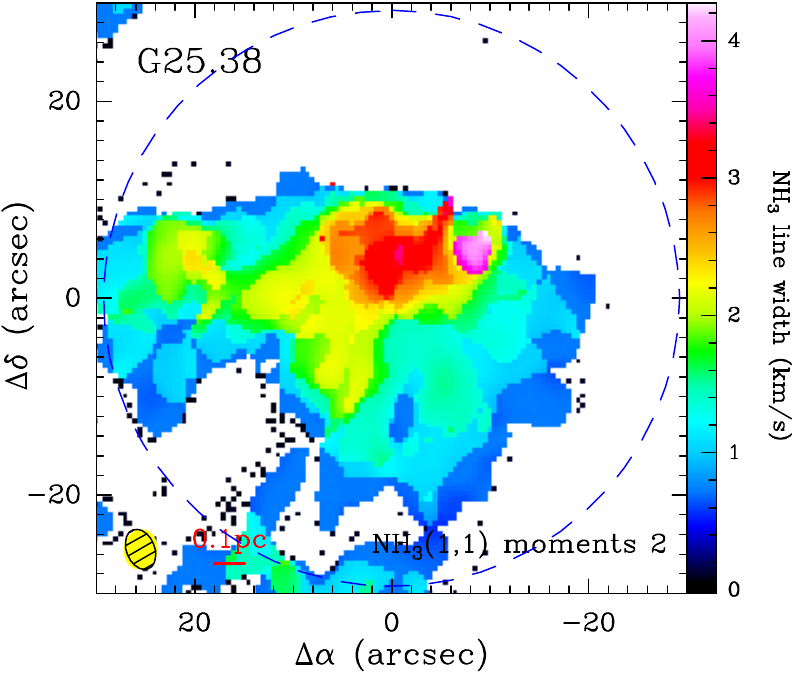}}
\subfigure[]{\includegraphics[width=0.33\textwidth,angle=0]{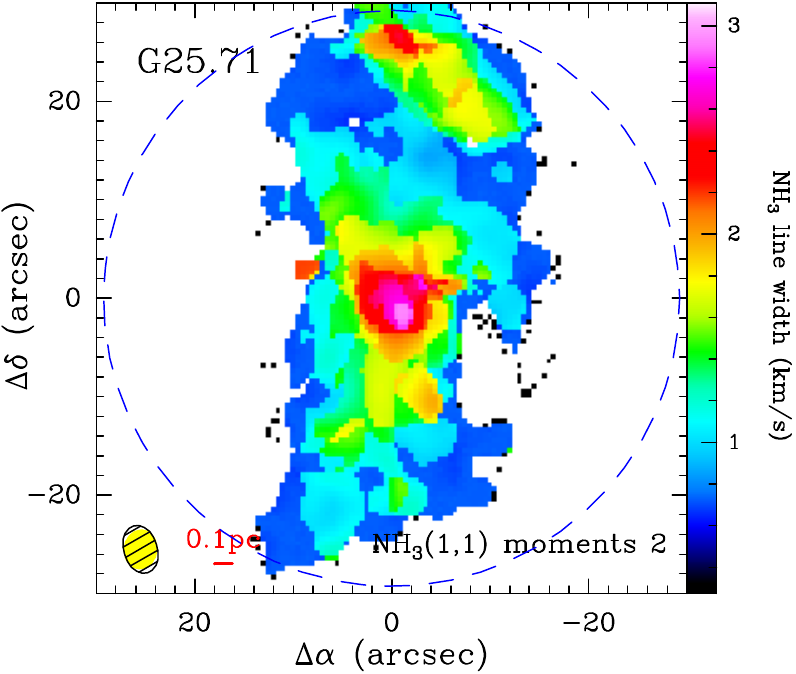}}
\caption{Velocity dispersion (moment\,2) of NH$_3$ (1,\,1) line. The synthesized beam size of each subfigure is indicated at the bottom-left corner.}
\label{Fig_mom2_appendix}
\end{figure*}

\begin{table*}[ht]
\caption{Clump parameters derived from 450, 850, and 870\,$\mu$m observations.}
\label{tab_atlasgal} \centering \scriptsize
\setlength{\tabcolsep}{1.8mm}{
\begin{tabular}{cccccccccccccc}
\hline \hline
Source	& No. &	Offsets	& FWHM & $R_{\rm eff}$ 	&  $T_{\rm kin}$ &	$I_{\rm
peak}$	&	$S_{\rm int}$	& ${\rm Mass}$	& $N_{\rm H_2}$	& $\Sigma$  & $n_{\rm H_2}$ & $\Delta v_{\rm NH_3\,(1,1)}$ & $M_{\rm vir}$
\\
&		&	$''\,,\,''$	& $''$  & pc  & K	&	$\jyb$	&	Jy
& $\Msun$ & $\rm 10^{23}cm^{-2}$  & $\rm g\,cm^{-2}$	& $\rm 10^{5}cm^{-3}$ & $\kms$ & $\Msun$	\\
\hline
\multicolumn{2}{r}{450\,$\mu$m} \\
G18.17 & 1 &$ ( -0.74 , 0.55 ) $&$ 25.54 $&$ 0.277 $&$ 16.9 \pm 1.3 $&$ 10.46 $&$ 128.75 $&$ 1793.7 \pm 303.3 $&$ 5.8 \pm 1.0 $&$ 1.5 \pm 0.3 $&$ 6.1 \pm 1.0 $&$ 1.40 \pm 0.06 $&$ 48.7 \pm 2.0 $ \\
G18.21 & 1 &$ ( -0.58 , -0.40 ) $&$ 17.87 $&$ 0.187 $&$ 13.1 \pm 0.8 $&$ 7.80 $&$ 47.35 $&$ 1133.7 \pm 176.4 $&$ 8.1 \pm 1.3 $&$ 2.1 \pm 0.3 $&$ 12.6 \pm 2.0 $&$ 1.35 \pm 0.06 $&$ 31.9 \pm 1.4 $ \\
G23.97N & 1 &$ ( 2.29 , -3.20 ) $&$ 29.07 $&$ 0.396 $&$ 14.8 \pm 1.0 $&$ 10.02 $&$ 157.61 $&$ 4667.5 \pm 806.8 $&$ 7.4 \pm 1.3 $&$ 2.0 \pm 0.3 $&$ 5.5 \pm 0.9 $&$ 1.49 \pm 0.06 $&$ 74.1 \pm 3.2 $ \\
G23.98 & 1 &$ ( 0.13 , -12.17 ) $&$ 37.51 $&$ 0.511 $&$ 13.4 \pm 0.6 $&$ 6.98 $&$ 213.77 $&$ 8208.8 \pm 1028.3 $&$ 7.9 \pm 1.0 $&$ 2.1 \pm 0.3 $&$ 4.5 \pm 0.6 $&$ 1.37 \pm 0.04 $&$ 87.7 \pm 2.8 $ \\
G23.44 & 1 &$ ( -0.02 , -2.10 ) $&$ 20.88 $&$ 0.357 $&$ 26.4 \pm 2.7 $&$ 41.24 $&$ 329.82 $&$ 4741.1 \pm 834.4 $&$ 9.3 \pm 1.6 $&$ 2.5 \pm 0.4 $&$ 7.6 \pm 1.3 $&$ 2.86 \pm 0.02 $&$ 128.5 \pm 0.9 $ \\
G23.44 & 2 &$ ( -3.59 , 11.96 ) $&$ 9.86 $&$ 0.169 $&$ 30.7 \pm 4.4 $&$ 12.13 $&$ 26.50 $&$ 297.1 \pm 68.4 $&$ 2.6 \pm 0.6 $&$ 0.7 \pm 0.2 $&$ 4.5 \pm 1.0 $&$ 2.66 \pm 0.07 $&$ 56.5 \pm 1.5 $ \\
G23.97S & 1 &$ ( -0.98 , -1.90 ) $&$ 15.72 $&$ 0.215 $&$ 23.4 \pm 4.6 $&$ 14.05 $&$ 65.80 $&$ 751.3 \pm 271.7 $&$ 4.1 \pm 1.5 $&$ 1.1 \pm 0.4 $&$ 5.5 \pm 2.0 $&$ 2.93 \pm 0.07 $&$ 79.4 \pm 1.9 $ \\
G25.38 & 1 &$ ( -2.06 , -0.30 ) $&$ 21.96 $&$ 0.358 $&$ 23.2 \pm 2.0 $&$ 32.12 $&$ 282.88 $&$ 4639.2 \pm 740.4 $&$ 9.0 \pm 1.4 $&$ 2.4 \pm 0.4 $&$ 7.4 \pm 1.2 $&$ 2.31 \pm 0.04 $&$ 103.9 \pm 2.0 $ \\
G25.71 & 1 &$ ( -1.75 , -0.91 ) $&$ 15.89 $&$ 0.440 $&$ 19.6 \pm 1.1 $&$ 18.73 $&$ 92.51 $&$ 6083.6 \pm 689.9 $&$ 7.9 \pm 0.9 $&$ 2.1 \pm 0.2 $&$ 5.2 \pm 0.6 $&$ 1.88 \pm 0.03 $&$ 104.0 \pm 1.7 $ \\
G25.71 & 2 &$ ( -6.00 , 22.35 ) $&$ 14.10 $&$ 0.390 $&$ 19.1 \pm 2.2 $&$ 8.19 $&$ 33.97 $&$ 2336.9 \pm 564.5 $&$ 3.8 \pm 0.9 $&$ 1.0 \pm 0.2 $&$ 2.9 \pm 0.7 $&$ 1.84 \pm 0.04 $&$ 90.3 \pm 2.1 $ \\
\hline \multicolumn{2}{r}{850\,$\mu$m} \\
G18.17 & 1 &$ ( -0.58 , -0.39 ) $&$ 32.29 $&$ 0.351 $&$ 17.3 \pm 1.4 $&$ 2.10 $&$ 11.66 $&$ 1095.4 \pm 139.7 $&$ 2.2 \pm 0.3 $&$ 0.6 \pm 0.1 $&$ 1.8 \pm 0.2 $&$ 1.35 \pm 0.06 $&$ 59.2 \pm 2.4 $ \\
G18.21 & 1 &$ ( 0.32 , -0.51 ) $&$ 21.93 $&$ 0.230 $&$ 13.4 \pm 0.7 $&$ 1.60 $&$ 4.14 $&$ 554.5 \pm 48.5 $&$ 2.6 \pm 0.2 $&$ 0.7 \pm 0.1 $&$ 3.3 \pm 0.3 $&$ 1.29 \pm 0.06 $&$ 37.1 \pm 1.7 $ \\
G23.97N & 1 &$ ( 10.88 , -2.94 ) $&$ 21.90 $&$ 0.298 $&$ 14.1 \pm 0.4 $&$ 0.90 $&$ 2.37 $&$ 490.1 \pm 21.6 $&$ 1.4 \pm 0.1 $&$ 0.4 \pm 0.0 $&$ 1.3 \pm 0.1 $&$ 1.03 \pm 0.01 $&$ 38.5 \pm 0.3 $ \\
G23.98 & 1 &$ ( 0.65 , -6.65 ) $&$ 65.61 $&$ 0.894 $&$ 12.5 \pm 1.0 $&$ 1.07 $&$ 24.92 $&$ 6305.8 \pm 942.3 $&$ 2.0 \pm 0.3 $&$ 0.5 \pm 0.1 $&$ 0.6 \pm 0.1 $&$ 1.57 \pm 0.12 $&$ 176.4 \pm 13.2 $ \\
G23.44 & 1 &$ ( 0.07 , -0.72 ) $&$ 26.64 $&$ 0.456 $&$ 26.3 \pm 2.3 $&$ 7.20 $&$ 26.66 $&$ 3376.4 \pm 399.9 $&$ 4.1 \pm 0.5 $&$ 1.1 \pm 0.1 $&$ 2.6 \pm 0.3 $&$ 2.90 \pm 0.03 $&$ 166.2 \pm 1.5 $ \\
G23.97S & 1 &$ ( -0.74 , -0.01 ) $&$ 21.48 $&$ 0.294 $&$ 23.4 \pm 2.8 $&$ 3.32 $&$ 8.36 $&$ 792.7 \pm 133.6 $&$ 2.3 \pm 0.4 $&$ 0.6 \pm 0.1 $&$ 2.3 \pm 0.4 $&$ 3.16 \pm 0.07 $&$ 117.1 \pm 2.7 $ \\
G23.97S & 2 &$ ( 9.60 , -25.16 ) $&$ 23.19 $&$ 0.317 $&$ 15.5 \pm 1.2 $&$ 0.67 $&$ 2.41 $&$ 426.9 \pm 55.5 $&$ 1.1 \pm 0.1 $&$ 0.3 \pm 0.0 $&$ 1.0 \pm 0.1 $&$ 1.34 \pm 0.04 $&$ 53.2 \pm 1.5 $ \\
G25.38 & 1 &$ ( 0.11 , -0.39 ) $&$ 28.01 $&$ 0.457 $&$ 22.0 \pm 1.8 $&$ 4.92 $&$ 19.70 $&$ 2904.9 \pm 332.4 $&$ 3.5 \pm 0.4 $&$ 0.9 \pm 0.1 $&$ 2.2 \pm 0.3 $&$ 2.22 \pm 0.05 $&$ 127.2 \pm 2.8 $ \\
G25.71 & 1 &$ ( 0.22 , -2.80 ) $&$ 22.97 $&$ 0.635 $&$ 19.0 \pm 0.9 $&$ 2.58 $&$ 8.30 $&$ 4351.0 \pm 322.2 $&$ 2.7 \pm 0.2 $&$ 0.7 \pm 0.1 $&$ 1.2 \pm 0.1 $&$ 1.74 \pm 0.02 $&$ 138.9 \pm 1.8 $ \\
G25.71 & 2 &$ ( -6.01 , 24.50 ) $&$ 21.49 $&$ 0.594 $&$ 20.0 \pm 2.9 $&$ 0.76 $&$ 2.51 $&$ 1223.5 \pm 261.0 $&$ 0.9 \pm 0.2 $&$ 0.2 \pm 0.0 $&$ 0.4 \pm 0.1 $&$ 2.02 \pm 0.05 $&$ 151.0 \pm 3.8 $ \\
\hline \multicolumn{2}{r}{870\,$\mu$m} \\
G18.17 & 1 &$ ( 2.15 , 0.47 ) $&$ 27.47 $&$ 0.298 $&$ 16.9 \pm 1.2 $&$ 1.91 $&$ 3.91 $&$ 387.3 \pm 43.9 $&$ 1.1 \pm 0.1 $&$ 0.3 \pm 0.0 $&$ 1.1 \pm 0.1 $&$ 1.39 \pm 0.05 $&$ 52.2 \pm 2.0 $ \\
G18.21 & 1 &$ ( 2.36 , 1.64 ) $&$ 28.28 $&$ 0.296 $&$ 13.4 \pm 0.5 $&$ 1.70 $&$ 3.70 $&$ 501.8 \pm 31.5 $&$ 1.4 \pm 0.1 $&$ 0.4 \pm 0.0 $&$ 1.4 \pm 0.1 $&$ 1.19 \pm 0.04 $&$ 44.1 \pm 1.7 $ \\
G23.97N & 1 &$ ( 1.27 , -2.44 ) $&$ 23.95 $&$ 0.326 $&$ 15.1 \pm 1.1 $&$ 0.78 $&$ 1.22 $&$ 227.5 \pm 27.3 $&$ 0.5 \pm 0.1 $&$ 0.1 \pm 0.0 $&$ 0.5 \pm 0.1 $&$ 1.47 \pm 0.06 $&$ 60.4 \pm 2.6 $ \\
G23.98 & 1 &$ ( 0.23 , -3.16 ) $&$ 25.89 $&$ 0.353 $&$ 14.3 \pm 0.6 $&$ 0.70 $&$ 1.26 $&$ 258.1 \pm 19.2 $&$ 0.5 \pm 0.0 $&$ 0.1 \pm 0.0 $&$ 0.4 \pm 0.0 $&$ 1.40 \pm 0.02 $&$ 61.8 \pm 0.9 $ \\
G23.44 & 1 &$ ( -3.76 , -0.61 ) $&$ 27.59 $&$ 0.472 $&$ 27.1 \pm 2.7 $&$ 5.85 $&$ 12.09 $&$ 1498.6 \pm 196.1 $&$ 1.7 \pm 0.2 $&$ 0.4 \pm 0.1 $&$ 1.0 \pm 0.1 $&$ 2.94 \pm 0.02 $&$ 174.9 \pm 1.4 $ \\
G23.97S & 1 &$ ( 0.66 , -0.59 ) $&$ 28.41 $&$ 0.389 $&$ 22.7 \pm 2.1 $&$ 3.20 $&$ 7.00 $&$ 708.3 \pm 94.3 $&$ 1.2 \pm 0.2 $&$ 0.3 \pm 0.0 $&$ 0.9 \pm 0.1 $&$ 3.02 \pm 0.06 $&$ 148.0 \pm 2.7 $ \\
G25.38 & 1 &$ ( 0.78 , -1.71 ) $&$ 27.56 $&$ 0.449 $&$ 21.5 \pm 1.6 $&$ 3.39 $&$ 6.98 $&$ 1080.6 \pm 115.2 $&$ 1.3 \pm 0.1 $&$ 0.4 \pm 0.0 $&$ 0.9 \pm 0.1 $&$ 2.20 \pm 0.05 $&$ 124.5 \pm 2.9 $ \\
G25.71 & 1 &$ ( -0.99 , 2.06 ) $&$ 29.66 $&$ 0.820 $&$ 18.8 \pm 1.2 $&$ 1.76 $&$ 4.19 $&$ 2279.1 \pm 225.6 $&$ 0.8 \pm 0.1 $&$ 0.2 \pm 0.0 $&$ 0.3 \pm 0.0 $&$ 1.75 \pm 0.03 $&$ 180.9 \pm 3.4 $ \\
\hline
\end{tabular}}
\end{table*}

\begin{table*}[ht]
\caption{Fragment parameters derived from PdBI 3.5\,mm observations.}
\label{tab_3.5mm} \centering \scriptsize
\setlength{\tabcolsep}{1.6mm}{
\begin{tabular}{cccccccccccccc}
\hline \hline
Source	& No. &	Offsets	& FWHM & $R_{\rm eff}$ 	&  $T_{\rm kin}$ &	$I_{\rm
peak}$	&	$S_{\rm int}\,^\star$	& ${\rm Mass}$	& $N_{\rm H_2}$	& $\Sigma$  & $n_{\rm H_2}$ & $\Delta v_{\rm NH_3\,(1,1)}$ & $M_{\rm vir}$	\\
&		&	$''\,,\,''$	& $''$  & pc  & K	&	$\mjyb$	&	mJy
& $\Msun$ & $\rm 10^{23}cm^{-2}$  & $\rm g\,cm^{-2}$	& $\rm 10^{6}cm^{-3}$ & $\kms$ & $\Msun$	\\
\hline
\multicolumn{3}{r}{CD configuration observations} \\
G18.17 & 1 &$ ( 3.96 , 1.14 ) $&$ 4.13 $&$ 0.045 $&$ 16.2 \pm 1.9 $&$ 1.99 $&$ 3.44 $&$ 35.4 \pm 4.7 $&$ 4.4 \pm 0.6 $&$ 1.2 \pm 0.2 $&$ 2.9 \pm 0.4 $&$ 1.32 \pm 0.04 $&$ 7.4 \pm 0.2 $ \\
G18.17 & 2 &$ ( -1.05 , -0.69 ) $&$ 4.13 $&$ 0.045 $&$ 18.8 \pm 2.2 $&$ 1.73 $&$ 3.41 $&$ 29.7 \pm 4.0 $&$ 3.7 \pm 0.5 $&$ 1.0 \pm 0.1 $&$ 2.4 \pm 0.3 $&$ 1.40 \pm 0.06 $&$ 7.9 \pm 0.3 $ \\
G18.17 & 3 &$ ( 8.06 , 10.63 ) $&$ 3.25 $&$ 0.035 $&$ 16.6 \pm 1.3 $&$ 1.55 $&$ 1.76 $&$ 17.6 \pm 1.5 $&$ 3.5 \pm 0.3 $&$ 0.9 \pm 0.1 $&$ 2.9 \pm 0.3 $&$ 1.07 \pm 0.07 $&$ 4.7 \pm 0.3 $ \\
G18.21 & 1 &$ ( 0.39 , 5.62 ) $&$ 4.98 $&$ 0.052 $&$ 14.6 \pm 1.9 $&$ 1.21 $&$ 2.51 $&$ 27.0 \pm 4.0 $&$ 2.5 \pm 0.4 $&$ 0.7 \pm 0.1 $&$ 1.4 \pm 0.2 $&$ 1.49 \pm 0.06 $&$ 9.8 \pm 0.4 $ \\
G18.21 & 2 &$ ( 3.06 , 1.77 ) $&$ 4.29 $&$ 0.045 $&$ 14.4 \pm 1.4 $&$ 0.85 $&$ 1.39 $&$ 15.3 \pm 1.7 $&$ 1.9 \pm 0.2 $&$ 0.5 \pm 0.1 $&$ 1.2 \pm 0.1 $&$ 1.65 \pm 0.04 $&$ 9.3 \pm 0.2 $ \\
G23.97N & 1 &$ ( -0.74 , -0.49 ) $&$ 3.57 $&$ 0.049 $&$ 17.7 \pm 1.4 $&$ 2.19 $&$ 2.64 $&$ 38.7 \pm 3.4 $&$ 4.1 \pm 0.4 $&$ 1.1 \pm 0.1 $&$ 2.4 \pm 0.2 $&$ 1.28 \pm 0.06 $&$ 7.8 \pm 0.4 $ \\
G23.44 & 1$^{*}$ &$ ( 0.09 , -2.43 ) $&$ 4.08 $&$ 0.070 $&$ 40.1 \pm 17.3 $&$ 7.83 $&$ 11.01^{40.8\%} $&$ 42.9 \pm 19.4 $&$ 2.2 \pm 1.0 $&$ 0.6 \pm 0.3 $&$ 0.9 \pm 0.4 $&$ 3.66 \pm 0.03 $&$ 32.2 \pm 0.3 $ \\
G23.44 & 2 &$ ( -0.80 , 10.11 ) $&$ 4.55 $&$ 0.078 $&$ 31.7 \pm 5.6 $&$ 4.65 $&$ 8.46 $&$ 103.7 \pm 19.6 $&$ 4.3 \pm 0.8 $&$ 1.1 \pm 0.2 $&$ 1.6 \pm 0.3 $&$ 2.66 \pm 0.07 $&$ 26.0 \pm 0.7 $ \\
G23.44 & 3 &$ ( -4.15 , -3.94 ) $&$ 4.08 $&$ 0.070 $&$ 30.1 \pm 10.0 $&$ 1.67 $&$ 2.26 $&$ 29.3 \pm 10.4 $&$ 1.5 \pm 0.5 $&$ 0.4 \pm 0.1 $&$ 0.6 \pm 0.2 $&$ 2.60 \pm 0.10 $&$ 22.9 \pm 0.9 $ \\
G23.44 & 4 &$ ( 2.73 , -4.45 ) $&$ 3.61 $&$ 0.062 $&$ 46.7 \pm 18.1 $&$ 1.42 $&$ 1.75 $&$ 14.2 \pm 5.8 $&$ 0.9 \pm 0.4 $&$ 0.2 \pm 0.1 $&$ 0.4 \pm 0.2 $&$ 3.83 \pm 0.10 $&$ 29.8 \pm 0.8 $ \\
G23.44 & 5 &$ ( -0.11 , 2.10 ) $&$ 4.89 $&$ 0.084 $&$ 27.3 \pm 7.2 $&$ 1.43 $&$ 2.96 $&$ 42.6 \pm 12.0 $&$ 1.5 \pm 0.4 $&$ 0.4 \pm 0.1 $&$ 0.5 \pm 0.1 $&$ 3.23 \pm 0.04 $&$ 34.0 \pm 0.4 $ \\
G23.44 & 6 &$ ( -4.79 , 9.20 ) $&$ 3.62 $&$ 0.062 $&$ 27.5 \pm 4.5 $&$ 1.29 $&$ 1.52 $&$ 21.7 \pm 3.8 $&$ 1.4 \pm 0.2 $&$ 0.4 \pm 0.1 $&$ 0.7 \pm 0.1 $&$ 2.66 \pm 0.04 $&$ 20.7 \pm 0.3 $ \\
G23.97S & 1$^{*}$ &$ ( 1.13 , 4.45 ) $&$ 4.04 $&$ 0.055 $&$ 44.4 \pm 15.3 $&$ 9.43 $&$ 14.71^{21.7\%} $&$ 17.5 \pm 6.3 $&$ 1.4 \pm 0.5 $&$ 0.4 \pm 0.1 $&$ 0.8 \pm 0.3 $&$ 5.16 \pm 0.32 $&$ 35.9 \pm 2.2 $ \\
G23.97S & 2 &$ ( 2.60 , -17.24 ) $&$ 3.25 $&$ 0.044 $&$ 15.8 \pm 1.5 $&$ 1.76 $&$ 1.81 $&$ 30.4 \pm 3.3 $&$ 3.8 \pm 0.4 $&$ 1.0 \pm 0.1 $&$ 2.5 \pm 0.3 $&$ 1.43 \pm 0.13 $&$ 8.0 \pm 0.7 $ \\
G23.97S & 3 &$ ( -1.07 , -3.21 ) $&$ 3.52 $&$ 0.048 $&$ 18.2 \pm 3.3 $&$ 1.50 $&$ 1.83 $&$ 26.3 \pm 5.4 $&$ 2.8 \pm 0.6 $&$ 0.8 \pm 0.2 $&$ 1.7 \pm 0.3 $&$ 1.70 \pm 0.09 $&$ 10.3 \pm 0.6 $ \\
G23.97S & 4 &$ ( 3.04 , 0.42 ) $&$ 3.89 $&$ 0.053 $&$ 20.6 \pm 4.1 $&$ 1.44 $&$ 2.19 $&$ 27.4 \pm 6.0 $&$ 2.4 \pm 0.5 $&$ 0.6 \pm 0.1 $&$ 1.3 \pm 0.3 $&$ 3.00 \pm 0.04 $&$ 20.1 \pm 0.3 $ \\
G25.38 & 1 &$ ( -0.81 , 1.26 ) $&$ 4.74 $&$ 0.077 $&$ 32.2 \pm 6.1 $&$ 11.29 $&$ 22.36 $&$ 244.5 \pm 49.5 $&$ 10.2 \pm 2.1 $&$ 2.7 \pm 0.6 $&$ 3.9 \pm 0.8 $&$ 2.53 \pm 0.04 $&$ 24.5 \pm 0.4 $ \\
G25.38 & 2 &$ ( -4.01 , 21.42 ) $&$ 5.41 $&$ 0.088 $&$ 18.7 \pm 4.9 $&$ 7.88 $&$ 20.04 $&$ 396.3 \pm 115.8 $&$ 12.7 \pm 3.7 $&$ 3.4 \pm 1.0 $&$ 4.2 \pm 1.2 $&$ 2.53 \pm 0.32 $&$ 28.1 \pm 3.6 $ \\
G25.38 & 3 &$ ( 1.40 , 23.66 ) $&$ 4.62 $&$ 0.075 $&$ 19.0 \pm 1.9 $&$ 5.80 $&$ 11.57 $&$ 223.8 \pm 24.3 $&$ 9.9 \pm 1.1 $&$ 2.6 \pm 0.3 $&$ 3.8 \pm 0.4 $&$ 1.03 \pm 0.14 $&$ 9.7 \pm 1.3 $ \\
G25.38 & 4 &$ ( -4.68 , -21.99 ) $&$ 4.44 $&$ 0.072 $&$ 14.2 \pm 0.8 $&$ 4.25 $&$ 8.86 $&$ 239.1 \pm 14.8 $&$ 11.4 \pm 0.7 $&$ 3.0 \pm 0.2 $&$ 4.6 \pm 0.3 $&$ 1.03 \pm 0.14 $&$ 9.3 \pm 1.3 $ \\
G25.38 & 5 &$ ( -2.28 , -3.24 ) $&$ 4.00 $&$ 0.065 $&$ 17.9 \pm 1.9 $&$ 4.13 $&$ 5.85 $&$ 121.2 \pm 14.4 $&$ 7.1 \pm 0.8 $&$ 1.9 \pm 0.2 $&$ 3.2 \pm 0.4 $&$ 1.24 \pm 0.04 $&$ 10.1 \pm 0.4 $ \\
G25.38 & 6 &$ ( 2.08 , 18.57 ) $&$ 3.86 $&$ 0.063 $&$ 24.4 \pm 9.6 $&$ 3.37 $&$ 4.32 $&$ 63.7 \pm 27.4 $&$ 4.0 \pm 1.7 $&$ 1.1 \pm 0.5 $&$ 1.9 \pm 0.8 $&$ 1.24 \pm 0.04 $&$ 9.7 \pm 0.4 $ \\
G25.38 & 7 &$ ( -3.94 , 16.02 ) $&$ 3.80 $&$ 0.062 $&$ 19.6 \pm 4.3 $&$ 3.08 $&$ 3.89 $&$ 72.9 \pm 17.9 $&$ 4.7 \pm 1.2 $&$ 1.3 \pm 0.3 $&$ 2.2 \pm 0.5 $&$ 2.02 \pm 0.37 $&$ 15.8 \pm 2.9 $ \\
G25.38 & 8 &$ ( -5.76 , -7.39 ) $&$ 4.70 $&$ 0.077 $&$ 17.4 \pm 1.7 $&$ 2.10 $&$ 4.21 $&$ 90.4 \pm 10.1 $&$ 3.8 \pm 0.4 $&$ 1.0 \pm 0.1 $&$ 1.5 \pm 0.2 $&$ 1.20 \pm 0.04 $&$ 11.5 \pm 0.3 $ \\
G25.71 & 1$^{*}$ &$ ( -5.76 , 27.18 ) $&$ 3.66 $&$ 0.101 $&$ 33.0 \pm 8.5 $&$ 11.47 $&$ 13.70^{11.6\%} $&$ 48.6 \pm 13.4 $&$ 1.2 \pm 0.3 $&$ 0.3 \pm 0.1 $&$ 0.3 \pm 0.1 $&$ 1.63 \pm 0.17 $&$ 20.6 \pm 2.2 $ \\
G25.71 & 2$^{*}$ &$ ( -0.27 , -0.65 ) $&$ 3.67 $&$ 0.102 $&$ 23.6 \pm 4.4 $&$ 11.04 $&$ 13.56^{16.1\%} $&$ 96.1 \pm 19.4 $&$ 2.3 \pm 0.5 $&$ 0.6 \pm 0.1 $&$ 0.7 \pm 0.1 $&$ 2.30 \pm 0.04 $&$ 29.3 \pm 0.6 $ \\
G25.71 & 3 &$ ( -4.05 , 23.64 ) $&$ 3.61 $&$ 0.100 $&$ 23.5 \pm 3.3 $&$ 2.11 $&$ 2.98 $&$ 131.5 \pm 20.3 $&$ 3.3 \pm 0.5 $&$ 0.9 \pm 0.1 $&$ 1.0 \pm 0.1 $&$ 1.69 \pm 0.06 $&$ 21.2 \pm 0.8 $ \\
\hline
\multicolumn{3}{r}{BCD configuration observations} \\
G23.44 & 1$^{*}$ &$ ( -0.05 , -2.81 ) $&$ 2.59 $&$ 0.044 $&$ 58.9 \pm 53.9 $&$ 4.80 $&$ 10.58^{40.8\%} $&$ 27.6 \pm 26.1 $&$ 3.5 \pm 3.3 $&$ 0.9 \pm 0.9 $&$ 2.3 \pm 2.2 $&$ 3.71 \pm 0.04 $&$ 20.7 \pm 0.2 $ \\
G23.44 & 2 &$ ( -0.80 , 10.48 ) $&$ 2.22 $&$ 0.038 $&$ 32.9 \pm 6.6 $&$ 2.89 $&$ 4.69 $&$ 55.2 \pm 11.8 $&$ 9.6 \pm 2.0 $&$ 2.5 \pm 0.5 $&$ 7.3 \pm 1.6 $&$ 2.52 \pm 0.08 $&$ 12.0 \pm 0.4 $ \\
G23.44 & 3 &$ ( -1.10 , 7.91 ) $&$ 2.04 $&$ 0.035 $&$ 29.5 \pm 4.5 $&$ 1.21 $&$ 1.60 $&$ 21.1 \pm 3.5 $&$ 4.3 \pm 0.7 $&$ 1.2 \pm 0.2 $&$ 3.6 \pm 0.6 $&$ 2.63 \pm 0.08 $&$ 11.5 \pm 0.4 $ \\
G23.44 & 4 &$ ( 0.86 , -0.13 ) $&$ 2.01 $&$ 0.034 $&$ 27.6 \pm 5.5 $&$ 1.28 $&$ 1.59 $&$ 22.6 \pm 4.8 $&$ 4.8 \pm 1.0 $&$ 1.3 \pm 0.3 $&$ 4.0 \pm 0.9 $&$ 3.66 \pm 0.03 $&$ 15.8 \pm 0.1 $ \\
G23.44 & 5 &$ ( -4.03 , 8.96 ) $&$ 2.21 $&$ 0.038 $&$ 27.8 \pm 3.8 $&$ 1.09 $&$ 1.71 $&$ 24.0 \pm 3.5 $&$ 4.2 \pm 0.6 $&$ 1.1 \pm 0.2 $&$ 3.2 \pm 0.5 $&$ 2.79 \pm 0.06 $&$ 13.3 \pm 0.3 $ \\
G23.44 & 6 &$ ( -0.10 , 2.04 ) $&$ 2.15 $&$ 0.037 $&$ 28.8 \pm 7.6 $&$ 0.82 $&$ 1.23 $&$ 16.7 \pm 4.8 $&$ 3.1 \pm 0.9 $&$ 0.8 \pm 0.2 $&$ 2.4 \pm 0.7 $&$ 3.35 \pm 0.05 $&$ 15.5 \pm 0.2 $ \\
G23.97S & 1$^{*}$ &$ ( 1.09 , 4.35 ) $&$ 2.67 $&$ 0.037 $&$ 37.3 \pm 9.7 $&$ 5.56 $&$ 12.22^{21.7\%} $&$ 17.5 \pm 4.8 $&$ 3.3 \pm 0.9 $&$ 0.9 \pm 0.2 $&$ 2.6 \pm 0.7 $&$ 5.36 \pm 0.33 $&$ 24.7 \pm 1.5 $ \\
G23.97S & 2 &$ ( 2.66 , 6.29 ) $&$ 1.96 $&$ 0.027 $&$ 34.8 \pm 7.9 $&$ 1.46 $&$ 1.72 $&$ 12.2 \pm 2.9 $&$ 4.2 \pm 1.0 $&$ 1.1 \pm 0.3 $&$ 4.6 \pm 1.1 $&$ 4.57 \pm 0.41 $&$ 15.4 \pm 1.4 $ \\
G23.97S & 3 &$ ( 3.10 , 0.10 ) $&$ 2.43 $&$ 0.033 $&$ 20.0 \pm 3.6 $&$ 0.81 $&$ 1.45 $&$ 18.7 \pm 3.7 $&$ 4.2 \pm 0.8 $&$ 1.1 \pm 0.2 $&$ 3.7 \pm 0.7 $&$ 2.95 \pm 0.04 $&$ 12.3 \pm 0.2 $ \\
G25.38 & 1 &$ ( -0.34 , 1.97 ) $&$ 2.44 $&$ 0.040 $&$ 34.8 \pm 7.4 $&$ 6.60 $&$ 11.78 $&$ 118.6 \pm 26.8 $&$ 18.7 \pm 4.2 $&$ 5.0 \pm 1.1 $&$ 13.7 \pm 3.1 $&$ 2.68 \pm 0.04 $&$ 13.4 \pm 0.2 $ \\
G25.38 & 2 &$ ( 2.09 , 24.60 ) $&$ 2.45 $&$ 0.040 $&$ 19.1 \pm 2.3 $&$ 4.32 $&$ 7.71 $&$ 149.0 \pm 20.3 $&$ 23.3 \pm 3.2 $&$ 6.2 \pm 0.8 $&$ 17.0 \pm 2.3 $&$ 1.11 \pm 0.10 $&$ 5.6 \pm 0.5 $ \\
G25.38 & 3 &$ ( -3.46 , 21.19 ) $&$ 3.72 $&$ 0.061 $&$ 18.3 \pm 1.2 $&$ 2.87 $&$ 11.87 $&$ 239.4 \pm 17.3 $&$ 16.3 \pm 1.2 $&$ 4.3 \pm 0.3 $&$ 7.8 \pm 0.6 $&$ 1.12 \pm 0.09 $&$ 8.5 \pm 0.7 $ \\
G25.38 & 4 &$ ( -0.10 , 27.37 ) $&$ 2.32 $&$ 0.038 $&$ 17.7 \pm 3.8 $&$ 2.71 $&$ 4.36 $&$ 91.7 \pm 22.1 $&$ 16.0 \pm 3.9 $&$ 4.3 \pm 1.0 $&$ 12.3 \pm 3.0 $&$ 1.14 \pm 0.07 $&$ 5.4 \pm 0.4 $ \\
G25.38 & 5 &$ ( -1.96 , -1.50 ) $&$ 3.46 $&$ 0.056 $&$ 23.4 \pm 3.5 $&$ 2.70 $&$ 9.99 $&$ 153.9 \pm 24.8 $&$ 12.1 \pm 1.9 $&$ 3.2 \pm 0.5 $&$ 6.2 \pm 1.0 $&$ 1.64 \pm 0.05 $&$ 11.6 \pm 0.4 $ \\
G25.38 & 6 &$ ( 0.43 , 21.97 ) $&$ 2.18 $&$ 0.036 $&$ 18.5 \pm 2.7 $&$ 2.62 $&$ 3.82 $&$ 76.5 \pm 12.6 $&$ 15.1 \pm 2.5 $&$ 4.0 \pm 0.7 $&$ 12.4 \pm 2.0 $&$ 1.15 \pm 0.07 $&$ 5.1 \pm 0.3 $ \\
G25.38 & 7 &$ ( 2.77 , 19.17 ) $&$ 2.49 $&$ 0.041 $&$ 17.7 \pm 2.0 $&$ 2.45 $&$ 4.60 $&$ 96.2 \pm 12.4 $&$ 14.6 \pm 1.9 $&$ 3.9 \pm 0.5 $&$ 10.5 \pm 1.4 $&$ 1.16 \pm 0.07 $&$ 5.9 \pm 0.4 $ \\
G25.38 & 8 &$ ( -3.31 , 3.02 ) $&$ 2.34 $&$ 0.038 $&$ 23.5 \pm 4.5 $&$ 2.46 $&$ 4.31 $&$ 66.2 \pm 13.7 $&$ 11.4 \pm 2.4 $&$ 3.0 \pm 0.6 $&$ 8.7 \pm 1.8 $&$ 2.47 \pm 0.04 $&$ 11.9 \pm 0.2 $ \\
G25.38 & 9 &$ ( -1.22 , 19.79 ) $&$ 2.17 $&$ 0.035 $&$ 17.7 \pm 3.1 $&$ 1.75 $&$ 2.59 $&$ 54.1 \pm 10.6 $&$ 10.8 \pm 2.1 $&$ 2.9 \pm 0.6 $&$ 8.9 \pm 1.7 $&$ 1.17 \pm 0.06 $&$ 5.2 \pm 0.2 $ \\
G25.38 & 10 &$ ( 6.51 , 15.51 ) $&$ 2.08 $&$ 0.034 $&$ 17.5 \pm 1.0 $&$ 1.57 $&$ 2.05 $&$ 43.7 \pm 2.7 $&$ 9.5 \pm 0.6 $&$ 2.5 \pm 0.2 $&$ 8.2 \pm 0.5 $&$ 1.18 \pm 0.04 $&$ 5.0 \pm 0.2 $ \\
G25.38 & 11 &$ ( -7.06 , 18.99 ) $&$ 2.49 $&$ 0.041 $&$ 17.7 \pm 3.5 $&$ 1.50 $&$ 3.33 $&$ 69.6 \pm 15.4 $&$ 10.6 \pm 2.3 $&$ 2.8 \pm 0.6 $&$ 7.6 \pm 1.7 $&$ 1.19 \pm 0.04 $&$ 6.0 \pm 0.2 $ \\
G25.38 & 12 &$ ( -5.11 , 23.95 ) $&$ 2.88 $&$ 0.047 $&$ 17.5 \pm 3.1 $&$ 1.45 $&$ 3.81 $&$ 81.0 \pm 16.1 $&$ 9.2 \pm 1.8 $&$ 2.4 \pm 0.5 $&$ 5.7 \pm 1.1 $&$ 1.14 \pm 0.07 $&$ 6.7 \pm 0.4 $ \\
G25.38 & 13 &$ ( 1.08 , -0.43 ) $&$ 2.27 $&$ 0.037 $&$ 26.8 \pm 3.2 $&$ 1.40 $&$ 2.14 $&$ 28.4 \pm 3.6 $&$ 5.2 \pm 0.7 $&$ 1.4 \pm 0.2 $&$ 4.1 \pm 0.5 $&$ 2.18 \pm 0.07 $&$ 10.1 \pm 0.3 $ \\
G25.38 & 14 &$ ( -4.51 , 18.23 ) $&$ 2.21 $&$ 0.036 $&$ 18.8 \pm 2.1 $&$ 1.33 $&$ 2.14 $&$ 41.9 \pm 5.3 $&$ 8.1 \pm 1.0 $&$ 2.1 \pm 0.3 $&$ 6.5 \pm 0.8 $&$ 1.12 \pm 0.04 $&$ 5.1 \pm 0.2 $ \\
G25.38 & 15 &$ ( -4.54 , -6.68 ) $&$ 2.23 $&$ 0.036 $&$ 16.7 \pm 1.8 $&$ 1.16 $&$ 1.76 $&$ 39.4 \pm 4.8 $&$ 7.5 \pm 0.9 $&$ 2.0 \pm 0.2 $&$ 6.0 \pm 0.7 $&$ 1.26 \pm 0.06 $&$ 5.8 \pm 0.3 $ \\
G25.38 & 16 &$ ( -4.95 , -3.56 ) $&$ 2.14 $&$ 0.035 $&$ 18.2 \pm 1.2 $&$ 1.32 $&$ 1.87 $&$ 38.1 \pm 2.8 $&$ 7.8 \pm 0.6 $&$ 2.1 \pm 0.2 $&$ 6.5 \pm 0.5 $&$ 1.18 \pm 0.09 $&$ 5.1 \pm 0.4 $ \\
G25.71 & 1$^{*}$ &$ ( -0.58 , -0.92 ) $&$ 1.89 $&$ 0.052 $&$ 25.3 \pm 5.9 $&$ 9.72 $&$ 10.86^{16.1\%} $&$ 71.2 \pm 17.8 $&$ 6.5 \pm 1.6 $&$ 1.7 \pm 0.4 $&$ 3.6 \pm 0.9 $&$ 2.45 \pm 0.06 $&$ 16.1 \pm 0.4 $ \\
G25.71 & 2$^{*}$ &$ ( -5.86 , 26.88 ) $&$ 2.18 $&$ 0.060 $&$ 31.6 \pm 13.7 $&$ 9.20 $&$ 13.10^{11.6\%} $&$ 48.8 \pm 22.6 $&$ 3.4 \pm 1.6 $&$ 0.9 \pm 0.4 $&$ 1.6 \pm 0.7 $&$ 1.77 \pm 0.16 $&$ 13.4 \pm 1.2 $ \\
G25.71 & 3 &$ ( -1.50 , -13.11 ) $&$ 2.06 $&$ 0.057 $&$ 17.7 \pm 1.8 $&$ 1.19 $&$ 1.52 $&$ 91.6 \pm 10.7 $&$ 7.1 \pm 0.8 $&$ 1.9 \pm 0.2 $&$ 3.6 \pm 0.4 $&$ 1.03 \pm 0.00 $&$ 7.3 \pm 0.0 $ \\
G25.71 & 4 &$ ( 1.73 , 0.31 ) $&$ 2.72 $&$ 0.075 $&$ 23.2 \pm 3.0 $&$ 1.17 $&$ 2.88 $&$ 128.9 \pm 18.4 $&$ 5.7 \pm 0.8 $&$ 1.5 \pm 0.2 $&$ 2.2 \pm 0.3 $&$ 1.98 \pm 0.04 $&$ 18.7 \pm 0.3 $ \\
\hline
\end{tabular}}
\begin{flushleft}
$\rm ^{(\star)}$ {The contributions from dust emission are following a measured total flux, e.g., 40.8\% of the 3.5\,mm flux for G23.44-1 is from dust emission. This was estimated by SED fitting (see Figure \ref{Fig_SED}).\\}
$\rm ^{(*)}$ {The sources G23.44-1, G23.97S-1, G25.71-1, and G25.71-2 are clearly associated with compact H\,II regions.}
\end{flushleft}
\end{table*}

\begin{table*}[ht]
\caption{Fragment parameters derived from PdBI 1.3\,mm observations.}
\label{tab_1.3mm} \centering \scriptsize
\setlength{\tabcolsep}{1.8mm}{
\begin{tabular}{cccccccccccccc}
\hline \hline
Source	& No. &	Offsets	& FWHM & $R_{\rm eff}$ 	&  $T_{\rm kin}$ &	$I_{\rm
peak}$	&	$S_{\rm int}$	& ${\rm Mass}$	& $N_{\rm H_2}$	& $\Sigma$  & $n_{\rm H_2}$ & $\Delta v_{\rm NH_3\,(1,1)}$ & $M_{\rm vir}$		 \\
&		&	$''\,,\,''$	& $''$  & pc  & K	&	$\mjyb$	&	mJy
& $\Msun$ & $\rm 10^{23}cm^{-2}$  & $\rm g\,cm^{-2}$	& $\rm 10^{6}cm^{-3}$ & $\kms$ & $\Msun$ 	\\
\hline
\multicolumn{3}{r}{CD configuration observations} \\
G18.17 & 1 &$ ( -1.79 , -4.01 ) $&$ 1.29 $&$ 0.014 $&$ 19.1 \pm 1.8 $&$ 10.81 $&$ 11.01 $&$ 3.8 \pm 0.5 $&$ 4.9 \pm 0.6 $&$ 1.3 \pm 0.2 $&$ 10.1 \pm 1.3 $&$ 1.52 \pm 0.07 $&$ 2.7 \pm 0.1 $ \\
G18.21 & 1 &$ ( 1.46 , 2.60 ) $&$ 1.55 $&$ 0.016 $&$ 15.4 \pm 1.9 $&$ 11.61 $&$ 13.72 $&$ 5.9 \pm 1.0 $&$ 5.6 \pm 0.9 $&$ 1.5 \pm 0.2 $&$ 10.0 \pm 1.7 $&$ 1.65 \pm 0.07 $&$ 3.4 \pm 0.1 $ \\
G23.97N & 1 &$ ( -1.19 , -0.99 ) $&$ 1.38 $&$ 0.019 $&$ 17.9 \pm 1.3 $&$ 12.39 $&$ 13.58 $&$ 8.1 \pm 0.8 $&$ 5.7 \pm 0.6 $&$ 1.5 \pm 0.2 $&$ 8.9 \pm 0.9 $&$ 1.25 \pm 0.06 $&$ 2.9 \pm 0.2 $ \\
G23.97N & 2 &$ ( 7.51 , -2.45 ) $&$ 1.67 $&$ 0.023 $&$ 25.2 \pm 1.9 $&$ 11.12 $&$ 16.07 $&$ 6.2 \pm 0.6 $&$ 3.0 \pm 0.3 $&$ 0.8 \pm 0.1 $&$ 3.8 \pm 0.4 $&$ 1.03 \pm 0.01 $&$ 2.9 \pm 0.0 $ \\
G23.97N & 3 &$ ( 3.02 , -0.99 ) $&$ 1.38 $&$ 0.019 $&$ 18.0 \pm 1.6 $&$ 8.64 $&$ 8.66 $&$ 5.1 \pm 0.6 $&$ 3.6 \pm 0.4 $&$ 1.0 \pm 0.1 $&$ 5.6 \pm 0.7 $&$ 1.03 \pm 0.01 $&$ 2.4 \pm 0.0 $ \\
G23.98 & 1 &$ ( 4.52 , -2.66 ) $&$ 1.31 $&$ 0.018 $&$ 16.6 \pm 2.1 $&$ 10.05 $&$ 10.07 $&$ 6.7 \pm 1.1 $&$ 5.2 \pm 0.9 $&$ 1.4 \pm 0.2 $&$ 8.5 \pm 1.4 $&$ 1.12 \pm 0.06 $&$ 2.5 \pm 0.1 $ \\
G23.44 & 1 &$ ( -0.07 , -3.26 ) $&$ 2.59 $&$ 0.044 $&$ 62.0 \pm 63.2 $&$ 33.35 $&$ 121.65 $&$ 26.5 \pm 29.3 $&$ 3.4 \pm 3.7 $&$ 0.9 \pm 1.0 $&$ 2.2 \pm 2.5 $&$ 3.73 \pm 0.04 $&$ 20.8 \pm 0.2 $ \\
G23.44 & 2 &$ ( 0.03 , 9.16 ) $&$ 2.00 $&$ 0.034 $&$ 33.2 \pm 8.6 $&$ 25.92 $&$ 54.78 $&$ 24.1 \pm 7.3 $&$ 5.1 \pm 1.6 $&$ 1.4 \pm 0.4 $&$ 4.4 \pm 1.3 $&$ 2.49 \pm 0.07 $&$ 10.7 \pm 0.3 $ \\
G23.44 & 3 &$ ( -2.88 , -3.21 ) $&$ 1.42 $&$ 0.024 $&$ 52.9 \pm 31.8 $&$ 18.85 $&$ 28.00 $&$ 7.2 \pm 4.8 $&$ 3.1 \pm 2.0 $&$ 0.8 \pm 0.5 $&$ 3.7 \pm 2.4 $&$ 2.90 \pm 0.09 $&$ 8.8 \pm 0.3 $ \\
G23.97S & 1 &$ ( 1.06 , 4.22 ) $&$ 2.15 $&$ 0.029 $&$ 38.0 \pm 10.4 $&$ 43.37 $&$ 105.43 $&$ 25.3 \pm 7.9 $&$ 7.3 \pm 2.3 $&$ 1.9 \pm 0.6 $&$ 7.2 \pm 2.3 $&$ 5.46 \pm 0.34 $&$ 20.2 \pm 1.3 $ \\
G23.97S & 2 &$ ( -0.60 , -7.25 ) $&$ 1.63 $&$ 0.022 $&$ 15.9 \pm 1.6 $&$ 9.59 $&$ 23.73 $&$ 16.8 \pm 2.3 $&$ 8.4 \pm 1.2 $&$ 2.2 \pm 0.3 $&$ 11.0 \pm 1.5 $&$ 1.67 \pm 0.09 $&$ 4.7 \pm 0.3 $ \\
G25.38 & 1 &$ ( -0.45 , 2.15 ) $&$ 1.62 $&$ 0.026 $&$ 36.2 \pm 9.1 $&$ 42.94 $&$ 65.93 $&$ 23.8 \pm 6.9 $&$ 8.5 \pm 2.5 $&$ 2.3 \pm 0.7 $&$ 9.4 \pm 2.7 $&$ 2.68 \pm 0.04 $&$ 8.9 \pm 0.1 $ \\
G25.38 & 2 &$ ( 0.58 , 1.33 ) $&$ 1.29 $&$ 0.021 $&$ 33.9 \pm 4.0 $&$ 16.87 $&$ 18.88 $&$ 7.3 \pm 1.0 $&$ 4.2 \pm 0.6 $&$ 1.1 \pm 0.2 $&$ 5.7 \pm 0.8 $&$ 2.69 \pm 0.06 $&$ 7.1 \pm 0.1 $ \\
G25.38 & 3 &$ ( -1.94 , 2.75 ) $&$ 1.45 $&$ 0.024 $&$ 28.6 \pm 10.3 $&$ 12.64 $&$ 18.91 $&$ 9.0 \pm 3.9 $&$ 4.0 \pm 1.7 $&$ 1.1 \pm 0.5 $&$ 4.9 \pm 2.1 $&$ 2.54 \pm 0.04 $&$ 7.6 \pm 0.1 $ \\
G25.38 & 4 &$ ( -3.77 , 3.19 ) $&$ 1.39 $&$ 0.023 $&$ 22.1 \pm 3.7 $&$ 11.89 $&$ 14.75 $&$ 9.6 \pm 2.0 $&$ 4.7 \pm 1.0 $&$ 1.2 \pm 0.3 $&$ 6.0 \pm 1.2 $&$ 2.40 \pm 0.05 $&$ 6.8 \pm 0.1 $ \\
G25.71 & 1 &$ ( -0.81 , -1.07 ) $&$ 1.49 $&$ 0.041 $&$ 26.1 \pm 6.7 $&$ 28.47 $&$ 34.99 $&$ 53.5 \pm 16.8 $&$ 7.9 \pm 2.5 $&$ 2.1 \pm 0.7 $&$ 5.6 \pm 1.7 $&$ 2.51 \pm 0.06 $&$ 13.0 \pm 0.3 $ \\
G25.71 & 2 &$ ( 0.91 , -0.47 ) $&$ 1.60 $&$ 0.044 $&$ 24.4 \pm 4.7 $&$ 11.04 $&$ 18.89 $&$ 31.3 \pm 7.4 $&$ 4.0 \pm 0.9 $&$ 1.1 \pm 0.3 $&$ 2.6 \pm 0.6 $&$ 2.13 \pm 0.04 $&$ 11.8 \pm 0.2 $ \\
G25.71 & 3 &$ ( 2.57 , 0.09 ) $&$ 1.45 $&$ 0.040 $&$ 22.9 \pm 2.7 $&$ 7.82 $&$ 9.59 $&$ 17.2 \pm 2.5 $&$ 2.7 \pm 0.4 $&$ 0.7 \pm 0.1 $&$ 1.9 \pm 0.3 $&$ 1.78 \pm 0.05 $&$ 9.0 \pm 0.3 $ \\
\hline \multicolumn{3}{r}{BCD configuration observations} \\
G23.44 & 1 &$ ( 0.05 , -2.78 ) $&$ 0.92 $&$ 0.016 $&$ 63.1 \pm 70.1 $&$ 17.72 $&$ 33.12 $&$ 7.1 \pm 8.5 $&$ 7.1 \pm 8.6 $&$ 1.9 \pm 2.3 $&$ 13.2 \pm 15.9 $&$ 3.75 \pm 0.05 $&$ 7.4 \pm 0.1 $ \\
G23.44 & 2 &$ ( -0.97 , 10.89 ) $&$ 0.75 $&$ 0.013 $&$ 32.5 \pm 5.9 $&$ 10.08 $&$ 12.57 $&$ 5.6 \pm 1.2 $&$ 8.6 \pm 1.8 $&$ 2.3 \pm 0.5 $&$ 19.4 \pm 4.1 $&$ 2.44 \pm 0.09 $&$ 3.9 \pm 0.1 $ \\
G23.44 & 3 &$ ( -3.09 , -3.44 ) $&$ 1.00 $&$ 0.017 $&$ 55.8 \pm 45.7 $&$ 9.06 $&$ 20.24 $&$ 4.9 \pm 4.4 $&$ 4.2 \pm 3.8 $&$ 1.1 \pm 1.0 $&$ 7.2 \pm 6.4 $&$ 2.81 \pm 0.13 $&$ 6.0 \pm 0.3 $ \\
G23.44 & 4 &$ ( -0.97 , -3.24 ) $&$ 0.90 $&$ 0.015 $&$ 34.4 \pm 11.7 $&$ 9.11 $&$ 16.95 $&$ 7.1 \pm 2.8 $&$ 7.5 \pm 3.0 $&$ 2.0 \pm 0.8 $&$ 14.2 \pm 5.6 $&$ 3.48 \pm 0.06 $&$ 6.8 \pm 0.1 $ \\
G23.44 & 5 &$ ( 0.09 , 8.82 ) $&$ 0.98 $&$ 0.017 $&$ 32.4 \pm 8.6 $&$ 7.57 $&$ 21.74 $&$ 9.8 \pm 3.1 $&$ 8.7 \pm 2.7 $&$ 2.3 \pm 0.7 $&$ 15.1 \pm 4.7 $&$ 2.52 \pm 0.07 $&$ 5.3 \pm 0.1 $ \\
G23.44 & 6 &$ ( 0.33 , -3.72 ) $&$ 1.11 $&$ 0.019 $&$ 58.3 \pm 58.7 $&$ 8.07 $&$ 23.06 $&$ 5.4 \pm 5.9 $&$ 3.7 \pm 4.1 $&$ 1.0 \pm 1.1 $&$ 5.7 \pm 6.3 $&$ 3.83 \pm 0.06 $&$ 9.2 \pm 0.2 $ \\
G23.44 & 7 &$ ( -2.16 , -2.53 ) $&$ 0.93 $&$ 0.016 $&$ 37.8 \pm 18.1 $&$ 8.19 $&$ 15.51 $&$ 5.9 \pm 3.2 $&$ 5.8 \pm 3.2 $&$ 1.5 \pm 0.8 $&$ 10.6 \pm 5.8 $&$ 3.13 \pm 0.09 $&$ 6.3 \pm 0.2 $ \\
G23.44 & 8 &$ ( -1.54 , 9.44 ) $&$ 0.84 $&$ 0.014 $&$ 33.5 \pm 6.7 $&$ 7.31 $&$ 12.15 $&$ 5.3 \pm 1.2 $&$ 6.4 \pm 1.5 $&$ 1.7 \pm 0.4 $&$ 12.9 \pm 3.0 $&$ 2.79 \pm 0.08 $&$ 5.1 \pm 0.1 $ \\
G23.44 & 9 &$ ( 0.51 , 10.26 ) $&$ 0.73 $&$ 0.012 $&$ 35.0 \pm 9.6 $&$ 6.65 $&$ 8.54 $&$ 3.5 \pm 1.1 $&$ 5.6 \pm 1.8 $&$ 1.5 \pm 0.5 $&$ 13.1 \pm 4.2 $&$ 2.21 \pm 0.08 $&$ 3.5 \pm 0.1 $ \\
G23.44 & 10 &$ ( 0.86 , -2.51 ) $&$ 0.75 $&$ 0.013 $&$ 42.5 \pm 21.3 $&$ 7.13 $&$ 9.08 $&$ 3.0 \pm 1.7 $&$ 4.6 \pm 2.6 $&$ 1.2 \pm 0.7 $&$ 10.3 \pm 5.9 $&$ 3.73 \pm 0.04 $&$ 6.0 \pm 0.1 $ \\
G23.44 & 11 &$ ( -0.70 , 8.86 ) $&$ 0.76 $&$ 0.013 $&$ 32.6 \pm 7.2 $&$ 6.54 $&$ 10.49 $&$ 4.7 \pm 1.2 $&$ 6.9 \pm 1.8 $&$ 1.8 \pm 0.5 $&$ 15.5 \pm 4.0 $&$ 2.64 \pm 0.08 $&$ 4.3 \pm 0.1 $ \\
G23.44 & 12 &$ ( -1.84 , 5.79 ) $&$ 0.90 $&$ 0.015 $&$ 27.1 \pm 5.0 $&$ 5.22 $&$ 9.05 $&$ 5.0 \pm 1.1 $&$ 5.3 \pm 1.2 $&$ 1.4 \pm 0.3 $&$ 10.1 \pm 2.2 $&$ 2.47 \pm 0.09 $&$ 4.8 \pm 0.2 $ \\
G23.44 & 13 &$ ( -0.54 , -4.29 ) $&$ 0.72 $&$ 0.012 $&$ 43.3 \pm 34.1 $&$ 4.85 $&$ 5.64 $&$ 1.8 \pm 1.6 $&$ 3.0 \pm 2.7 $&$ 0.8 \pm 0.7 $&$ 7.1 \pm 6.3 $&$ 3.63 \pm 0.09 $&$ 5.6 \pm 0.1 $ \\
G23.44 & 14 &$ ( -0.15 , 0.14 ) $&$ 0.77 $&$ 0.013 $&$ 34.8 \pm 11.9 $&$ 4.12 $&$ 5.33 $&$ 2.2 \pm 0.9 $&$ 3.2 \pm 1.3 $&$ 0.8 \pm 0.3 $&$ 7.0 \pm 2.8 $&$ 3.73 \pm 0.03 $&$ 6.2 \pm 0.1 $ \\
G23.44 & 15 &$ ( 1.66 , -2.46 ) $&$ 0.82 $&$ 0.014 $&$ 32.8 \pm 9.7 $&$ 4.23 $&$ 7.93 $&$ 3.5 \pm 1.2 $&$ 4.5 \pm 1.5 $&$ 1.2 \pm 0.4 $&$ 9.3 \pm 3.2 $&$ 3.57 \pm 0.07 $&$ 6.3 \pm 0.1 $ \\
G23.44 & 16 &$ ( 0.69 , -4.30 ) $&$ 0.74 $&$ 0.013 $&$ 57.0 \pm 42.8 $&$ 4.04 $&$ 5.04 $&$ 1.2 \pm 1.0 $&$ 1.9 \pm 1.5 $&$ 0.5 \pm 0.4 $&$ 4.3 \pm 3.5 $&$ 4.00 \pm 0.08 $&$ 6.4 \pm 0.1 $ \\
G23.97S & 1 &$ ( 0.94 , 4.09 ) $&$ 1.08 $&$ 0.015 $&$ 35.9 \pm 9.1 $&$ 19.29 $&$ 47.91 $&$ 12.3 \pm 3.6 $&$ 14.1 \pm 4.1 $&$ 3.7 \pm 1.1 $&$ 27.7 \pm 8.1 $&$ 5.58 \pm 0.29 $&$ 10.4 \pm 0.5 $ \\
G23.97S & 2 &$ ( 1.67 , 3.99 ) $&$ 0.82 $&$ 0.011 $&$ 37.5 \pm 12.3 $&$ 9.90 $&$ 18.76 $&$ 4.6 \pm 1.7 $&$ 9.1 \pm 3.4 $&$ 2.4 \pm 0.9 $&$ 23.5 \pm 8.8 $&$ 5.36 \pm 0.37 $&$ 7.6 \pm 0.5 $ \\
G23.97S & 3 &$ ( 0.56 , 5.09 ) $&$ 0.86 $&$ 0.012 $&$ 54.9 \pm 24.5 $&$ 9.53 $&$ 17.99 $&$ 2.9 \pm 1.4 $&$ 5.2 \pm 2.5 $&$ 1.4 \pm 0.7 $&$ 12.8 \pm 6.3 $&$ 4.76 \pm 0.29 $&$ 7.0 \pm 0.4 $ \\
G23.97S & 4 &$ ( 2.15 , 5.37 ) $&$ 0.78 $&$ 0.011 $&$ 38.3 \pm 9.8 $&$ 4.64 $&$ 6.15 $&$ 1.5 \pm 0.4 $&$ 3.2 \pm 0.9 $&$ 0.9 \pm 0.2 $&$ 8.7 \pm 2.6 $&$ 4.96 \pm 0.40 $&$ 6.7 \pm 0.5 $ \\
G23.97S & 5 &$ ( -0.55 , -7.14 ) $&$ 0.86 $&$ 0.012 $&$ 16.1 \pm 2.9 $&$ 4.32 $&$ 6.52 $&$ 4.5 \pm 1.1 $&$ 8.1 \pm 2.0 $&$ 2.2 \pm 0.5 $&$ 20.2 \pm 4.9 $&$ 1.65 \pm 0.10 $&$ 2.4 \pm 0.1 $ \\
G23.97S & 6 &$ ( 2.03 , 3.21 ) $&$ 0.71 $&$ 0.010 $&$ 40.2 \pm 22.9 $&$ 4.27 $&$ 4.84 $&$ 1.1 \pm 0.7 $&$ 2.9 \pm 1.9 $&$ 0.8 \pm 0.5 $&$ 8.6 \pm 5.6 $&$ 4.65 \pm 0.28 $&$ 5.7 \pm 0.3 $ \\
G23.97S & 7 &$ ( 5.06 , -0.80 ) $&$ 0.91 $&$ 0.012 $&$ 23.6 \pm 6.2 $&$ 3.86 $&$ 6.74 $&$ 2.8 \pm 0.9 $&$ 4.6 \pm 1.5 $&$ 1.2 \pm 0.4 $&$ 10.7 \pm 3.5 $&$ 2.75 \pm 0.17 $&$ 4.3 \pm 0.3 $ \\
G25.38 & 1 &$ ( -0.53 , 2.10 ) $&$ 1.19 $&$ 0.019 $&$ 36.9 \pm 10.5 $&$ 23.19 $&$ 73.52 $&$ 25.9 \pm 8.5 $&$ 17.2 \pm 5.7 $&$ 4.6 \pm 1.5 $&$ 25.8 \pm 8.5 $&$ 2.72 \pm 0.05 $&$ 6.6 \pm 0.1 $ \\
G25.38 & 2 &$ ( -3.51 , 3.15 ) $&$ 0.83 $&$ 0.014 $&$ 23.2 \pm 5.2 $&$ 11.11 $&$ 16.52 $&$ 10.1 \pm 2.8 $&$ 13.9 \pm 3.9 $&$ 3.7 \pm 1.0 $&$ 29.8 \pm 8.3 $&$ 2.45 \pm 0.05 $&$ 4.2 \pm 0.1 $ \\
G25.38 & 3 &$ ( 0.81 , 1.41 ) $&$ 0.94 $&$ 0.015 $&$ 33.6 \pm 4.6 $&$ 6.12 $&$ 14.81 $&$ 5.8 \pm 0.9 $&$ 6.2 \pm 1.0 $&$ 1.6 \pm 0.3 $&$ 11.7 \pm 1.9 $&$ 2.71 \pm 0.06 $&$ 5.2 \pm 0.1 $ \\
G25.38 & 4 &$ ( -7.31 , 1.91 ) $&$ 0.76 $&$ 0.012 $&$ 20.1 \pm 3.7 $&$ 5.85 $&$ 7.80 $&$ 5.7 \pm 1.3 $&$ 9.3 \pm 2.2 $&$ 2.5 \pm 0.6 $&$ 21.9 \pm 5.1 $&$ 1.96 \pm 0.07 $&$ 3.0 \pm 0.1 $ \\
G25.38 & 5 &$ ( -1.91 , 2.48 ) $&$ 0.74 $&$ 0.012 $&$ 29.5 \pm 12.5 $&$ 5.93 $&$ 7.33 $&$ 3.4 \pm 1.7 $&$ 5.8 \pm 2.9 $&$ 1.5 \pm 0.8 $&$ 13.9 \pm 7.0 $&$ 2.50 \pm 0.04 $&$ 3.8 \pm 0.1 $ \\
G25.38 & 6 &$ ( 0.28 , 2.46 ) $&$ 0.70 $&$ 0.011 $&$ 36.3 \pm 7.1 $&$ 4.08 $&$ 4.35 $&$ 1.6 \pm 0.4 $&$ 3.0 \pm 0.7 $&$ 0.8 \pm 0.2 $&$ 7.6 \pm 1.7 $&$ 2.88 \pm 0.05 $&$ 4.1 \pm 0.1 $ \\
G25.71 & 1 &$ ( -0.66 , -1.02 ) $&$ 0.77 $&$ 0.021 $&$ 25.1 \pm 6.4 $&$ 25.61 $&$ 32.77 $&$ 52.5 \pm 16.5 $&$ 28.9 \pm 9.1 $&$ 7.7 \pm 2.4 $&$ 39.5 \pm 12.4 $&$ 2.51 \pm 0.06 $&$ 6.7 \pm 0.2 $ \\
G25.71 & 2 &$ ( 0.27 , 6.90 ) $&$ 0.74 $&$ 0.020 $&$ 17.1 \pm 1.5 $&$ 5.15 $&$ 6.25 $&$ 16.3 \pm 2.0 $&$ 9.7 \pm 1.2 $&$ 2.6 \pm 0.3 $&$ 13.9 \pm 1.7 $&$ 1.25 \pm 0.06 $&$ 3.2 \pm 0.1 $ \\
G25.71 & 3 &$ ( 0.67 , -0.25 ) $&$ 0.86 $&$ 0.024 $&$ 24.7 \pm 5.8 $&$ 3.75 $&$ 6.03 $&$ 9.8 \pm 2.8 $&$ 4.3 \pm 1.2 $&$ 1.2 \pm 0.3 $&$ 5.3 \pm 1.5 $&$ 2.16 \pm 0.04 $&$ 6.5 \pm 0.1 $ \\
G25.71 & 4 &$ ( 2.06 , -0.21 ) $&$ 0.94 $&$ 0.026 $&$ 22.1 \pm 3.6 $&$ 3.74 $&$ 7.94 $&$ 14.8 \pm 3.1 $&$ 5.5 \pm 1.1 $&$ 1.5 \pm 0.3 $&$ 6.1 \pm 1.3 $&$ 1.84 \pm 0.05 $&$ 6.0 \pm 0.2 $ \\
G25.71 & 5 &$ ( -1.60 , -1.23 ) $&$ 0.86 $&$ 0.024 $&$ 23.7 \pm 5.4 $&$ 3.87 $&$ 6.37 $&$ 10.9 \pm 3.1 $&$ 4.8 \pm 1.4 $&$ 1.3 \pm 0.4 $&$ 5.9 \pm 1.7 $&$ 2.52 \pm 0.07 $&$ 7.5 \pm 0.2 $ \\
G25.71 & 6 &$ ( -0.69 , -0.14 ) $&$ 0.73 $&$ 0.020 $&$ 24.9 \pm 6.5 $&$ 3.90 $&$ 4.37 $&$ 7.1 \pm 2.3 $&$ 4.3 \pm 1.4 $&$ 1.2 \pm 0.4 $&$ 6.2 \pm 2.0 $&$ 2.43 \pm 0.04 $&$ 6.2 \pm 0.1 $ \\
\hline
\end{tabular}}
\end{table*}

\label{lastpage}
\end{document}